\documentclass[%
 reprint,
nofootinbib,
nobibnotes,
 amsmath,amssymb,
 aps,
floatfix,
]{revtex4-1}

\pdfoutput=1 

\usepackage{graphicx}
\DeclareGraphicsExtensions{.pdf}

\usepackage{dcolumn}
\usepackage{bm}
\usepackage{xspace}
\usepackage{multirow}
\usepackage{bigstrut}

\usepackage{subfigure}

\usepackage{lineno}

\usepackage{atlasphysics}
\usepackage{preprintcover}

\usepackage[hyperindex,breaklinks]{hyperref} 

\usepackage{preprintcover}  
\PreprintCoverPaperTitle{Measurement of Higgs boson production in the diphoton decay channel in $pp$ collisions 
 at center-of-mass energies of 7 and 8~TeV with the ATLAS detector}
\PreprintIdNumber{CERN-PH-EP-2014-198}  
\PreprintCoverAbstract{A 
measurement of the production processes of the recently discovered Higgs boson 
is performed in the two-photon final state using \lumiseven~\ifb\ of proton--proton 
collisions data at $\sqrt{s}=7 \TeV$ and \lumieight~\ifb\ 
at $\sqrt{s}=8 \TeV$ collected by the ATLAS detector at the Large Hadron Collider. 
The number of observed Higgs boson decays to diphotons divided by the corresponding Standard Model prediction, 
called the signal strength, 
is found to be ${\combmucompactS}$ at the \combmassname, ${\mH = \combmass}$ \GeV.
The analysis is optimized to measure the signal strengths for individual Higgs boson production processes at
this value of \mH. They are found to be ${\mugghC}$, ${\muvbfC}$, ${\muwhC}$, ${\muzhC}$, and ${\mutthC}$,
for Higgs boson production through gluon fusion, vector-boson fusion, and in association with a $W$ or $Z$ boson 
or a top-quark pair, respectively.
Compared with the previously published ATLAS analysis, the results 
reported here 
also benefit from a new energy calibration
procedure for photons and the subsequent
reduction of the systematic uncertainty on the diphoton mass resolution.  
No significant deviations from the predictions of the Standard Model are found.
}  
\PreprintJournalName{Phys.~Rev.~D}

\newcommand{\invfb}{\ensuremath{\mathrm{fb^{-1}}}}
\renewcommand{\ifb}{\invfb}

\newcommand{\ilumi}{\ensuremath{\mathrm{cm}^{-2}~\mathrm{s}^{-1}}}

\newcommand{\combmass}{\ensuremath{125.4}}
\newcommand{\combmassname}{value of the Higgs boson mass measured by ATLAS}

\newcommand{\lumieight}{20.3}
\newcommand{\lumiseven}{4.5}

\newcommand{\mugghS}{\ensuremath{ \mu_{\ggH}  &  = 1.32 \ \pm 0.32  \ \mathrm{(stat.)} \ ^{+0.13}_{-0.09} \  \mathrm{(syst.)} \ ^{+0.19}_{-0.11} \ \mathrm{(theory)} }}
\newcommand{\muvbfS}{\ensuremath{ \mu_{\VBF}  &  = 0.8   \ \pm 0.7  \ \mathrm{(stat.)} \ ^{+0.2}_{-0.1} \ \mathrm{(syst.)} \ ^{+0.2}_{-0.3} \ \mathrm{(theory)} }}
\newcommand{\muwhS}{\ensuremath{ \mu_{\WH}   & =  1.0   \pm 1.5  \ \mathrm{(stat.)} \ ^{+0.3}_{-0.1} \ \mathrm{(syst.)} \ ^{+0.2}_{-0.1} \ \mathrm{(theory)} }}
\newcommand{\muzhS}{\ensuremath{ \mu_{\ZH}     & = 0.1   \ ^{+3.6}_{-0.1} \ \mathrm{(stat.)} \ ^{+0.7}_{-0.0} \ \mathrm{(syst.)} \ ^{+0.1}_{-0.0}  \ \mathrm{(theory)} }} 
\newcommand{\mutthS}{\ensuremath{ \mu_{\ttH}     & = 1.6   \ ^{+2.6}_{-1.8} \ \mathrm{(stat.)} \ ^{+0.6}_{-0.4} \ \mathrm{(syst.)} \ ^{+0.5}_{-0.2}  \ \mathrm{(theory)} }}

\newcommand{\mugghSS}{\ensuremath{ & = 1.32 \pm 0.38 }}
\newcommand{\muvbfSS}{\ensuremath{ & = 0.8   \pm 0.7  }}
\newcommand{\muwhSS}{\ensuremath{  &= 1.0    \pm 1.6 }}
\newcommand{\muzhSS}{\ensuremath{  &= 0.1    \ ^{+3.7}_{-0.1} }}
\newcommand{\mutthSS}{\ensuremath{ &= 1.6    \ ^{+2.7}_{-1.8} }}

\newcommand{\mugghT}{\ensuremath{ \mu_{\ggH} &= 1.32\pm 0.38}}
\newcommand{\muvbfT}{\ensuremath{ \mu_{\VBF} &= 0.8 \pm 0.7}}
\newcommand{\muwhT}{\ensuremath{ \mu_{\WH}   &= 1.0 \pm 1.6 }}
\newcommand{\muzhT}{\ensuremath{ \mu_{\ZH}   &= 0.1 \ ^{+3.7}_{-0.1} }}
\newcommand{\mutthT}{\ensuremath{ \mu_{\ttH} &= 1.6 \ ^{+2.7}_{-1.8} }}

\newcommand{\mugghC}{\ensuremath{ \mu_{\ggH} = 1.32\pm 0.38}}
\newcommand{\muvbfC}{\ensuremath{ \mu_{\VBF} = 0.8 \pm 0.7}}
\newcommand{\muwhC}{\ensuremath{ \mu_{\WH}   = 1.0 \pm 1.6 }}
\newcommand{\muzhC}{\ensuremath{ \mu_{\ZH}   = 0.1 \ ^{+3.7}_{-0.1} }}
\newcommand{\mutthC}{\ensuremath{ \mu_{\ttH} = 1.6 \ ^{+2.7}_{-1.8} }}

\newcommand{\combmu}{\ensuremath{ \mu & =  1.17 \pm 0.23 \ (\mathrm{stat.})  \ ^{+0.10}_{-0.08} \ (\mathrm{syst.}) \ ^{+0.12}_{-0.08} \ (\mathrm{theory}) }}
\newcommand{\pubcombmu}{\ensuremath{ \mu = 1.55 \ ^{+0.33}_{-0.28} }}

\newcommand{\muttHpub}{\ensuremath{ \mu_{\ttH} = 1.3 \ ^{+2.5}_{-1.7} \ (\mathrm{stat.}) \ ^{+0.8}_{-0.4} \ (\mathrm{syst.}) }}

\newcommand{\combmucompactSS}{\ensuremath{  & =  1.17 \pm 0.27 }}
\newcommand{\combmucompactS}{\ensuremath{ \mu = 1.17 \pm 0.27 }}

\newcommand{\muVBFomuggF}{\ensuremath{0.6 \ ^{+0.8}_{-0.5}}}
\newcommand{\muVHomuggF}{\ensuremath{0.6 \ ^{+1.1}_{-0.6} }}
\newcommand{\muttHomuggF}{\ensuremath{1.2 \ ^{+2.2}_{-1.4}}}

\newcommand{\VBF}{\text{VBF}}
\newcommand{\ggH}{\text{ggF}}    
\newcommand{\ggF}{\text{ggF}}    

\newcommand{\WH}{\ensuremath{WH}}
\newcommand{\ZH}{\ensuremath{ZH}}
\newcommand{\VH}{\ensuremath{VH}}
\newcommand{\tH}{\ensuremath{tH}}
\newcommand{\ttH}{\ensuremath{t\bar{t}H}}
\newcommand{\bbH}{\ensuremath{b\bar{b}H}}
\newcommand{\tHbj}{\ensuremath{tHbj}}
\newcommand{\WtH}{\ensuremath{tHW}}
\newcommand{\tHW}{\ensuremath{tHW}}

\newcommand{\hgg}{\ensuremath{H\to\gamma\gamma}}
\newcommand{\hfourl}{\ensuremath{H \rightarrow ZZ^{(*)} \rightarrow 4\ell}}
\newcommand{\mgg}{\ensuremath{m_{\gamma\gamma}}}

\newcommand{\DR}{\ensuremath{\Delta R}}
\newcommand{\dr}{\DR}

\def\LL   {\ensuremath{\mathcal{L}}}
\def\vecth   {\ensuremath{\bm{\theta}}\xspace}

\newcommand{\pTt}{\ensuremath{p_{\mathrm{Tt}}}}

\newcommand{\mjj}{\ensuremath{m_{jj}}}

\newcommand{\thetaspurc}{\ensuremath{\theta_{\mathrm{spur},c}}}

\begin{document}

\preprint{APS/123-QED}

%\linenumbers
\title{Measurement of Higgs boson production in the diphoton decay channel in $pp$ collisions 
 at center-of-mass energies of 7 and 8~TeV with the ATLAS detector}

\date{\today}

\begin{abstract}

A 
measurement of the production processes of the recently discovered Higgs boson 
is performed in the two-photon final state using \lumiseven~\ifb\ of proton--proton 
collisions data at $\sqrt{s}=7 \TeV$ and \lumieight~\ifb\ 
at $\sqrt{s}=8 \TeV$ collected by the ATLAS detector at the Large Hadron Collider. 
The number of observed Higgs boson decays to diphotons divided by the corresponding Standard Model prediction, 
called the signal strength, 
is found to be ${\combmucompactS}$ at the \combmassname, ${\mH = \combmass}$ \GeV.
The analysis is optimized to measure the signal strengths for individual Higgs boson production processes at
this value of \mH. They are found to be ${\mugghC}$, ${\muvbfC}$, ${\muwhC}$, ${\muzhC}$, and ${\mutthC}$,
for Higgs boson production through gluon fusion, vector-boson fusion, and in association with a $W$ or $Z$ boson 
or a top-quark pair, respectively.
Compared with the previously published ATLAS analysis, the results 
reported here 
also benefit from a new energy calibration
procedure for photons and the subsequent
reduction of the systematic uncertainty on the diphoton mass resolution.  
No significant deviations from the predictions of the Standard Model are found.

\end{abstract}

\maketitle

\section{Introduction}
\label{sec:introduction}
	
In July 2012, the ATLAS and CMS collaborations independently reported observations of a new 
particle~\cite{ObservationATLAS,ObservationCMS} compatible with the Standard Model (SM) Higgs 
boson~\cite{Englert:1964et,Higgs:1964ia,Higgs:1964pj,Guralnik:1964eu,Higgs:1966ev,Kibble:1967sv}.
Since then, measurements of the properties of this new boson have been carried out to further 
elucidate its role in electroweak symmetry breaking and the mechanism of fermion mass generation.
In addition to measurements of its mass~\cite{Aad:2014aba,ref:cmsmass} and its spin and 
parity~\cite{atlas-spin,cms-spin}, 
the strengths of the couplings of the Higgs boson to fermions and vector bosons are of primary
interest~\cite{atlas-couplings-diboson,ref:cmsmass}. These couplings, which are predicted to depend on the value of \mH, can be tested by measurements
of the ratios of the number of observed Higgs bosons 
produced 
through gluon fusion (\ggH), weak vector-boson fusion (\VBF) and associated production
with a \Wboson\ boson (\WH), a \Zboson\ boson (\ZH) or a top-quark pair (\ttH) to the corresponding SM
predictions. The good diphoton invariant mass resolution
of the ATLAS detector makes it possible to measure these ratios, or \emph{signal strengths} $\mu$, in the diphoton final state,
separating the small, narrow Higgs boson signal from the large continuum background.

Measurements of the individual signal strengths of the production processes listed above are
presented in this article. They probe both the Higgs boson production and the \hgg\
decay rate: in order to test the production through VBF and associated production with 
a \Wboson\ or \Zboson\ boson or a $t\bar{t}$ pair independently of the ${\hgg}$ branching ratio, signal
strengths of these processes relative to ggF production are also presented.
A combination of \lumiseven\ \ifb\ of $pp$ collision data recorded at $\sqrt{s} = 7\TeV$ 
and \lumieight\ \ifb\ of data recorded at $\sqrt{s} = 8\TeV$ (the LHC Run 1 data)
is analyzed.
The analysis is designed to maximize the sensitivity to the signal strengths 
while using the same event selection as the measurement of the Higgs boson mass discussed in Ref.~\cite{Aad:2014aba}.
This is achieved by defining categories of diphoton candidate events that exploit
the characteristic features
of the final states of the different production modes.

The signal strengths are extracted from maximum likelihood fits 
to unbinned invariant mass distributions of diphoton candidates observed in the different event categories,
modeled by a narrow Higgs boson resonance on continuum 
backgrounds.
All the results presented in this article are
obtained for a Higgs boson mass ${\mH=\combmass}$~\GeV\ measured by ATLAS using the combination
of results from the decay channels ${\hgg}$ and ${\hfourl}$~\cite{Aad:2014aba}.
The CMS collaboration has recently updated its measurements of the Higgs properties in 
the diphoton channel as discussed in Ref.~\cite{newCMSHgg}. 

Compared with the previous results obtained with the same dataset~\cite{atlas-couplings-diboson},
this new analysis profits from a refined energy calibration 
procedure that improves the expected mass resolution of the signal in the inclusive diphoton sample by approximately 
10\%~\cite{calibration_paper}.
In addition, the uncertainty on the photon energy resolution is reduced by approximately a factor of two. 
Furthermore, experimental 
uncertainties on the integrated luminosity, photon identification, and photon isolation are reduced.
Two new categories enriched in \ttH\ events and a dedicated dilepton category that distinguishes
\ZH\ from \WH\ production have been added. Finally, the event selection and categorization
are tuned to improve the sensitivity of the analysis. 
The above refinements contribute almost equally to an overall improvement
of about 10\% in the expected uncertainty on the combined signal strength.

The article is organized in the following way.
The \mbox{ATLAS} detector is briefly described in Sec.~\ref{sec:atlas}. 
The data and Monte Carlo (MC) samples used for this analysis are presented in Sec.~\ref{sec:dataset} while
details of the reconstruction of photons, electrons, muons, jets and missing transverse
momentum are given in Sec.~\ref{sec:objects}. 
The diphoton event selection is discussed in Sec.~\ref{sec:selection} followed by a description of the event 
categorization in Sec.~\ref{sec:categorisation}.
The models of the signal and background distributions used to fit the data are presented in 
Sec.~\ref{sec:signalandbackgroundmodels}.
The systematic uncertainties are described in 
Sec.~\ref{sec:systematics}.
Using the statistical procedure briefly outlined in Sec.~\ref{sec:statistic},
the results of the combination of the $\sqrt{s} = 7\TeV$ and $\sqrt{s} = 8\TeV$ 
data for the Higgs boson signal strengths 
are extracted and presented in Sec.~\ref{sec:results}.
The conclusions of this study are summarized in Sec.~\ref{sec:conclusion}.

\section{The ATLAS detector}
\label{sec:atlas}

The ATLAS experiment \cite{ATLAS_detector} is a multipurpose particle physics 
detector with a forward-backward symmetric cylindrical geometry and nearly 4$\pi$ 
coverage in solid angle.\footnote{ATLAS uses a right-handed coordinate system with 
its origin at the nominal 
interaction point (IP) in the center of the detector and the $z$-axis along the
beam pipe. The $x$-axis points from the IP to the center of the LHC ring, and the
$y$-axis points upward. Cylindrical coordinates ($r,\phi$) are used in the transverse
plane, $\phi$ being the azimuthal angle around the beam pipe. The pseudorapidity
is defined in terms of the polar $\theta$ angle as ${\eta = -\ln\left[\tan(\theta/2)\right]}$.}

The inner tracking detector (ID) covers the pseudorapidity range ${|\eta|<2.5}$ and 
consists of a silicon pixel detector, a silicon microstrip detector, and a transition 
radiation tracker (TRT) in the range ${|\eta| < 2.0}$. The ID is surrounded by a 
superconducting 
solenoid providing a 2~T magnetic field. The ID allows an accurate 
reconstruction of charged-particle tracks originating from the proton--proton collision region
as well as from secondary vertices, which permits an efficient reconstruction of photons 
interacting in the ID through $e^{+}e^{-}$ pair
production up to a radius in the transverse plane of about 80~cm.

The electromagnetic (EM) calorimeter is a lead/liquid-argon (LAr) sampling 
calorimeter with an accordion geometry. It is divided into two barrel 
sections
that
cover the pseudorapidity region ${|\eta| < 1.475}$ and two end-cap sections
that cover the pseudorapidity regions ${1.375 < |\eta| < 3.2}$. It consists 
of three (two) longitudinal layers in shower depth in the region ${|\eta|<2.5}$ (${2.5 < |\eta| < 3.2}$).
The first one has a thickness of approximately four radiation lengths and, in the ranges ${|\eta|<1.4}$ and 
${1.5<|\eta|<2.4}$, is segmented into high-granularity strips in the $\eta$ direction, 
typically
${0.003\times 0.1}$ in ${\eta\times\phi}$ in the barrel regions. The first-layer sampling strips
provide 
event-by-event discrimination between
prompt photon showers and two overlapping showers coming from a $\pi^{0} \rightarrow \gamma\gamma$ decay. 
The second layer, which collects most of the energy 
deposited in the calorimeter by photons and electrons, has a thickness of 
about 17 radiation 
lengths and a granularity of ${0.025\times 0.025}$ in ${\eta\times\phi}$. 
The third layer, which has a thickness ranging from two to twelve radiation lengths as a function of $\eta$, is used 
to account for longitudinal fluctuations of high-energy electromagnetic showers.
A thin presampler layer located in front of the EM calorimeter in the pseudorapidity 
interval $|\eta| < 1.8$ is used to correct 
for energy loss upstream of the calorimeter. 

The hadronic calorimeter, which surrounds the EM calorimeter, consists of a steel/scintillator-tile 
calorimeter in the range ${|\eta| < 1.7}$ and two copper/LAr calorimeters 
spanning ${1.5 < |\eta| < 3.2}$. The acceptance is extended to ${|\eta| = 4.9}$
by two sampling calorimeters longitudinally segmented in shower depth into three sections using LAr as active 
material and copper (first section) or tungsten (second and third sections) as absorber.

The muon spectrometer (MS), located outside the calorimeters, consists of three 
large air-core superconducting toroid systems with precision tracking chambers that provide
accurate muon tracking for ${|\eta| < 2.7}$ and fast detectors for triggering for ${|\eta| < 2.4}$.

A three-level trigger system is used to select events containing two photon candidates. 
The first-level trigger is hardware-based: using a cell granularity 
(${0.1 \times 0.1}$ in ${\eta \times \phi}$) that is coarser than that
of the EM calorimeter, it searches for electromagnetic deposits 
with a transverse energy \ET\
above a programmable threshold. The second- and third-level triggers (collectively 
referred to as the \emph{high-level trigger}) are implemented in software and exploit the 
full granularity and accurate energy calibration of the calorimeter.

\section{Data and Monte Carlo samples}
\label{sec:dataset}

Events from $pp$ collisions were recorded using a
diphoton trigger with \ET\ thresholds of  
35~\gev\ and 25~\gev\ for the leading and sub-leading photon candidates, respectively, in the 8~TeV 
data and 20~\gev\ for both photon candidates in the 7 TeV data~\cite{atlas-trigger}. 
In the high-level trigger, clusters of energy in the EM calorimeter were
reconstructed and required to satisfy
loose criteria according to expectations for EM showers initiated by photons. 
This trigger has a signal efficiency above 99\% for events fulfilling the final 
event selection. 
After application of data quality requirements, the 8~TeV (7~TeV) 
data sample corresponds to a total integrated luminosity of \lumieight~\invfb\ (\lumiseven~\invfb).
The instantaneous luminosity is typically about 
$6\cdot 10^{33}$~\ilumi\ ($3\cdot 10^{33}$~\ilumi) in the analyzed 8~TeV (7~TeV) data,
resulting in an average number of $pp$ collisions 
per bunch crossing of about 21 (9) in the 8~TeV (7~TeV) data. 

Simulated samples of Higgs bosons decaying into two photons were generated separately for the five
production modes whose signal strengths are measured here (\ggH, \VBF, \WH, \ZH, and \ttH)
and for Higgs boson
masses from 100~\GeV\ to 160~\GeV\ (115~\GeV\ to 135 ~\GeV\ for the \ttH\ samples) in 5~\GeV\ steps. 
Samples 
of Higgs boson events produced in association with a single top quark, \tH, which is predicted 
to make a small contribution to the selection of candidates from \ttH\ production, were also generated. 

The AU2~\cite{atlasmctunes} tuning of \textsc{Pythia8} ~\cite{pythia8} is used to simulate
the minimum-bias events and the underlying event.
The normalizations of the production mode samples are performed following the
recommendations of the LHC Higgs cross-section working group~\cite{lhcxs} as described below.

Gluon fusion events are generated with
\textsc{Powheg-box}~\cite{Nason:2004rx,Frixione:2007vw,powhegbox,powheg_ggF,Bagnaschi:2011tu} interfaced with
\textsc{Pythia8} for the underlying event, parton showering and
hadronization. 
The
overall normalization of the \ggH\
process used to estimate the expected event rate is taken from a calculation at next-to-next-to-leading order
(NNLO)~\cite{Djouadi:1991tka, Dawson:1990zj, Spira:1995rr,
Harlander:2002wh, Anastasiou:2002yz, Ravindran:2003um} in QCD. 
Next-to-leading-order (NLO) electroweak (EW) corrections are also 
included~\cite{Aglietti:2004nj,Actis:2008ug}. 
The effect of the interference of ${gg \rightarrow H \rightarrow \gamma \gamma}$
with the continuum ${gg \rightarrow \gamma\gamma}$ background induced by quark loops 
is taken into account using an averaging procedure~\cite{Dixon_pc} that combines
LO~\cite{dixon_interference} and NLO corrections~\cite{Dixon:2013haa}:
the destructive interference causes 
a ${\sim 1\%}$ reduction of the \ggH\ cross section.

The \VBF\ samples are generated using
\textsc{Powheg-box}~\cite{powheg_VBF} interfaced with \textsc{Pythia8} and normalized to a cross
section calculated with full NLO QCD and EW
corrections~\cite{Ciccolini:2007jr, Ciccolini:2007ec, Arnold:2008rz}
with an approximate NNLO QCD correction applied~\cite{Bolzoni:2010xr}.

Higgs bosons produced in association with a \Zboson\ boson or a \Wboson\ boson (collectively referred to as \VH)
are generated with \textsc{Pythia8}.
The predictions for \VH\ are normalized to cross sections calculated at
NNLO~\cite{Brein:2003wg} with NLO EW radiative 
corrections~\cite{Ciccolini:2003jy}
applied.
 
The \ttH\ samples are generated using the \textsc{Powhel} generator, a combination of the 
\textsc{Powheg-box}  and \textsc{Helac-NLO}~\cite{HelacNLO} generators, interfaced with \textsc{Pythia8}.
The full NLO QCD corrections are
included~\cite{Beenakker:2001rj, Beenakker:2002nc, Dawson:2002tg, Dawson:2003zu} in 
the \ttH\ normalization. A sample of events from \tH\ production in the t-channel in association with a $b$-jet and a light jet $j$
(\tHbj) are generated with \textsc{MadGraph}~\cite{Maltoni:2002qb} interfaced with
\textsc{Pythia8}; the normalization of the production cross section is taken from 
Refs.~\cite{Maltoni:2001hu,Barger:2009ky,Farina:2012xp,Biswas:2012bd,Agrawal:2012ga}.
A sample of \tH\ events produced in association with a \Wboson\ boson (\tHW) is generated 
using \textsc{MadGraph5\_aMC@NLO}~\cite{Alwall:2014hca}
interfaced to \textsc{Herwig++}~\cite{Bahr:2008pv}.

The 
branching ratio  for ${\hgg}$ and its uncertainty~\cite{Djouadi:1997yw, Actis:2008ts} are compiled in Ref.~\cite{lhcxs}.
The \textsc{CT10}~\cite{cteq10} parton distribution function (PDF) set is used for 
the \textsc{Powheg-box} samples while 
\textsc{CTEQ6l1}~\cite{cteq6} is used
for the \textsc{Pythia8} samples.

Additional corrections to the shape of
the generated \pt\ distribution of Higgs bosons produced by gluon fusion are applied to match the distribution 
from a calculation at NNLO+NNLL provided by \textsc{HRes2.1}, 
which includes exact calculations of the effects of the top and bottom quark masses~\cite{hres_1,hres_2} 
as well as dynamical 
renormalization and factorization scales.
Calculations based on \textsc{HRes} predict a lower rate of events at high \pt\ compared with the
nominal \textsc{Powheg-box} samples and thus
the contribution from events with two or more jets, which mostly populate the high-\pt\ region, is affected.
To simultaneously 
reproduce the inclusive Higgs
\pt\ distribution as well as the $\geq 2$ jet component, the \ggH\ events with two or more
jets are first normalized to a NLO calculation~\cite{Boughezal:2013oha}. 
Then, Higgs boson \pt-dependent weighting functions are determined using an iterative procedure.
First, the events with two or more jets are 
weighted in order to match the Higgs boson \pt\ distribution from \textsc{MiNLO HJJ}
predictions~\cite{minlo_hjj}. As a second step, the inclusive
spectrum is weighted to match the \textsc{HRes} distribution. 
These two steps are iteratively repeated until the inclusive Higgs \pt\ spectrum 
agrees well with the \textsc{HRes} prediction while 
preserving the normalization of the $\geq 2$ jet component.
The events simulated for \VBF, \WH, and \ZH\ production are re-weighted so that the \pt\ distributions
of the Higgs bosons match the ones predicted by \textsc{Hawk}~\cite{hawk_1,hawk_2,hawk_3}.

The contribution from Higgs boson production in association
with a \bbbar\ pair
(\bbH)
is accounted for in this analysis: the cross section of this process 
is calculated
 in a four-flavor PDF scheme (4FS) at NLO QCD~\cite{Dawson:2003kb,Dittmaier:2003ej,Dawson:2005vi} 
 and a five-flavor PDF scheme (5FS) at NNLO QCD~\cite{Harlander:2003ai}. 
 These two calculations are combined using the Santander matching procedure~\cite{Harlander:2011aa,LHCHiggsCrossSectionWorkingGroup:2012vm}.
Since 
the \pt\ spectrum of the $b$-jets                                                                 
is expected to be soft, the 
jet environments for \ggF\ and \bbH\ production are quite similar and thus 
the detection efficiency for \bbH\ is assumed to be the same as for \ggF.

The invariant mass distributions and normalizations of the backgrounds 
in the event categories are estimated by fits to the data.
However, the choices of the functional forms used to model the backgrounds and the uncertainties associated with these
choices are determined mostly by MC studies, as described in detail in Sec.~\ref{sec:background_model}.
For these studies $\gamma \gamma$ and $\gamma$--jet background samples were generated by 
\textsc{Sherpa}~\cite{sherpa,Hoeche:2009xc} and the  
jet--jet 
background samples by \textsc{Pythia8}. The normalizations of these samples are determined by
measurements of a data sample of preselected diphoton events as described in Sec.~\ref{sec:background_model}. More
details about the background control sample used for each category are also given in Sec.~\ref{sec:background_model}.

A summary of the event generators and PDF sets for the individual signal 
and background processes used in this analysis is reported in Table~\ref{mc_table}.
The normalization accuracy and SM cross sections with ${\mH=125.4}$~GeV for ${\sqrt{s}=7}$~TeV and ${\sqrt{s}=8}$~TeV are also given
for the different Higgs production modes.
\begin{table*}
  \caption{
    Summary of event generators and PDF sets used to model the signal and the main background processes.
    The SM cross sections $\sigma$ for the Higgs production processes with ${\mH=125.4}$~GeV are also given 
    separately for ${\sqrt{s} = 7}$~TeV and ${\sqrt{s} = 8}$~TeV, together with the orders of the calculations.
  }
  \label{mc_table}
  \begin{center}
    \begin{tabular}{lcccccc}
      \hline\hline
\multirow{2}{*}{Process} & \multirow{2}{*}{Generator} & \multirow{2}{*}{Showering} & \multirow{2}{*}{PDF set}  & \multirow{2}{*}{Order of calculation} & $\sigma [\mathrm{pb}] $ & $\sigma [\mathrm{pb}]$ \\
                     &                     &                     &                    &                          & $\sqrt{s}=7$ TeV & $\sqrt{s}=8$ TeV \\
      \hline
      \ggH  & \textsc{Powheg-box}                      & \textsc{Pythia8}  & \textsc{CT10}    & NNLO(QCD)+NLO(EW)           &  15.04  & 19.15 \\
      \VBF  & \textsc{Powheg-box}                      & \textsc{Pythia8}  & \textsc{CT10}    & NLO(QCD+EW)+app.NNLO(QCD)  &  1.22   & 1.57 \\
      \WH   & \textsc{Pythia8}                     & \textsc{Pythia8}  & \textsc{CTEQ6L1} & NNLO(QCD)+NLO(EW)           & 0.57    & 0.70  \\
      \ZH   & \textsc{Pythia8}                     & \textsc{Pythia8}  & \textsc{CTEQ6L1} & NNLO(QCD)+NLO(EW)           &  0.33   & 0.41  \\
      \ttH  & \textsc{Powhel}                      & \textsc{Pythia8}  & \textsc{CT10}    & NLO(QCD)                    & 0.09    & 0.13  \\
      \tHbj & \textsc{MadGraph}                    & \textsc{pythia8}  & \textsc{CT10}    &      NLO(QCD)               & 0.01    & 0.02 \\
      \WtH  & \textsc{MadGraph5\_aMC@NLO} & \textsc{herwig++} & \textsc{CT10}    & NLO(QCD)                             & $<$0.01 & $<$0.01\\ 
      \bbH  &                  -                   &              -    &          -       &  5FS(NNLO)+4FS(NLO)         & 0.15    & 0.20 \\
      \hline
      $\gamma\gamma$ & \textsc{Sherpa} & \textsc{Sherpa} & \textsc{CT10} \\
     $\gamma$--jet & \textsc{Sherpa} & \textsc{Sherpa} & \textsc{CT10} \\
      jet--jet & \textsc{Pythia8} & \textsc{Pythia8} & \textsc{CTEQ6L1} \\
      \hline \hline

    \end{tabular}
  \end{center}
\end{table*}

The stable particles, defined as the particles with a lifetime longer
than 10~ps, are passed through a full detector
simulation \cite{simuAtlas} based on \textsc{Geant4} \cite{geant4}.
Pileup effects are simulated by overlaying each MC event with a variable 
number of MC inelastic $pp$ collisions generated using \textsc{Pythia8}, 
taking into account in-time pileup 
(collisions in the same bunch crossing as the signal),
out-of-time pileup 
(collisions in other bunch crossings within the time-window of the detector sensitivity),
and the LHC bunch train structure. The MC events are weighted to 
reproduce the distribution of the average number of interactions per bunch crossing 
observed in the data.
The resulting detector signals are passed through the same event reconstruction algorithms
as used for the data.
Since the length of the beam spot along the beam axis 
is slightly wider in the MC samples
than in the data, a weighting procedure is applied to the 8~TeV (7~TeV) MC events to match the 4.8~cm (5.6~cm) RMS 
length observed in the 8~TeV (7~TeV) data.

In order to increase the number of available MC background events, 
especially for the optimization of the event categorization  (Sec.~\ref{sec:categorisation}) and 
background shape parameterization studies (Sec.~\ref{sec:background_model}),
MC samples based
on fast, simplified models of the detector response rather than full
simulation are used: the resolutions and reconstruction
efficiencies for photons and jets are tuned as functions of the transverse momentum and 
pseudorapidity to reproduce the ones obtained from fully simulated samples of $\gamma\gamma$ and $\gamma$--jet events.  
These samples are typically about 1000 times 
 larger than
 the corresponding collected data samples after analysis selections. 

\section{Physics object definitions}
\label{sec:objects}

The reconstruction and identification of the physics objects (photons, electrons, muons, jets)
and the measurement of missing transverse momentum are described here. Unless otherwise stated, the
descriptions apply to both the 7~TeV and the 8~TeV data.

\subsection{Photons}
\label{photon_reco}

The photon reconstruction is seeded by energy deposits (clusters) in the
EM calorimeter with ${\ET > 2.5}$~GeV in projective
towers of size ${0.075\times 0.125}$ in the ${\eta\times\phi}$ plane.
The reconstruction algorithm looks for possible matches between
energy clusters and tracks reconstructed in the
inner detector and extrapolated to the calorimeter. Well-reconstructed
tracks matched to clusters are classified as electron candidates while clusters
without matching tracks are classified as unconverted photon candidates. 
Clusters matched to pairs of tracks that are consistent with the hypothesis
of a 
${\gamma \rightarrow e^{+}e^{-}}$
conversion process are classified as converted photon candidates.
Due to the intrinsic ambiguity between electron and photon
signatures, clusters may be reconstructed both with electron and
photon hypotheses to maximize the reconstruction efficiency for both.
In particular, clusters matched to single tracks without hits in
an active region of the pixel layer nearest to the beam pipe are considered
both as converted photon candidates~\cite{photonreco} and electron candidates. 
The cluster reconstruction efficiency for photons with ${\ET > 25}$~\GeV\ is
estimated from simulation~\cite{photonreco} to be close to 100\% while the efficiency to actually
reconstruct them as photons is 96\%. In the remaining cases
these clusters are incorrectly reconstructed as electrons but not
as photons.
The probability for a real electron with ${\ET > 25}$~\GeV\ to be
reconstructed as a photon fulfilling the tight identification criteria described below is
measured in data to vary between 3\% and 10\%, depending on the
pseudorapidity and the conversion class of the candidate.

In the following, a brief review of the calibration procedure for photons is reported;
a detailed description can be found in Ref.~\cite{calibration_paper}.
The energy measurement is performed by summing the energies measured in the EM calorimeter cells 
belonging to the candidate cluster. The size of the cluster depends on the photon classification: in the barrel, a 
${\Delta \eta \times \Delta \phi = 0.075\times 0.125}$ cluster is used for unconverted photons and ${0.075\times 0.175}$ 
for converted photons to account for the opening of the $e^{+}e^{-}$ pair in the 
$\phi$ direction due to the magnetic field. In the end-cap, a cluster size of ${\Delta\eta\times\Delta\phi = 0.125\times 0.125}$
is used for all candidates. 
The cluster energy has to be corrected for energy losses in the inactive
materials in front of the calorimeter, for the fraction of energy deposited outside the area of the cluster 
in the \mbox{${\eta\phi}$-plane} and into the hadronic calorimeter in the direction of the shower propagation.
Finally, due to the finite cluster size in $\eta$ and $\phi$ coordinates and
the variation of the amount of absorber material crossed by incident
particles as a function of $\phi$, a correction has to account for
the variation of the energy response as a function of the impact
point on the calorimeter.
The calibration coefficients used to make this correction are 
obtained from a detailed simulation of the detector geometry
and are optimized with a boosted decision tree (BDT)~\cite{TMVA}.
The response is calibrated separately for converted and unconverted photon candidates. 
The inputs to the energy calibration algorithm are the measured energy per calorimeter layer,
including the presampler, the $\eta$ position 
of the cluster and the local position of the shower within the second-layer cell corresponding to the 
cluster centroid. In addition, the track transverse momenta and the conversion
radius for converted photons are used as input to the regression algorithm to further improve the energy resolution, especially at low energy.
This
new calibration procedure gives a 10\% improvement in the expected invariant mass resolution
for \Hgg\ events with respect to the calibration used in our previous publications.
The energy scales of the data and simulation are equalized by applying $\eta$-dependent
correction factors to match the invariant mass distributions
of \Zee\ events. In this procedure, the simulated width of the \Zboson\ boson resonance is matched to the one
observed in data by adding a contribution to the constant term of the electron energy resolution.
The photon energy scale uncertainty is 0.2--0.3\%\ for
${|\eta|<1.37}$ and ${|\eta|>1.82}$, and 0.6\%\ for
${1.52<|\eta|<1.82}$. A similar accuracy is achieved for converted and
unconverted photons, and the energy dependence of the uncertainty is weak. The uncertainties
in the photon energy scales are confirmed by an independent
analysis of 
radiative \Zboson~boson decays. The
relative uncertainty on the energy resolution is about 10\%\ for
photons with ${\et \sim 60}$~\gev.
The uncertainty on the photon energy resolution is reduced by approximately a factor of two with respect to our 
previous publications: 
this reduction comes from improvements on the detector simulation model, from a better knowledge of 
the material upstream of the calorimeter, and from more detailed calibration corrections applied 
to the data~\cite{calibration_paper}. These improvements lead to a better agreement between the 
$m_{ee}$ distributions in simulated \Zee\ events with the ones measured in data, that in turn prompt 
a reduced uncertainty of the energy resolution effective constant term. In addition, the new procedure to compute the 
photon energy resolution uncertainty is more effective at disentangling the contributions from the knowledge of the 
material in front of the calorimeter and of the intrinsic calorimeter energy resolution, as discussed
in Sec.~\ref{sec:unc_massres}.
The energy response of the calorimeter in 
data varies by less than
0.1\%\ over time. The simulation is found to describe the dependence of the response on pileup
conditions at the same accuracy level.

The photon identification algorithm is based on the lateral and longitudinal energy profiles 
of the shower measured in the calorimeter~\cite{PhotonID}. 
First, the fraction of energy in the hadronic calorimeter is used, together
with the  shape of the lateral profile of the shower 
as measured in the second layer of the electromagnetic calorimeter, to
reject photon candidates from jets with a large hadronic component. 
Then, observables built from measurements in
the high-granularity 
first layer of the calorimeter are used to discriminate prompt photons from overlapping photon pairs that
originate in the decays of neutral mesons produced in jet fragmentation.
Based on these discriminating variables, two sets of tight identification criteria, for converted and 
unconverted photon candidates, are applied to the 8~TeV data. 
The identification criteria are based on rectangular cuts optimized on simulated
electromagnetic showers in $\gamma$--jet events and simulated jets in QCD dijet events. 
The agreement between data and simulation for the individual discriminating variables is checked
using a pure sample of photons from radiative \Zll$\gamma$\ decays  (where $\ell$ is an electron or a muon) 
and an inclusive photon sample
after background subtraction. As a result,
small corrections are applied to the identification variables in the simulation
to account for the observed mis-modeling of lateral shower profiles in the calorimeter.
The photon identification cuts are carefully tuned to guarantee stability 
of the efficiency as a function of the in-time pileup within a few per cent.
The identification efficiency for unconverted (converted) photons is typically 
{83--95\%} {(87--99\%)} for \mbox{$30< \ET <100 $ \gev}.
Correction factors as a function of $\eta$, \ET\ and conversion class are derived to correct for the 
residual mismatch between the efficiency in the simulation and the efficiency measured in the data.
For the analysis of the 7~TeV data, 
the discriminating observables are combined into a single discriminant by a neural-network
(NN) algorithm~\cite{TMVA}:  with similar jet rejection power, the multivariate approach improves the
identification efficiency by 8--10\%\ with respect to the cut-based identification~\cite{PhotonID}.
For the analysis of the 8~TeV data, the re-optimized cut-based identification
has a similar jet rejection power for a given identification efficiency.

Two complementary isolation 
variables are used to further suppress the
number of jets in the photon candidate samples.
The first variable is the sum of the transverse energies of positive-energy topological
clusters~\cite{clustering} deposited in the calorimeter within a cone
of ${\Delta R \equiv \sqrt{(\Delta\eta)^2 + (\Delta\phi)^2} = 0.4}$ around each photon. 
The energy sum excludes the contribution due to the photon cluster and 
an estimate of the energy deposited by the photon outside its associated cluster. 
The median \et\ density for the event in question, caused
by the underlying event (UE) and additional
minimum-bias interactions occurring in the same or neighboring bunch crossings
(in-time and out-of-time pileup, respectively), is subtracted
on an event-by-event basis using an algorithm described in Ref.~\cite{Salam} and
implemented 
as described in Ref.~\cite{ATLASPromptPhoton}. 
Despite these corrections,  a residual dependence of the calorimetric isolation selection efficiency
$\epsilon_{\mathrm{iso}}$ on the number of primary vertices reconstructed
by the inner tracking detector~\cite{atlasprimaryvertexperf} is observed: an example is shown in
Fig.~\ref{fig:isol_eff} for a maximum allowed energy of 4 \gev\ in the isolation cone.
To improve the efficiency of the isolation selection for events with large pileup,
the calorimetric isolation is complemented by a track isolation defined as
the scalar sum of the transverse momenta of
all tracks with ${p_\mathrm{T}>1}$~\GeV\ (0.4~\GeV\ for the 7~TeV data) within a cone of size
${\Delta R = 0.2}$ around
each photon. The track isolation efficiency is insensitive to out-of-time pileup and its dependence
on the in-time pileup is reduced by selecting
only tracks 
consistent with originating from the diphoton production
vertex (defined in Sec.~\ref{sec:selection}) and not associated with converted photon candidates.
A track in the 7~TeV (8~TeV) data is considered to be associated with the diphoton production vertex
if the point of closest approach of its extrapolation is within 5~mm (15~mm) of the vertex along the $z$-axis
and within 0.5~mm (1.5~mm) of the vertex in the transverse plane.
For a given
sample purity, a reduction of the dependence of the selection efficiency on the in-time pileup is
obtained by combining a looser calorimeter isolation
selection with a track isolation requirement.
Photon candidates are required to have 
a calorimetric isolation less than 6~\GeV\  (5.5~\GeV\ for the 7 TeV data) 
and a track isolation less than 2.6~\GeV\ (2.2~\GeV\ for the 7 TeV data).
The efficiency of the isolation cuts in the simulation is corrected by a small
\pt -dependent 
factor extracted from measurements in data performed with a pure sample of 
photons from radiative \Zee$\gamma$\ decays and \Zee\ events. 

\begin{figure}[h]
  \centering
  \includegraphics[width=\columnwidth]{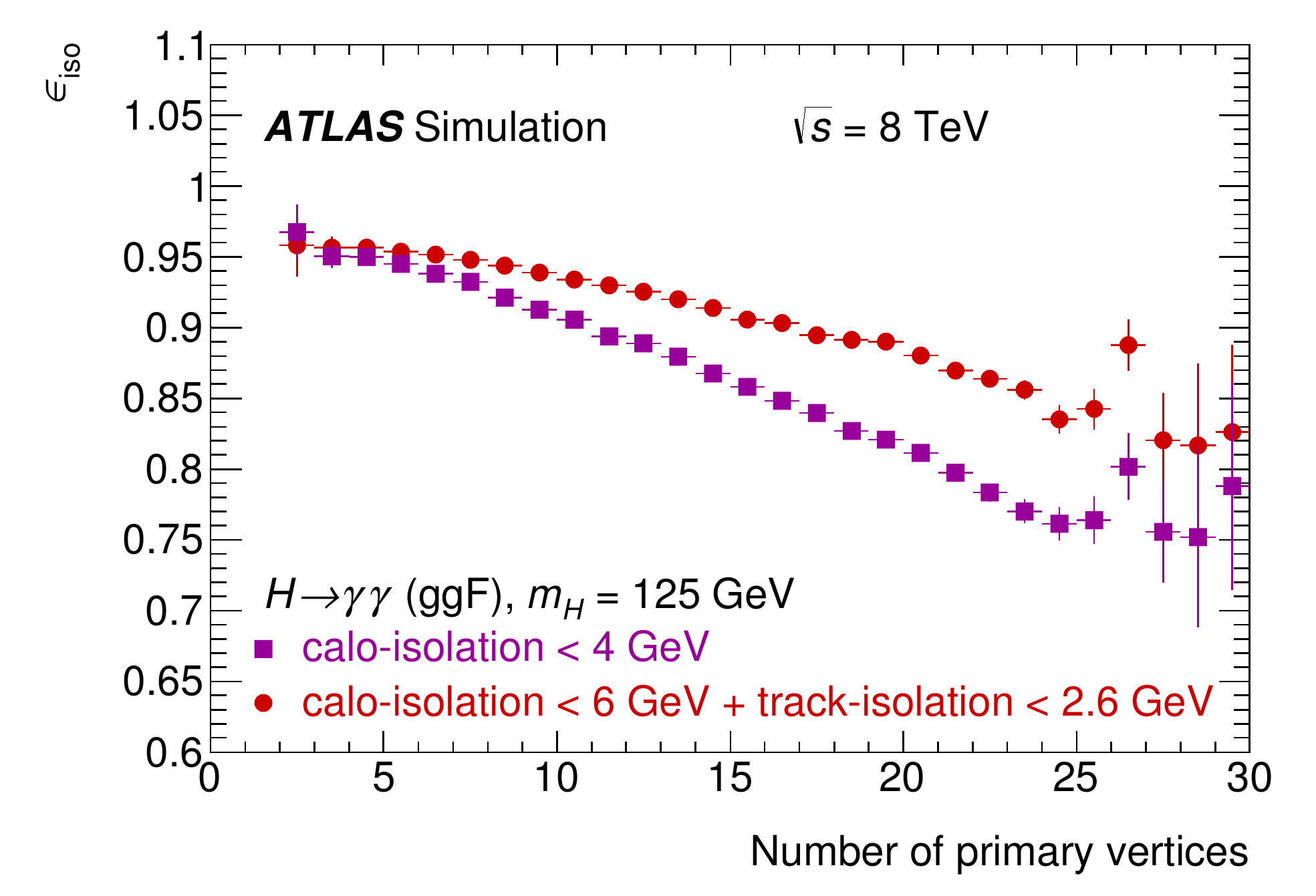} 
  \caption{Efficiency $\epsilon_\mathrm{iso}$ to fulfill the isolation requirement  	
    as a function of the number of primary vertices in each event, determined with a MC 
    sample of Higgs bosons decaying into two photons
    with \mH=125 \GeV\ and $\sqrt{s} = 8$ TeV. Events are required to satisfy the kinematic 
    selection described in
    Sec.~\ref{sec:selection}. The efficiency of the event selection obtained
    with a tight calorimetric isolation requirement \mbox{(4 \gev)} is compared 
    with the case in which a looser calorimetric isolation (6 \gev) is combined
    with a track isolation (2.6 \gev) selection.} 
  \label{fig:isol_eff}
\end{figure}

\subsection{Leptons}
\label{sec:leptons}

Electron candidates, as mentioned above, are built from clusters of energy deposited in the
electromagnetic calorimeter that are associated with at least one 
well-reconstructed track in the inner detector. 
In this analysis electron candidates are required to satisfy the loose identification
criterion of a likelihood-based discriminating variable~\cite{ElectronIDLikelihood}.
A cut-based identification selection is used in the 7 TeV analysis 
and the electrons are required to fulfill the medium criteria defined in Ref.~\cite{Aad:2014fxa}.
The determination of the energy of the electron candidate is performed using a 
${\Delta\eta\times\Delta\phi = 0.075\times 0.175}$ cluster in the barrel
to recover the energy 
spread in
$\phi$ 
from bremsstrahlung photons
while 
a ${0.125\times 0.125}$ cluster is used in the end-cap. The cluster energy is
calibrated as discussed in Sec.~\ref{photon_reco} with a dedicated set
of calibration coefficients optimized for electrons.
The transverse momentum \pT\ of an electron is computed from the cluster energy and
the track direction at the interaction point.
Electrons are required to be in the region ${|\eta|< 2.47}$ and to satisfy 
${\et > 15}$~GeV.  
Finally,  the  electron  candidates  must satisfy both the track-based  and
calorimetric isolation criteria relative to the \et\ of the candidate.
The calorimetric transverse isolation 
energy  within a ${\Delta  R =  0.4}$ cone is required to be 
less than 20\% of the electron candidate's \et,  whereas the sum of the transverse momenta of the tracks within a
cone of ${\Delta R = 0.2}$ around the track of the electron 
candidate is required to be less than
15\% of the electron candidate's \et.

Muon candidates are formed from tracks reconstructed independently in the MS and in the ID
and from energy deposits measured in the calorimeters~\cite{Aad:2014rra}.
Different types of muon candidates are built depending on the available information from
the different sub-detector systems: the main algorithm combines tracks reconstructed
separately by the ID and the MS. To extend the acceptance region beyond the ID limit to include $2.5 < |\eta| <2.7$,
tracks reconstructed in the MS stand-alone are used. 
Finally, to increase the acceptance for low-\pt\ muons or for muons that  pass through
uninstrumented regions of the MS, muon candidates are reconstructed from tracks in 
the ID associated with a track segment in the MS or to a calorimetric energy deposition
compatible with the one from a minimum-ionizing particle.  
In this analysis, muons from all different algorithms are used and required to have 
${|\eta|< 2.7}$ and ${\pT > 10}$~\GeV: the combination of the different
algorithms ensures a $\sim 99$\% efficiency to detect a muon over the full acceptance range.
A candidate is also required to satisfy exactly the same isolation criteria
(relative to its \pt) as for electrons.

\subsection{Jets}
\label{sec:jet_reco}

Jets are reconstructed using the anti-$k_t$ algorithm~\cite{JetAlgo} with radius parameter ${R=0.4}$
and are required to have 
${|\eta|<4.4}$ and satisfy (unless stated otherwise) ${\pt> 30}$~\GeV. Jets are discarded if they
are within ${\dr=0.2}$ of an isolated electron or within ${\dr=0.4}$ of an isolated photon.
The inputs to the jet-finding are topological calorimeter clusters~\cite{jetcal_2011} formed with the energy
calibration appropriate for electromagnetic showers. The jet energy is calibrated using scale factors extracted from 
simulated dijet events by matching the energies of the generator-level and reconstructed jets. In addition, for the 8 TeV data, the pileup dependence
of the jet response is suppressed by subtracting the median \ET\ density for the event multiplied by the
transverse area of the
jet~\cite{Cacciari:2007fd,jetPileupCorr}.  A residual pileup correction that is proportional to the number of reconstructed
primary vertices and to the average number of interactions per bunch crossing further reduces the pileup
dependence, in particular in the forward region. 
Finally, the jet energy is corrected by an absolute scale factor determined 
using $\gamma$+jet, $Z$+jet and multijet events in data, and a relative $\eta$-dependent factor
measured with dijet events in data.
In order to suppress jets produced by pileup, jets within the tracking acceptance  ($|\eta_{j}|<2.4$) are required to
have a jet vertex fraction\footnote{The jet vertex fraction (JVF) is defined as the sum of $\pT$ of the tracks associated with the jet that are
produced at the diphoton's primary vertex, divided by the sum of $\pT$ of the tracks associated with the jet
from all collision vertices.} (JVF)~\cite{jetPileupCorr} larger than 0.5 (0.25) for the 7~TeV (8~TeV) data,
respectively.

In order to identify jets containing a $b$-hadron ($b$-jets), a NN-based algorithm 
is used to combine information from the tracks in a jet: the network 
exploits the measurements of 
the impact parameters of the tracks, 
any secondary vertices, and the outputs of decay topology algorithms as discussed in Refs.~\cite{ATLAS-CONF-2011-102,ATLAS-CONF-2014-046}.
Four different working points with efficiencies for identifying $b$-jets
(rejection factors for light jets) of 60\%\ (450), 
70\%\ (140), 80\%\ (29), and 85\%\ (11) are used in the analysis.
The efficiencies and rejection factors at the working points are calibrated
using control samples of data.

\subsection{Missing transverse momentum}
\label{sec:met_definition}

The measurement of the magnitude of the missing transverse momentum \MET\ is based on the 
transverse energy of all photon, electron and muon candidates, all jets 
sufficiently isolated from the photon, electron and muon candidates, and all calorimeter 
energy clusters not associated with any of these objects (\emph{soft term})~\cite{met_perf}. 
In order to improve the discrimination of multijet events, where \MET\ arises mainly from energy resolution effects, from
events with a large fraction of \MET\ due to non-interacting particles, an \MET-significance 
is defined as ${\MET / \sigma_{\MET}}$, where the square root of the scalar sum of the transverse
energies of all objects $\Sigma \ET$ is used in the estimator of the
\MET\ resolution ${\sigma_{\MET} = 0.67\,[\mathrm{GeV}^{1/2}] \sqrt{\Sigma \ET}}$.
The proportionality factor ${0.67\,[\mathrm{GeV}^{1/2}]}$ is determined with
fully reconstructed ${\Zll}$ events by removing the 
leptons in the measurement of $\MET$~\cite{ATLAS:2012wna}.

\section{Event selection}
\label{sec:selection}

The measurement of the signal strengths of Higgs boson production is based on
the extraction of resonance signals in the diphoton invariant mass spectra
of 12 independent categories of events that are described in the next section. Common diphoton 
selection criteria 
are applied to all events.
At least two photon candidates are required to be in a fiducial
region of the EM calorimeter defined by ${|\eta| < 2.37}$, excluding the transition region between the barrel
and the end-cap calorimeters (${1.37 < |\eta| < 1.56}$). Photon candidates in this fiducial region are ordered
according to their \et\ and only the first two are considered: the leading and sub-leading photon candidates are
required to have ${\et / \mgg > 0.35}$ and 0.25, respectively, where \mgg\ is the invariant mass of the two selected
photons. Requirements on the \ET\ of the two selected photons relative to $\mgg$ 
are found to give 
$\mgg$ spectra 
that are described by simpler parameterizations
than for the constant cuts on \ET\ used in Ref.~\cite{atlas-couplings-diboson}, as discussed in 
Sec.~\ref{sec:background_model}.

The typical signal selection efficiency of the kinematic cuts described above ranges between
50\% (for events from \WH\ production) to 60\% (for events from \ttH\ production).

The invariant mass of the two photons is given by 
\begin{equation*}
\mgg = \sqrt{ 2 E_{1} E_{2}  (1- \mathrm{cos}\  \alpha)}, 
\end{equation*}
where $E_{1}$ and $E_{2}$ are the energies of the leading and sub-leading photons and $\alpha$ is the
opening angle of the two photons with respect to their production vertex.
The selection of the correct diphoton production vertex is important for the resolution of the $\alpha$ measurement and thus for
the precise measurement of \mgg.
A position resolution on the diphoton production vertex of about 15~mm in the $z$ direction with the photon trajectories
measured by the EM calorimeter alone is achieved, which is sufficient to keep the contribution from the opening angle to the mass resolution smaller
than the contribution from the energy resolution. 
However, an efficient procedure to select the diphoton production vertex 
among the primary vertex candidates reconstructed with the tracking detector is necessary. This selection allows the information
associated with the primary vertex to be used to compute the track-based quantities used in the object definitions, such as
the computation  of photon isolation with tracks (Sec.~\ref{photon_reco}) and the selection of jets associated with the hard 
interaction (Sec.~\ref{sec:jet_reco}).

The diphoton production vertex is selected from the reconstructed collision vertices 
using a neural-network algorithm. For each vertex the algorithm takes the following as input:
the combined $z$-position of the intersections of the extrapolated photon trajectories
(reconstructed by exploiting the longitudinal segmentation of the calorimeter) with the beam axis;
the sum of the squared transverse momenta $\sum p_\mathrm{T}^2$ and the 
scalar sum of the transverse momenta $\sum p_\mathrm{T}$ of the tracks associated with the
vertex; the difference in azimuthal angle $\Delta \phi$ between 
the direction defined by the vector sum of the track momenta and that of the diphoton system. 
The trajectory of each photon is measured using the longitudinal segmentation of the calorimeter
and a constraint from the average collision point of the proton beams. For converted photons, the
position of the conversion 
vertex is also used if tracks from the conversion have hits in the silicon detectors.

The production vertex selection is studied with ${\Zee}$ events in data and simulation 
by removing the electron tracks from the events and then measuring the efficiency for finding the 
vertex associated with the \Zboson\ boson production. The MC simulation is found to accurately describe 
the efficiency measured in data, as shown in Fig.~\ref{fig:primary_vertex}.
The efficiency for finding the reconstructed diphoton primary vertex $\epsilon_\mathrm{PV}$
in simulated ${\Hgg}$ events from ggF production within 0.3~mm (15~mm) of the
true vertex is around 85\% (93\%) over the typical range of the number of collision vertices per
event observed in the 8~TeV data. The efficiency $\epsilon_\mathrm{PV}$ increases for large diphoton \pT\ as the hadron system recoiling against the diphoton evolves 
into one or more jets, which in turn contain additional higher \pT\ tracks. These additional tracks make it more likely to reconstruct the diphoton vertex
as a primary vertex. Therefore, by re-weighting 
the simulated ${\Zee}$ events to approximate the harder \pT\ spectrum of the simulated
Higgs boson signal, $\epsilon_\mathrm{PV}$ is well reproduced. The corresponding efficiencies for the 7 TeV data and MC samples are slightly higher, due to less pileup, 
and the efficiencies are as consistent as those for the 8 TeV data and MC samples.

\begin{figure}[!htbp]
  \centering
  \includegraphics[width=\columnwidth]{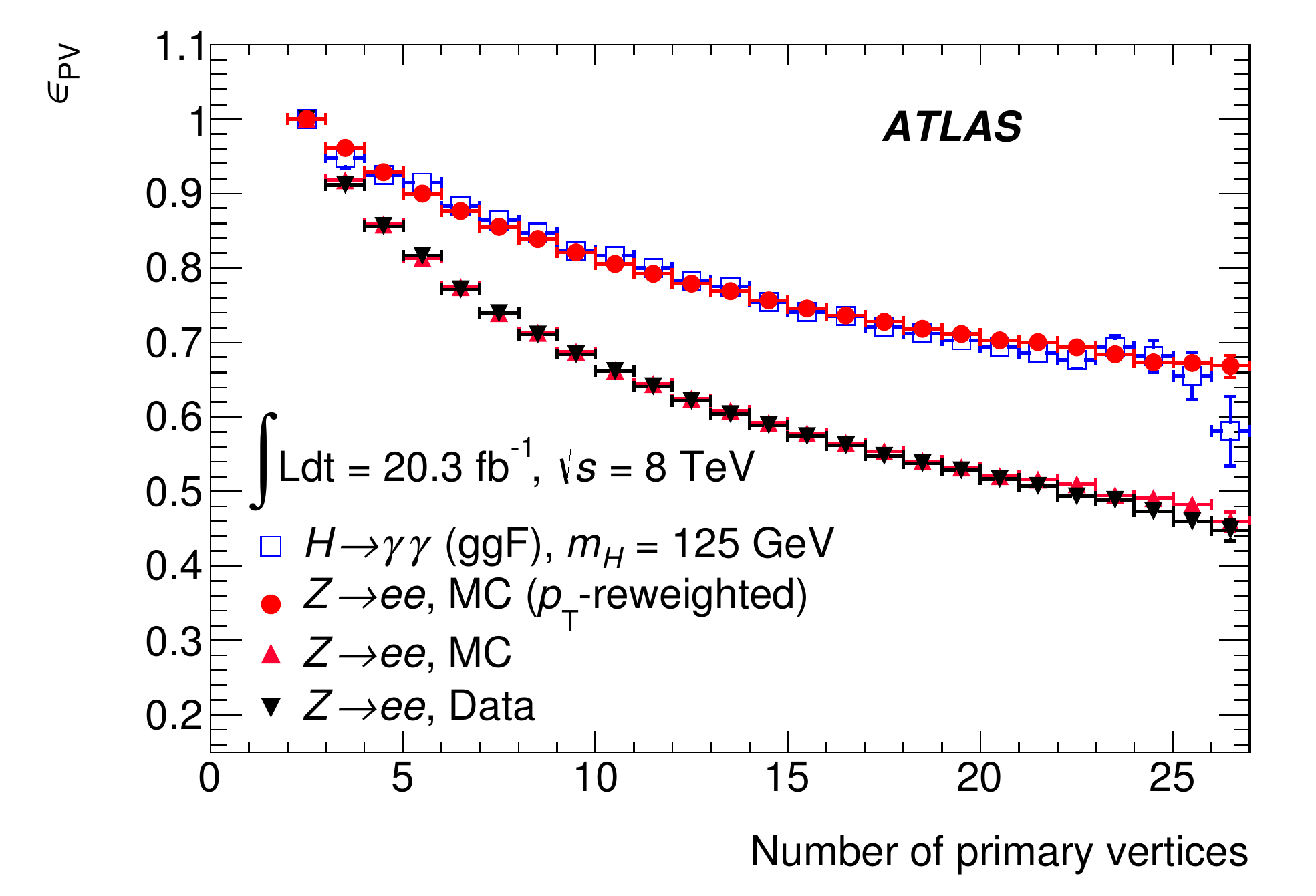} 
  \caption{Efficiency $\epsilon_\mathrm{PV}$ to select a diphoton vertex
within 0.3~mm of the production vertex as a function of the number 
of primary vertices in the event. The plot shows $\epsilon_\mathrm{PV}$ for simulated \ggF\
events (\mH=125 \GeV) with two unconverted photons (hollow blue squares), for \Zee\ events with the electron 
tracks removed for the neural network--based identification of the vertex, both in data 
(black triangles) and  simulation (red triangles), and the same simulated \Zee\ events
re-weighted to reproduce the \pT\ spectrum of simulated \ggF\ events (red circles).
}
  \label{fig:primary_vertex}
\end{figure}

A total of 94566 (17225) collision events at ${\sqrt{s}=8}$~\TeV\ (7~\TeV) were selected with a diphoton 
invariant mass between 105~\GeV\ and 160~\GeV. 
The efficiency to select ${\Hgg}$ events 
is estimated using MC samples and found to range between 32\% and 42\%, depending
on the production mode, as detailed in the following section.

\section{Event categorization}
\label{sec:categorisation}

Gluon fusion is expected to be the dominant production mode of Higgs bosons at the LHC, 
contributing about 87\% of the predicted total production cross section at ${\mH = \combmass}$~\GeV\ and
${\sqrt{s}=7}$--8~TeV, while  
\VBF\ and the associated production processes \VH\ and \ttH\ are predicted to
contribute only 7\%, 5\%, and 1\%, respectively.

Based on their properties, the selected diphoton events (Sec.~\ref{sec:selection}) 
are divided into 12 categories, separately for each of the 7~TeV and 8~TeV datasets, that are optimized
for sensitivity to the Higgs boson production modes studied here, for a Higgs boson mass of ${\mH=125}$~\GeV.
The event selections are applied to the initial diphoton sample in sequence,
as illustrated in Fig.~\ref{fig:categories_new}. Only events that fail all the previous event selections are candidates
for a given category, to ensure that the events are grouped into exclusive categories. The sequence of
categories 
is chosen to give precedence to
the production mechanisms that are expected to have the lowest signal yields. Each category is optimized
by adjusting the event selection criteria to minimize the expected uncertainty in the signal strength of the targeted production process.
Although the measurements are dominated by statistical uncertainties with the present dataset,
systematic uncertainties are taken into account during the optimization.

\begin{figure}[!htbp]
\centering
\includegraphics[height=0.7\textheight]{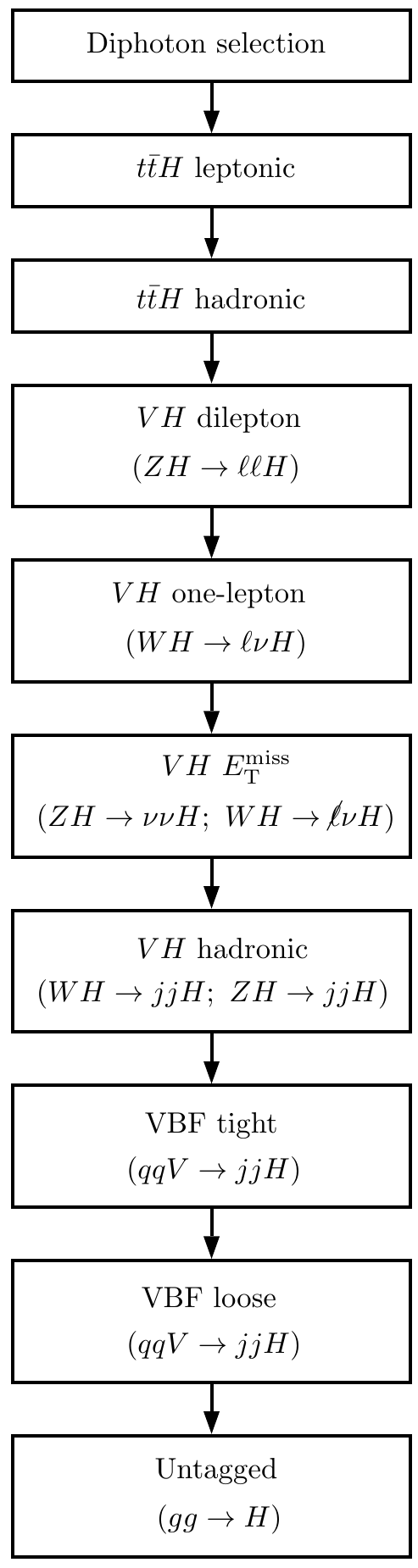}
\caption{Illustration of the order in which the criteria for the exclusive event categories are
applied to the selected diphoton events. The division of the last category, which is dominated by ggF
production, into four
sub-categories is described in Sec.~\ref{sec:untaggedcat}.}
\label{fig:categories_new}
\end{figure}

The 12 exclusive categories,
whose events have different signal invariant mass resolutions and signal-to-background ratios,
can be logically grouped into four sets depending on the production processes
they are expected to be most sensitive to, as described in the following subsections.
Comparisons between signal MC samples, background MC samples, 
and data in the sidebands of the \mgg\ distribution
are shown for the main kinematic quantities used to define several of the categories. 
The sidebands throughout this analysis consist of the relevant
candidate events with ${\mgg}$ in the ranges 105--120~GeV
or 130--160~GeV.

\subsection{Categories sensitive to \ttH}

The two first categories are designed to select data samples enriched
in leptonic and hadronic decays of top quark pairs, using the
event selection described in Ref.~\cite{ttHpaper}.
Events in the \emph{\ttH\ leptonic} category are required to contain
at least one electron or muon with ${\pt>15}$~GeV or ${\pt>10}$~GeV, respectively.
Events are retained if either two or more $b$-jets are found
or a single $b$-jet is found together with ${\MET\geq 20}$~\GeV.
The \mbox{$b$-jets} are required to have ${\pT \geq 25}$~GeV and to be
tagged using the 80\% (85\%) efficiency working point (WP) of the $b$-tagging algorithm~\cite{ATLAS-CONF-2014-046}
in the 8~TeV (7~TeV) data. 
In order to suppress the background contribution from \Zboson+jets with ${\Zee}$,
where a jet and an electron are misidentified as photons, events with an invariant electron--photon mass
of 84--94~GeV are rejected.

Events in the \emph{\ttH\ hadronic} category are required not to have a well-reconstructed and identified lepton
(electron or muon) passing the kinematic cuts described in Sec.~\ref{sec:leptons}.
Also, they are required to fulfill at least one of the following sets of criteria that are partly based on
the $b$-tagger, which is calibrated at several different working points of $b$-tagging efficiency 
\mbox{(Sec.~\ref{sec:jet_reco})}: 
\begin{enumerate}
\item{} at least six jets with \pt $> 25$ \GeV\ out of which two are $b$-tagged using the 80\%\ WP;
\item{} at least six jets with \pt $> 30$ \GeV\ out of which one is $b$-tagged using the 60\%\ WP;
\item{} at least five jets with \pt$ > 30$ \GeV\ out of which two are $b$-tagged using the 70\%\ WP.
\end{enumerate}
Only the first set of criteria above is applied to the 7~TeV data but with a working point efficiency of 85\%. 

The fraction of \ttH\ events relative to all signal production passing this selection
in the hadronic category is larger than 80\% while in the leptonic category it
ranges from 73\% to 84\% depending on the center-of-mass energy;  
the numbers are reported in \mbox{Tables \ref{tabefficiencies2011} and \ref{tabefficiencies2012}}. 
Contributions of about 10\% from \ggH\ events in the hadronic category and 10\% from \WH\ events in the 
leptonic category remain. The remaining 10\% in each of the two categories is accounted for by \WtH\ and \tHbj\ events. 

\subsection{Categories sensitive to \VH}
\label{sec:vhcategories}

In the second step of the categorization the
selection is optimized to 
identify 
events where a Higgs boson is produced in association with a \Zboson\ or \Wboson\ boson.
Compared with our previous studies, a new \VH\ \emph{dilepton} category is
added to separately measure the signal strength parameters for the \ZH\
and \WH\ production modes in order to better test the custodial
symmetry of the Higgs sector ~\cite{atlas-couplings-diboson}.
This new category exploits
the dilepton decay of the
\Zboson\ boson by requiring two same-flavor opposite-sign leptons (electrons
or muons) with ${\pt > 15}$~GeV and ${\pt > 10}$~GeV for electrons and muons,
respectively.
The invariant mass of the two leptons 
is required to be in the range 70--110~\GeV.
These requirements lead to a 99\%\ signal-only purity for 
$ZH$ production, the remaining 1\% coming from \ttH\ production \mbox{(Tables~\ref{tabefficiencies2011} and \ref{tabefficiencies2012})}.

The \mbox{\emph{\VH\ one-lepton}} category is optimized to select events with a leptonic decay of the
\Wboson\ boson by requiring the presence of one electron or muon with \pT\ greater than 15~GeV or
10~GeV, respectively. In order to exploit the missing transverse momentum signature of the
neutrino in the decay chain, the significance of the missing transverse momentum, as
defined in Sec.~\ref{sec:met_definition}, is required to be larger than 1.5.
For the optimization of the selection cuts in this category, the expected background contribution is derived
from data events in the sidebands.
Approximately 90\%\ of the signal events in this category are predicted to come
from \WH\ production, about 6\%\ from \ZH\ production, and 1--2\%\ from \ttH\ production.

The \emph{\VH\ \MET} category is optimized to be enriched in events
from \VH\ production with a leptonic decay of a \Wboson\ boson, where
the lepton is not detected or does not pass the selection for the
one-lepton category,
or with a \Zboson\ boson decay 
to two neutrinos. The minimal requirement on the significance of the missing transverse energy is 5.0, 
roughly equivalent to a direct requirement of ${\MET > 70}$--100~GeV, depending
on the value of $\sum \ET$. 
A further enrichment is obtained by requiring the magnitude \pTt\ \cite{pTtendnote} of the
component of the diphoton 
$\vec{p}_{\mathrm{T}}$
transverse to its thrust axis in the
transverse plane to be greater than 20~GeV. The \pTt\ is used as a discriminant, rather than
the \pT\ of the diphoton, because it is less affected by energy resolution and it is not
correlated with the invariant mass of the diphoton.
As for the \VH\ one-lepton category,
the background distributions for the cut optimizations are extracted from data events in the sidebands.
After the event selection approximately 50\% of the 
signal events in this category are predicted to come from \ZH\ production, 40\% from \WH\ production,
and the remaining 10\% mainly from \ttH\ production 
(Tables~\ref{tabefficiencies2011} and \ref{tabefficiencies2012}).

The \emph{\VH\ hadronic} category consists of events that include the signature of a hadronically
decaying vector boson. They are selected
by requiring the presence of two reconstructed jets with a dijet invariant mass \mjj\
in the range 60--110~GeV. 
The sensitivity is further enhanced
by requiring the difference between the pseudorapidities of the 
diphoton and the dijet systems ${|\eta_{\gamma\gamma} - \eta_{jj} |}$
to be less than one and the diphoton \pTt\ greater than 70~\GeV. 
The distributions
of the discriminating variables used to define the \VH\ hadronic category 
are shown in Fig.~\ref{fig:plotvar_VH} 
for signal events from different production modes and for
events from data and MC background.
The MC background is composed of a mixture of $\gamma\gamma$,
$\gamma$--jet and jet--jet samples normalized as discussed in Sec.~\ref{sec:background_model}.
Approximately 30\% (20\%) of the events in the \VH\ hadronic category  
come from \WH\ (\ZH) production after the selection, while the remaining fraction
is accounted for by \ggH\ events surviving the selection cuts.

\begin{figure}[!htbp]
\centering
\subfigure[]{\includegraphics[width=1.0\columnwidth]{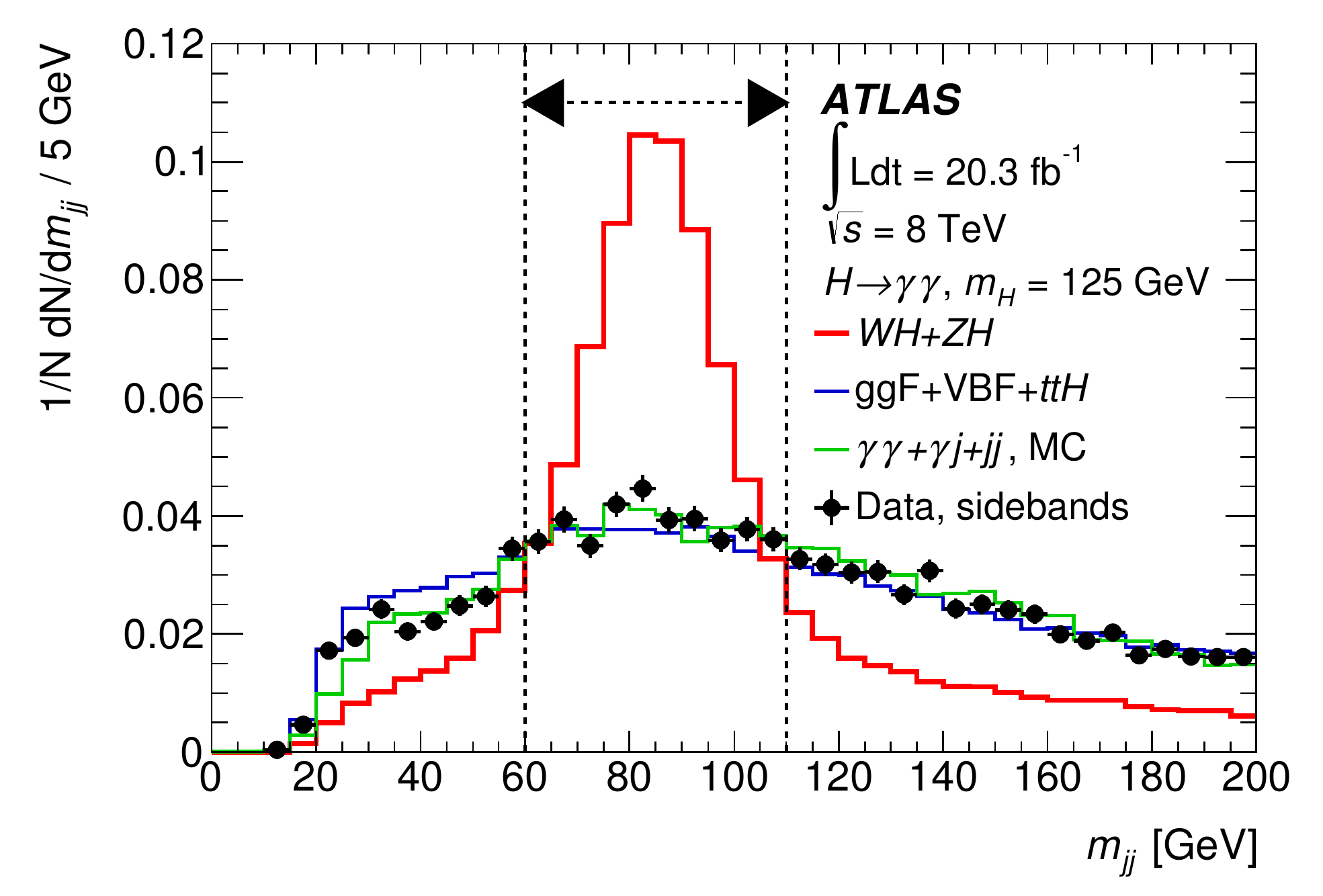}\label{VH_1} }
\subfigure[]{\includegraphics[width=1.0\columnwidth]{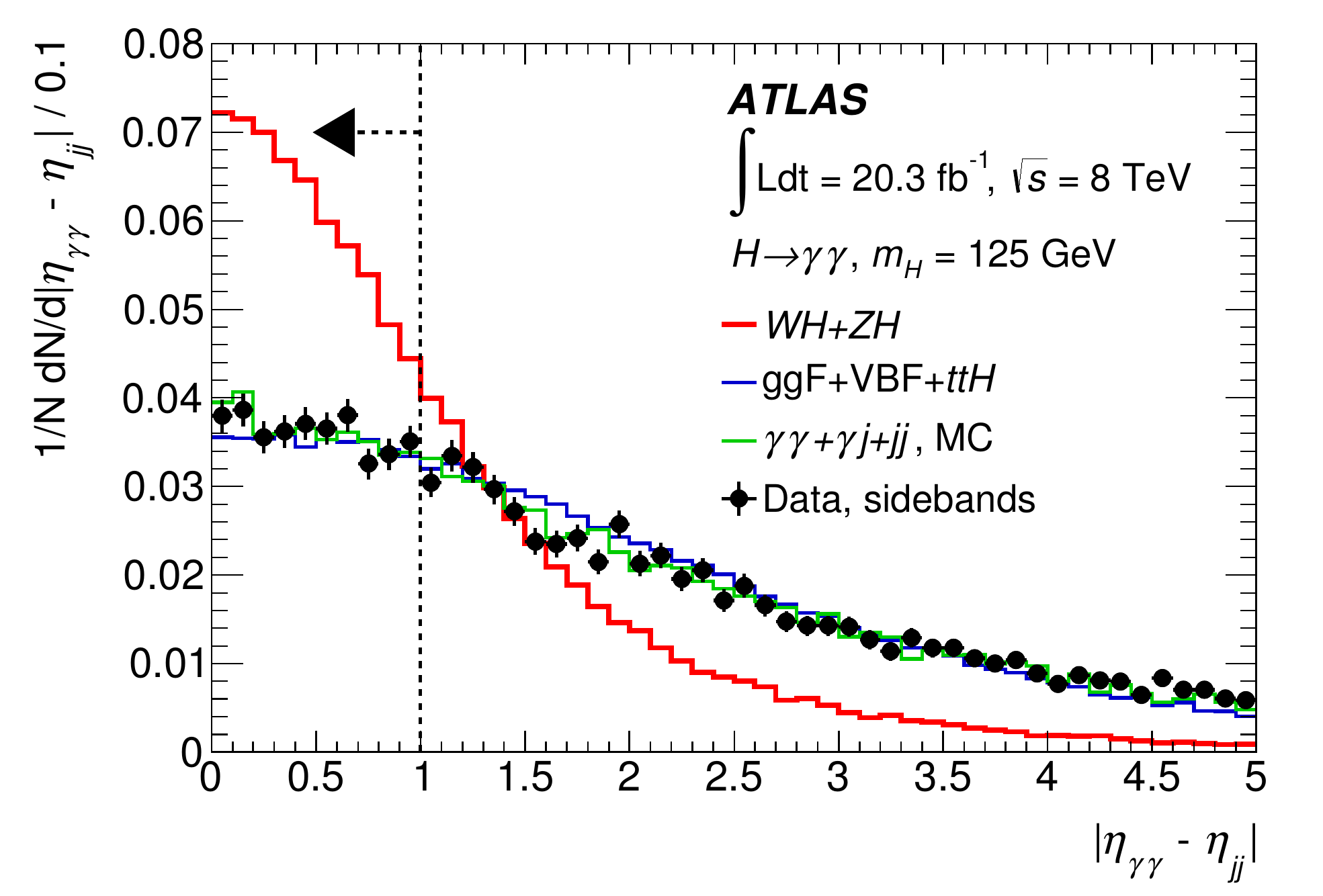}\label{VH_2}}
\subfigure[]{\includegraphics[width=1.0\columnwidth]{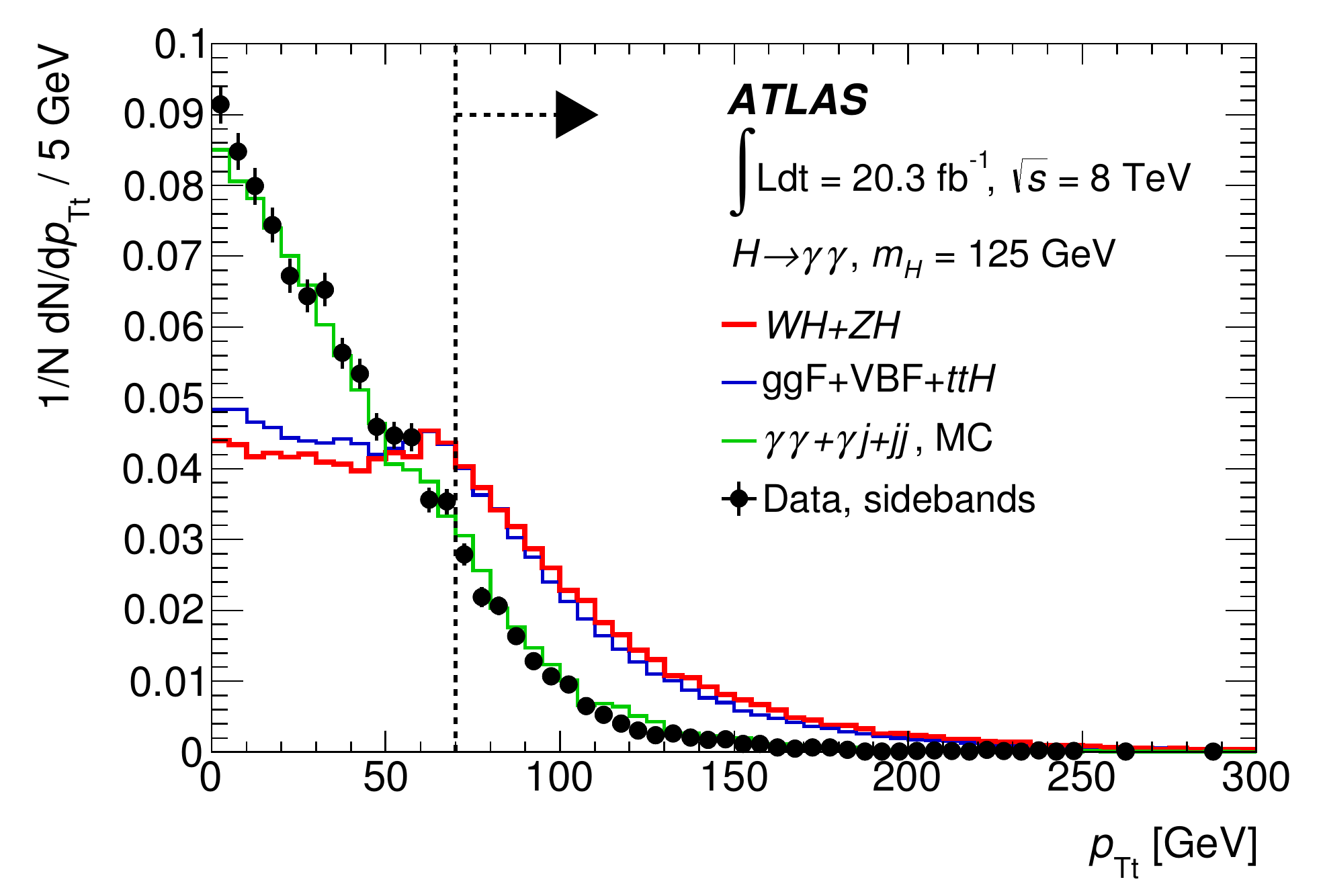}\label{VH_3}}
\caption{Normalized distributions of the variables described in the text (Sec.~\ref{sec:vhcategories})
used to sort diphoton events with
at least two reconstructed jets into the \VH\ hadronic category for the data in the sidebands (points), the predicted
sum of the \WH\ and \ZH\ signals (red histograms), the predicted signal feed-through from
ggF, VBF, and \ttH\   production modes (blue histograms), and the simulation
of the $\gamma\gamma$, $\gamma$--jet, and \mbox{jet--jet} background processes (green histograms).
The arrows indicate the selection criteria applied to these observables.
The mass of the Higgs boson in all signal samples is ${\mH=125}$~\GeV.
} 
\label{fig:plotvar_VH} 
\end{figure}

\subsection{Categories sensitive to \VBF}
\label{subsec:vbf_cat}

Signal events produced by the VBF mechanism are characterized by two well-separated jets with high transverse
momentum and little hadronic activity between them. Events are preselected by requiring at least two reconstructed 
jets. The two leading jets $j1$ and $j2$ (those with the highest \pT) are required to satisfy 
${\Delta\eta_{jj}\geq 2.0}$ and  ${|\eta^{*}|<5.0}$, where $\eta^{*}$ is the pseudorapidity of
the diphoton system relative to the average rapidity of the two leading jets 
${\eta^{*}\equiv\eta_{\gamma\gamma}- (\eta_{j1}+\eta_{j2})/2}$~\cite{1996PhRvD..54.6680R} and
$\Delta\eta_{jj}$ is the pseudorapidity separation between the two leading jets.
In order to optimize the sensitivity to VBF, a multivariate analysis exploits the full event topology
by combining six discriminating variables into a single discriminant 
that takes into account the correlations among them. For this purpose
a BDT is built with the following 
discriminating variables as input: 
\begin{enumerate}
\item{} \mjj, the invariant mass of the two leading jets $j_1$ and $j_2$;
\item{} $\Delta\eta_{jj}$; 
\item{} \pTt, the \pt\ of the diphoton with
respect to its thrust axis in the transverse plane; 
\item{} $\Delta\phi_{\gamma\gamma,jj}$, the azimuthal angle between the diphoton and the dijet systems;
\item{} $\Delta R^\mathrm{min}_{\gamma,j}$, the minimum separation between the leading/subleading photon and the leading/subleading
jet;
\item{} ${\eta^{*}}$.
\end{enumerate}
After the preselection, these variables are found to have little or no correlation to \mgg, thus ensuring that no
biases in the final diphoton mass fit are introduced.
The individual separation power between \VBF\ and  \ggH\ and prompt $\gamma\gamma$, \mbox{$\gamma$--jet} and \mbox{jet--jet} events 
is illustrated in Fig.~\ref{fig:plotvar_VBF} for each discriminating variable. 

The signal sample used to train the BDT is composed of simulated \VBF\ events,
while 
a mixture 
of samples is used for the background: a sample of simulated \ggH\ events, a sample of prompt
diphoton events generated with \textsc{Sherpa} for the irreducible background component,  and events from data in which one
or both photon candidates fail to satisfy the isolation criteria for the reducible $\gamma$--jet and jet--jet components.  
The contribution from \ggH\ to the background sample is normalized to
the rate predicted by the SM. The other background components are weighted
in order to reproduce the background composition measured in the data (see \mbox{Sec.~\ref{sec:background_model}}).

\begin{figure*}[!hp]
  \centering
	\subfigure[]{\includegraphics[width=1.0\columnwidth]{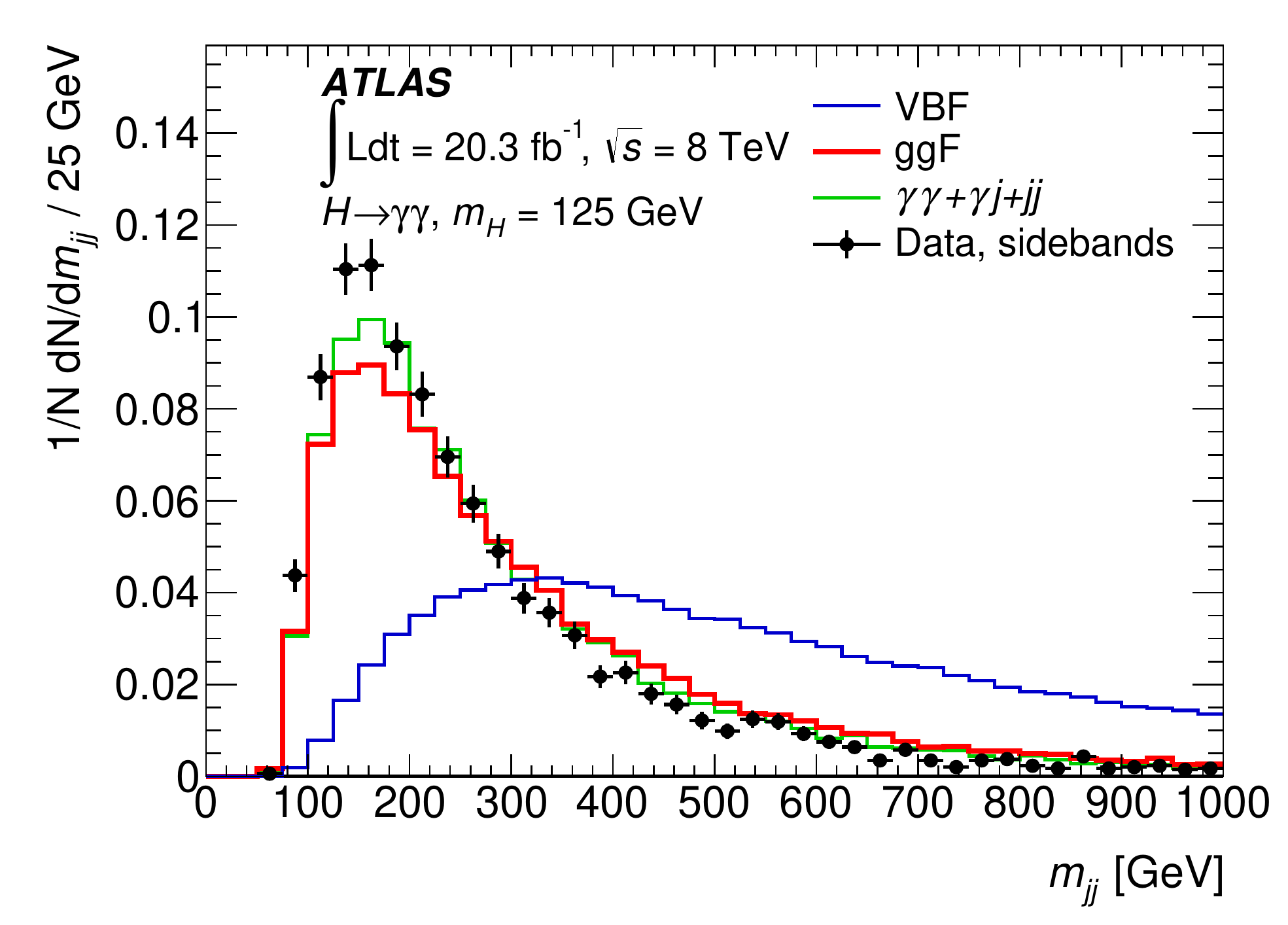}}
	\subfigure[]{\includegraphics[width=1.0\columnwidth]{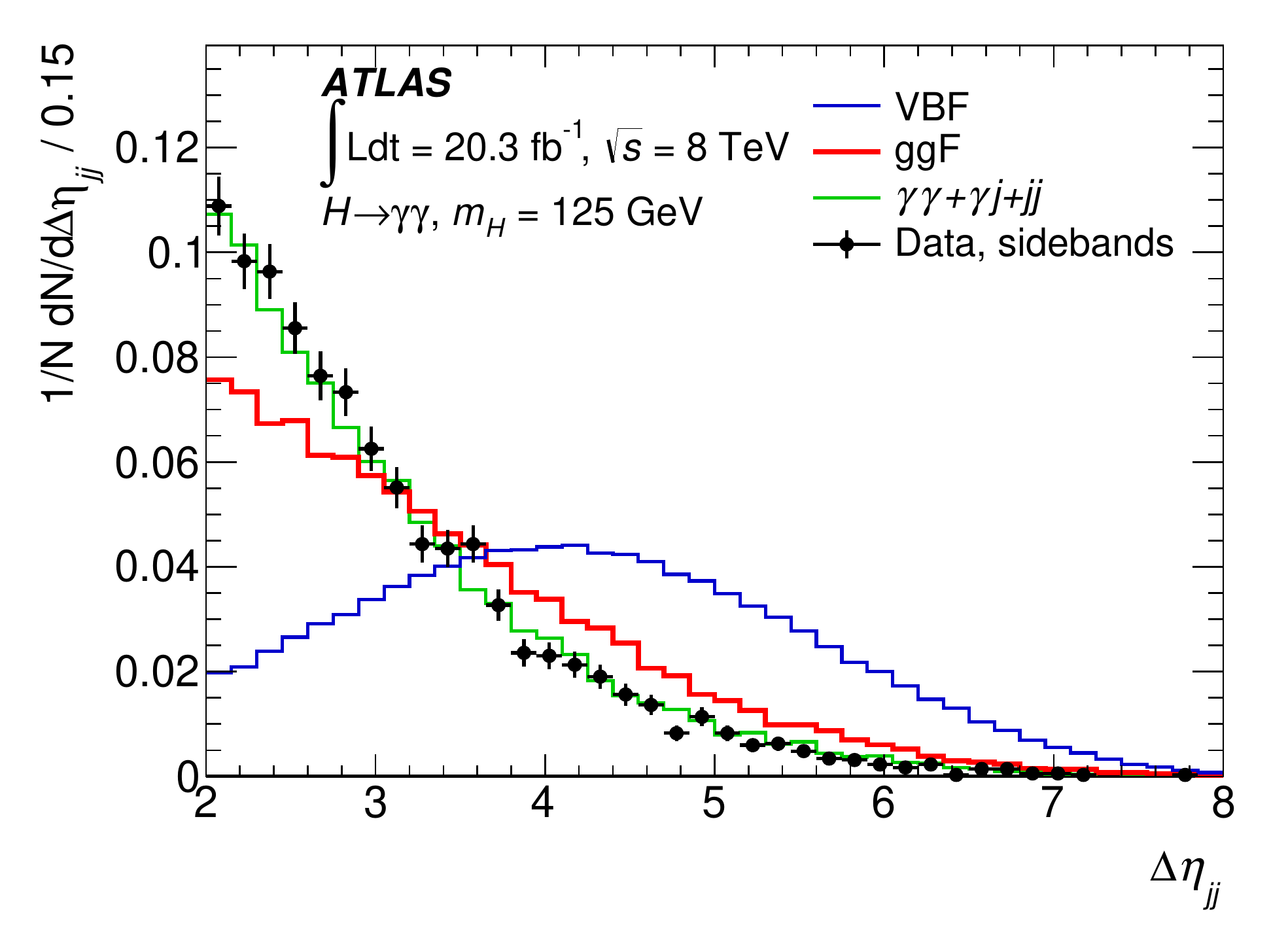}}
	\subfigure[]{\includegraphics[width=1.0\columnwidth]{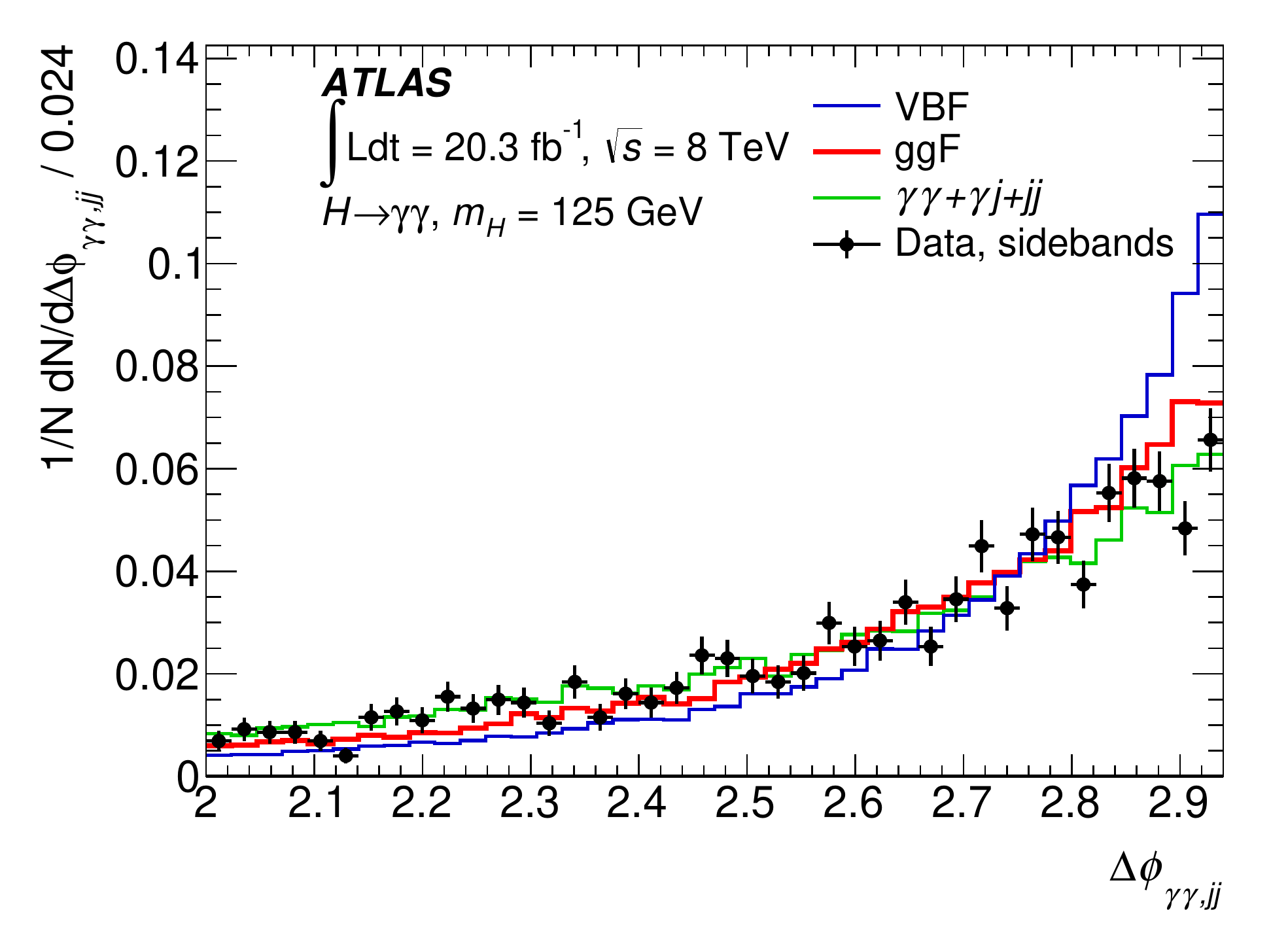}}
	\subfigure[]{\includegraphics[width=1.0\columnwidth]{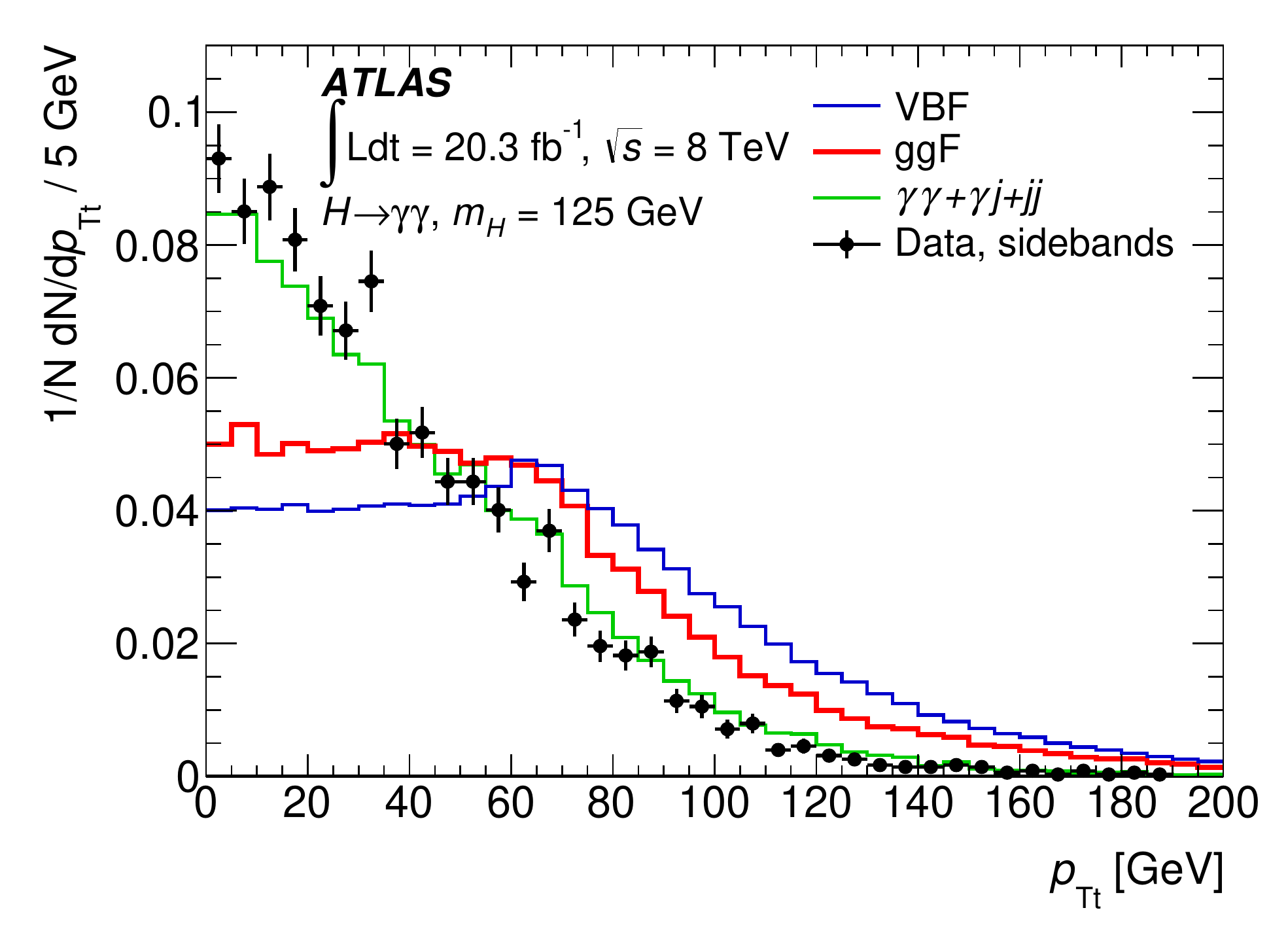}}
	\subfigure[]{\includegraphics[width=1.0\columnwidth]{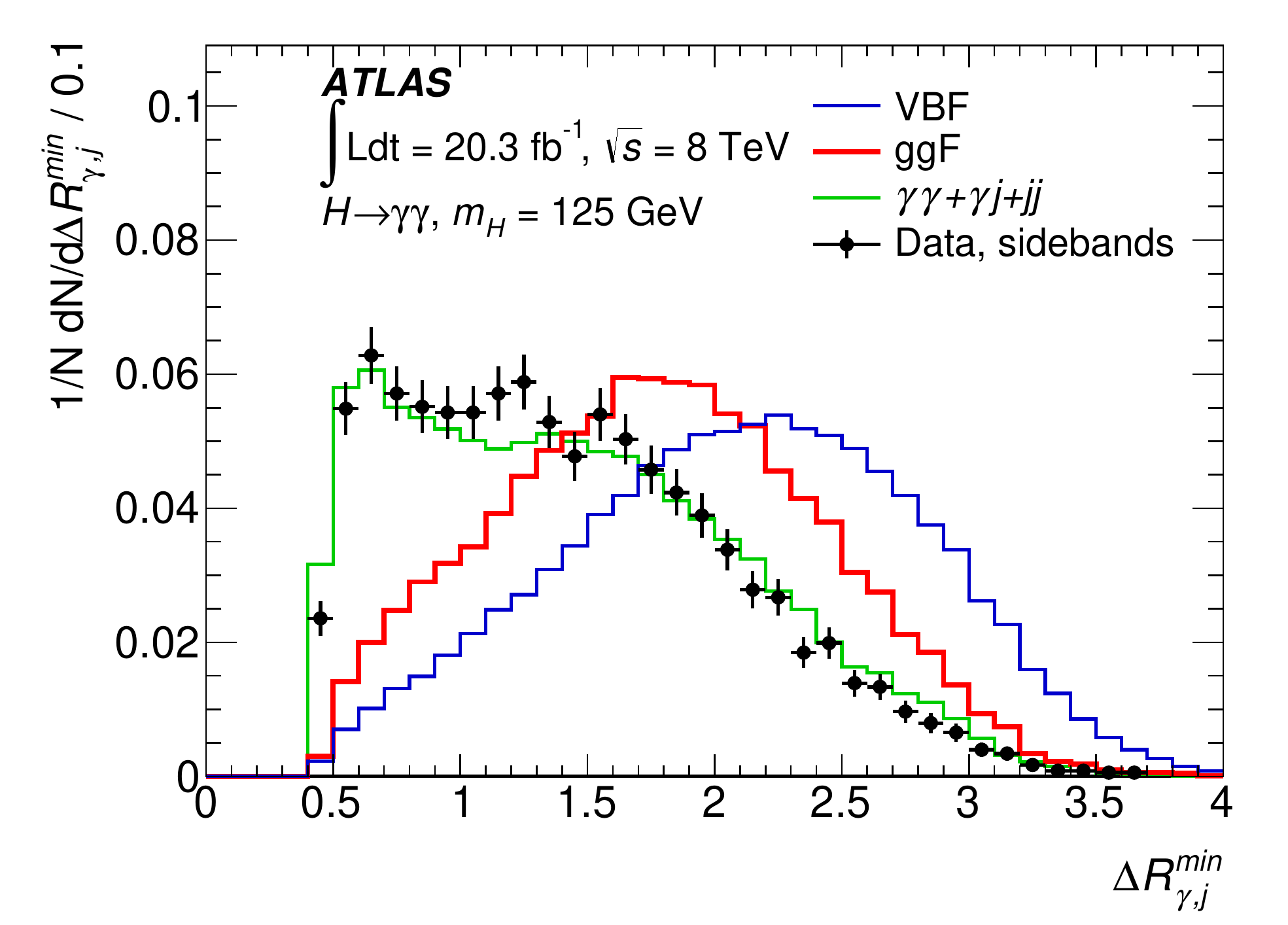}}
	\subfigure[]{\includegraphics[width=1.0\columnwidth]{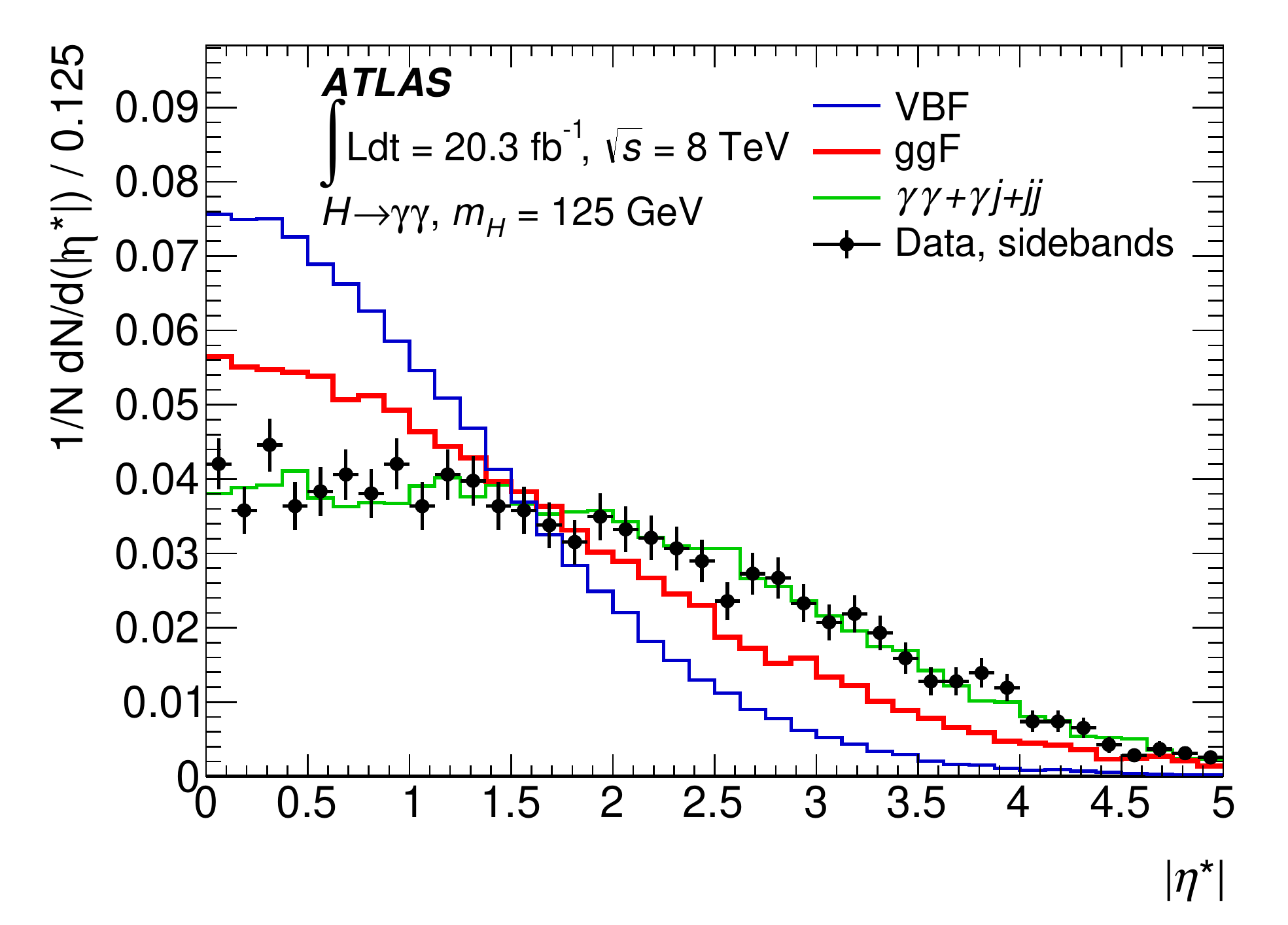}}
	\caption{Normalized kinematic distributions of the six variables describe in the text (Sec.~\ref{subsec:vbf_cat}) used to 
          build the BDT that assigns events to the \VBF\ categories, for diphoton candidates with 
          two well-separated jets (${\Delta\eta_{jj}\geq2.0}$ and 
          ${|\eta^{*}|<5.0}$).  
          Distributions are shown for data sidebands (points) and 
          simulation of the \VBF\ signal  (blue histograms),  feed-through from \ggH\ production (red histograms),
          and the continuum QCD background predicted by MC simulation and data control regions (green histograms) as described in the
          text. The signal \VBF\ and \ggF\ samples
          are generated with a Higgs boson mass ${\mH=125}$~\GeV.}
	\label{fig:plotvar_VBF}
\end{figure*}

Events are sorted into two categories with different VBF purities according to the output value of the BDT, $O_\mathrm{BDT}$:
\begin{enumerate}
\item{} \VBF\ tight: $O_\mathrm{BDT} \geq 0.83$;
\item{} \VBF\ loose: $0.3< O_\mathrm{BDT} <0.83$. 
\end{enumerate}
Figure~\ref{fig:VBF_MVA} shows the distributions of $O_\mathrm{BDT}$ for the \VBF\ signal, feed-through from \ggH\ production, the simulated continuum background, and data from the sidebands. The $O_\mathrm{BDT}$ distributions of the background MC prediction and the data in the sidebands are in good agreement. 
As an additional cross-check, the BDT is applied to a large sample of ${Z(\rightarrow ee)}$+jets in data and
MC samples. The resulting $O_\mathrm{BDT}$ distributions are found to be in excellent agreement.
The fraction of \VBF\ events in the \VBF\ tight (loose) category is approximately 80\% (60\%),  
the remaining 20\% (40\%) being contributed by \ggH\ events. 
An increase of about 6\% in the fraction of \VBF\ events assigned to the \VBF\ categories is obtained with the present
optimization with respect to our
previously published results~\cite{atlas-couplings-diboson}.

\begin{figure}[!htbp]
\centering
\includegraphics[width=\columnwidth]{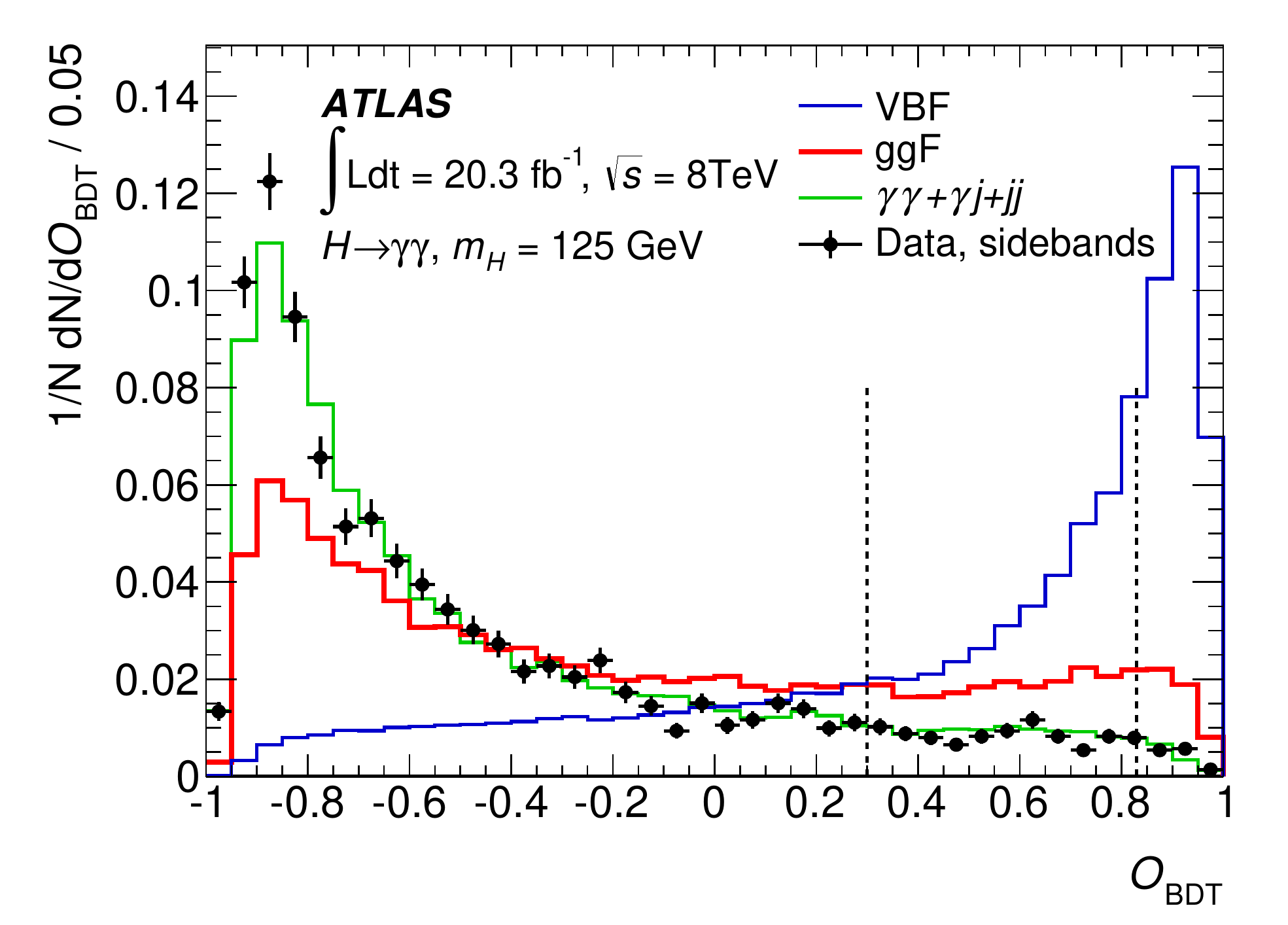}
\caption{Probability distributions of the output of the BDT $O_\mathrm{BDT}$ for the \VBF\ signal (blue), \ggH\ feed-through (red), continuum QCD background 
predicted by MC samples and data control regions (green) as described in the text,
and data sidebands (points). The two vertical dashed lines indicate the cuts on $O_\mathrm{BDT}$ that define the 
VBF loose and VBF tight categories. The signal \VBF\ and \ggF\ samples are generated with a Higgs boson mass ${\mH=125}$~\GeV.}
\label{fig:VBF_MVA} 
\end{figure} 

\subsection{Untagged categories}
\label{sec:untaggedcat}

Compared with our previously published analysis, the categorization of the 
events that are not assigned to the \ttH, \VH\ or \VBF\ categories 
is simplified by reducing the number of \emph{untagged} categories from nine 
to four with no increase in the signal strength uncertainty.
The category definition is based on the \pTt\ of the diphoton system
and the pseudorapidities of the photons:
\begin{enumerate}
\item{} \emph{Central - low  \pTt}: ${\pTt \leq 70}$~\GeV\ and both photons have ${|\eta|<0.95}$;
\item{} \emph{Central - high \pTt}: ${\pTt > 70}$~\GeV\ and both photons have ${|\eta|<0.95}$;
\item{} \emph{Forward - low  \pTt}: ${\pTt \leq 70}$~\GeV\ and at least one photon has ${|\eta|\geq 0.95}$;
\item{} \emph{Forward - high \pTt}: ${\pTt > 70}$~\GeV\ and at least one photon has ${|\eta|\geq 0.95}$.
\end{enumerate}
This categorization of the untagged events increases the signal-to-background ratio of the events with high \pTt\ with a gain of about a
factor of three (two) for central (forward) categories with respect to
low \pTt\ events, as illustrated in Fig.~\ref{fig:untagged_ptt}.
\begin{figure}[!htbp]
\vspace{2pt}
\centering
\subfigure[]{\includegraphics[width=\columnwidth]{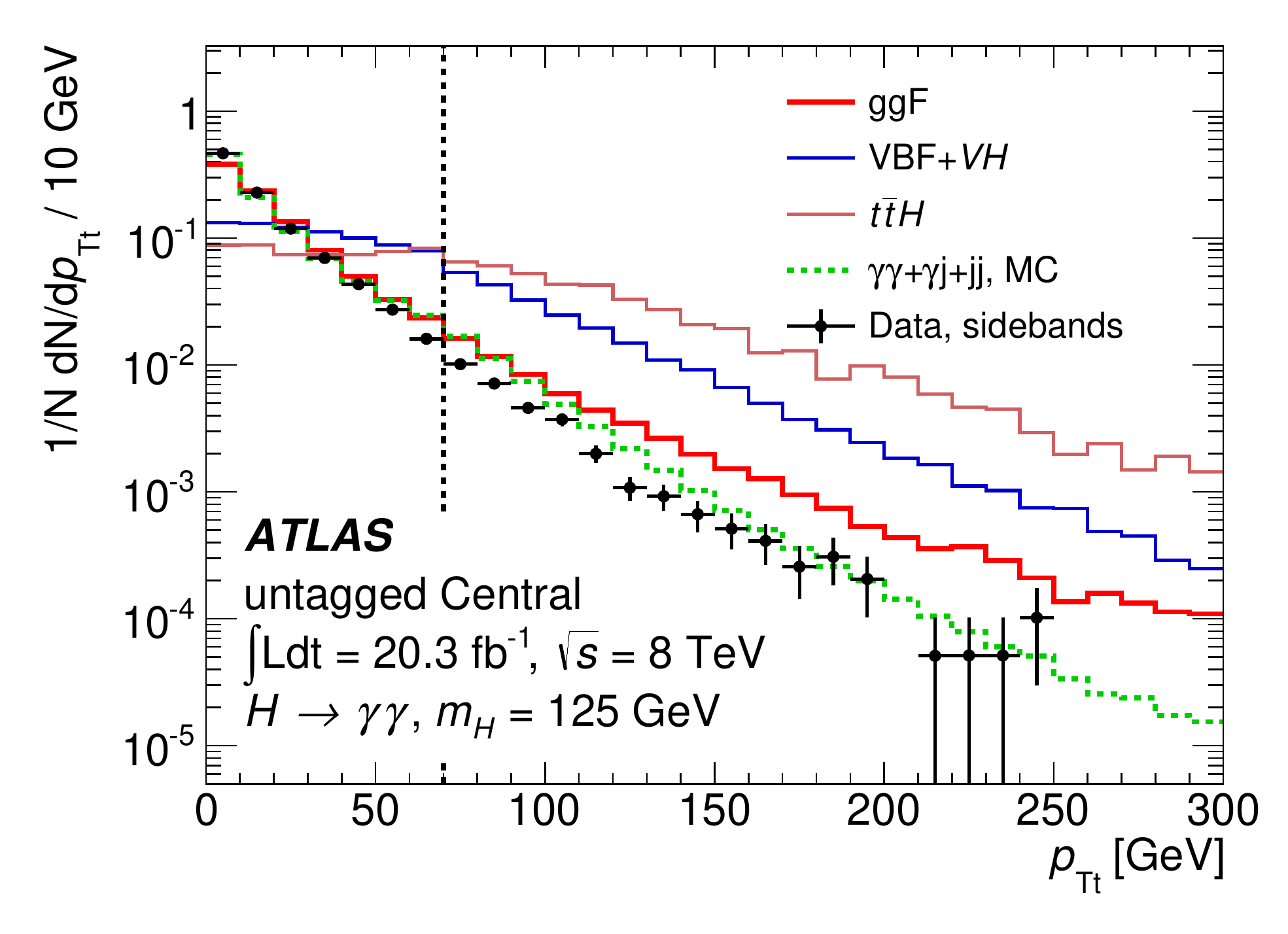}{\label{ptt_central}}}
\subfigure[]{\includegraphics[width=\columnwidth]{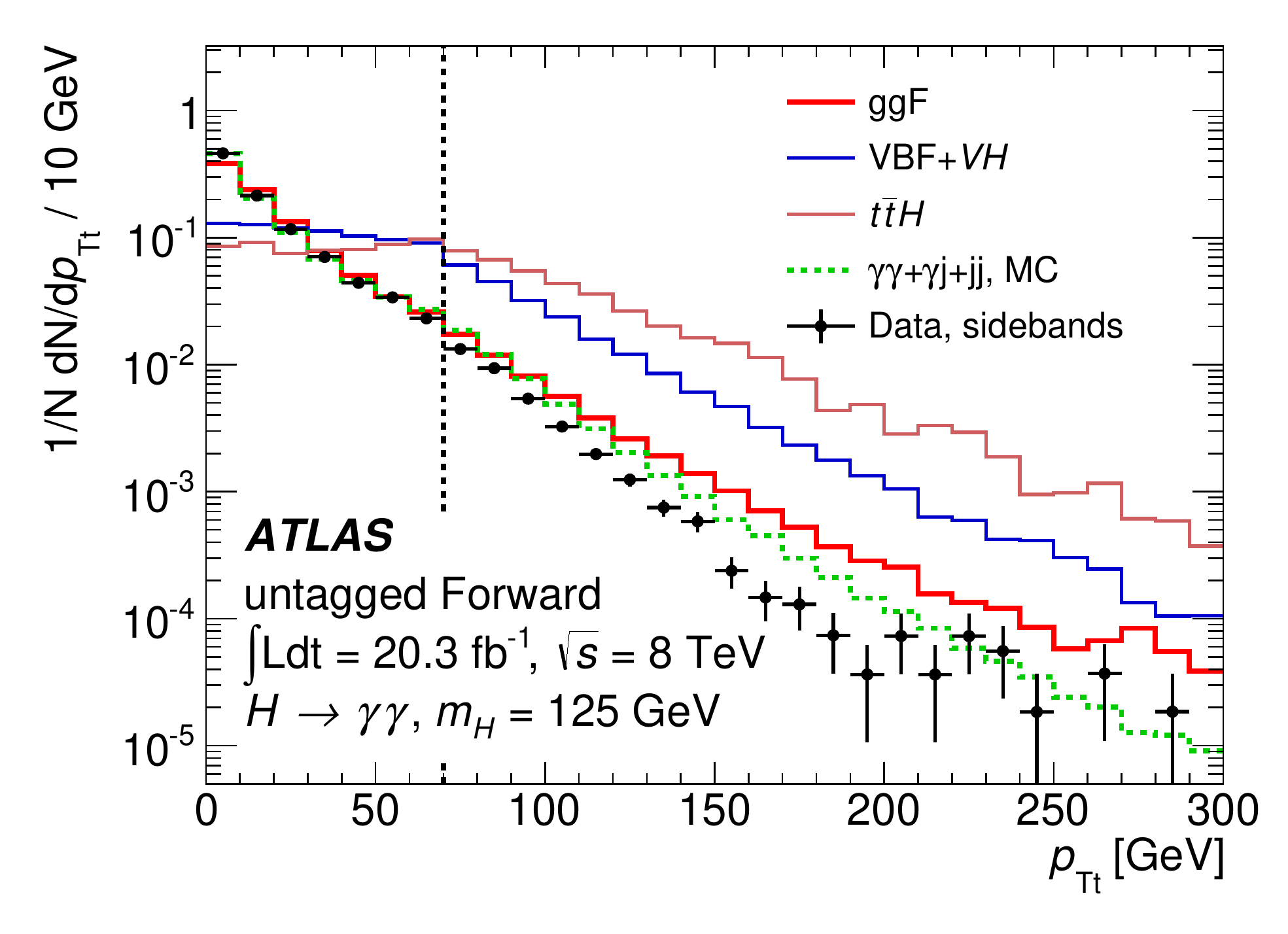}{\label{ptt_forward}}}
\caption{Distributions of the component of the diphoton $\vec{p}_{\mathrm{T}}$
transverse to its thrust axis in the transverse plane \pTt\ for diphoton candidates in the sidebands 
in the untagged (a) Central and (b) Forward categories for ${\sqrt{s}=8}$~TeV
for predicted Higgs boson production processes (solid histograms), the predicted sum of prompt $\gamma\gamma$, $\gamma$--jet
and jet--jet background processes (green histogram), and data (points). The
vertical dashed lines indicate the value used to classify events into the low- or high-\pTt\ categories.
The mass for all Higgs boson signal samples is ${\mH=125}$~\GeV.}
\label{fig:untagged_ptt} 
\end{figure} 
Since the MC background is not used directly in the analysis, the slight mis-modeling observed in the
high-\pTt\ region does not bias the signal measurement,
causing only a suboptimal choice of the discriminating cut.
The typical fraction of \ggH\ events in the low (high) \pTt\ categories is 90\% (70\%). The remaining 10\% (30\%) 
is equally accounted for by the contribution from \VBF\ events and the sum of all the remaining processes. 

\subsection{Summary of categories}
\label{sec:summarycat}

The predicted signal efficiencies, which include geometrical and kinematic acceptances, 
and event fractions per production mode in each event category for ${\mH=\combmass \GeV}$ are listed in
Tables~\ref{tabefficiencies2011} and \ref{tabefficiencies2012} for the 7~TeV and 8~TeV data, respectively.
The total expected numbers of signal events per event category $N_\mathrm{S}$ are also shown,
normalized as discussed in Sec.~\ref{sec:dataset}.

The dependence of the yield for each production process on the Higgs boson mass is parameterized
in each category with simple polynomials that are used to build the statistical model described in Sec.~\ref{sec:statistic}. 
As discussed in Sec.~\ref{sec:dataset},  the detection efficiency for \bbH\ events 
is assumed to be the same as for \ggH\ events. The expected contamination of \ggH\ and \VBF\ in the \VH\ \MET 
category is larger in 7~TeV data than in 8~TeV data
due to the poorer resolution of the \MET\ reconstruction algorithm used in the 7 TeV analysis.

\begin{table*}[h!]\footnotesize
\caption{Signal efficiencies $\epsilon$, which include geometrical and kinematic acceptances,  
 and expected signal event fractions $f$ per production mode in each event category for ${\sqrt{s} = 7}$~TeV and ${\mH =\combmass}$~\GeV. The second-to-last row shows the total efficiency per production process summed over the categories and
the overall average efficiency in the far right column. The total number of selected signal
events expected in each category $N_\mathrm{S}$ is reported in the last column while the total number of selected events expected from each production
mode is given in the last row. 
} 
\label{tabefficiencies2011}
\begin{center} 

\begin{tabular}{lccccccccccccccccc}

\hline\hline
         & \multicolumn{2}{c}{\ggH} & \multicolumn{2}{c}{\VBF} &\multicolumn{2}{c}{\WH} & \multicolumn{2}{c}{\ZH} & \multicolumn{2}{c}{\ttH} & \multicolumn{2}{c}{\bbH} & \multicolumn{2}{c}{\tHbj} & \multicolumn{2}{c}{\WtH} &  \\
Category &  $\epsilon$(\%) & $f$(\%) & $\epsilon$(\%) & $f$(\%) & $\epsilon$(\%)  & $f$(\%) &
$\epsilon$(\%) & $f$(\%) & $\epsilon$(\%) & $f$(\%) & $\epsilon$(\%) & $f$(\%) & $\epsilon$(\%)  & $f$(\%) & $\epsilon$(\%)  & $f$(\%)  & $N_\mathrm{S}$ \\
\hline
Central - low \pTt    &  15.5  &  92.2   &  8.5   &  4.1   &  7.2   &  1.6   &  7.9   &  1.0  &  3.4   &  0.1   &  15.5  &  1.0   &  -  &  -  &  -  &  -  &  26.0  \\
Central - high \pTt   &   1.0  &  71.8   &  2.7   & 16.4   &  2.1   &  6.1   &  2.3   &  3.7  &  2.9   &  1.2   &  1.0   &  0.7   &  -  &  -  &  -  &  -  &  2.1 \\
Forward - low \pTt    &  23.3  &  91.5   & 13.2   &  4.2   &  13.5  &  2.0   & 14.3   &  1.2  &  4.3   &  0.1   &  23.3  &  0.9   &  -  &  -  &  -  &  -  &  39.5 \\
Forward - high \pTt   &   1.3  &  70.6   &  4.0   & 16.7   &  3.5   &  6.9   &  3.6   &  4.1  &  2.9   &  0.9   &  1.3   &  0.7   &  -  &  -  &  -  &  -  &  3.0 \\
\VBF\ loose           &   0.4  &  38.6   &  7.9   & 60.0   &  0.2   &  0.6   &  0.2   &  0.3  &  0.2   &  0.1   &  0.4   &  0.4   &  -  &  -  &  -  &  -  &  1.7 \\
\VBF\ tight           &   0.1  &  18.1   &  6.3   & 81.5   & $<0.1$ &  0.1   & $<0.1$ &  0.1  &  0.1 & $<0.1$ &  0.1   &  0.2   &  -  &  -  &  -  &  -    &  1.0 \\
\VH\ hadronic         &   0.2  &  43.5   &  0.1   &  3.3   &  3.2   &  31.8  &  3.4   &  19.8 &  0.9   &  1.3   &  0.2   &  0.4   &  -  &  -  &  -  &  -  &  0.6 \\
\VH\ \MET             & $<0.1$ &  8.7    &  0.1   &  3.7   &  1.7   &  35.7  &  3.6   &  44.8 &  2.3   &  7.1   & $<0.1$ &  0.1   &  -  &  -  &  -  &  -  &  0.3 \\
\VH\ one-lepton       & $<0.1$ &  0.7    & $<0.1$ &  0.2   &  5.0   &  91.4  &  0.6   &  5.9  &  0.7   &  1.8   & $<0.1$ & $<0.1$ &  -  &  -  &  -  &  -  &  0.3 \\
\VH\ dilepton         & $<0.1$ &  $<0.1$ & $<0.1$ & $<0.1$ & $<0.1$ & $<0.1$ &  1.3 &  99.3 & $<0.1$ &  0.6   & $<0.1$ & $<0.1$ &  -  &  -  &  -  &  -    &  0.1 \\
\ttH\ hadronic        & $<0.1$ &  10.5   & $<0.1$ &  1.3   & $<0.1$ &  1.3   & $<0.1$ &  1.4  &  6.1   &  81.0  & $<0.1$ &  0.1   & 1.5 & 2.6 & 4.3 & 1.9 &  0.1 \\
\ttH\ leptonic        & $<0.1$ &  0.6    & $<0.1$ &  0.1   &  0.3   &  14.9  &  0.1   &  4.0  &  8.5   &  72.6  & $<0.1$ & $<0.1$ & 4.8 & 5.3 & 8.7 & 2.5 &  0.1 \\
\hline
 Total efficiency (\%)& 41.8   & -       & 42.9   & -      & 36.7   & -      & 37.3   & -     & 32.2   & -      & 41.8   & - & - & - & - &- & 41.6\%\\
\hline
Events  &  \multicolumn{2}{c}{64.8}  &  \multicolumn{2}{c}{5.4}  &\multicolumn{2}{c}{2.2} & \multicolumn{2}{c}{1.3} & \multicolumn{2}{c}{0.3} & \multicolumn{2}{c}{0.7} & \multicolumn{2}{c}{$<0.1$}& \multicolumn{2}{c}{$<0.1$}& 74.5 \\
\hline\hline

\end{tabular}
\end{center}

\end{table*}

\begin{table*}[h!]\footnotesize
\caption{Signal efficiencies $\epsilon$, which include geometrical and kinematic acceptances, 
 and expected signal event fractions $f$ per production mode in each event category for ${\sqrt{s} = 8}$~TeV and ${\mH =\combmass}$~\GeV. The second-to-last row shows the total efficiency per production process summed over the categories and
the overall average efficiency in the far right column. The total number of selected signal
events expected in each category $N_\mathrm{S}$ is reported in the last column while the total number of selected events from each production
mode is given in the last row. 
}
\label{tabefficiencies2012}
\begin{center}
\begin{tabular}{lccccccccccccccccc}
\hline\hline
         & \multicolumn{2}{c}{\ggH} & \multicolumn{2}{c}{\VBF} &\multicolumn{2}{c}{\WH} & \multicolumn{2}{c}{\ZH} & \multicolumn{2}{c}{\ttH} & \multicolumn{2}{c}{\bbH} & \multicolumn{2}{c}{\tHbj} & \multicolumn{2}{c}{\WtH} &   \\
Category &  $\epsilon$(\%) & $f$(\%) & $\epsilon$(\%) & $f$(\%) & $\epsilon$(\%)  & $f$(\%) &
$\epsilon$(\%) & $f$(\%) & $\epsilon$(\%) & $f$(\%) & $\epsilon$(\%) & $f$(\%) & $\epsilon$(\%)  & $f$(\%) & $\epsilon$(\%)  & $f$(\%)  & $N_\mathrm{S}$ \\
\hline
Central - low \pTt   &  14.1  &  92.3  &  7.5   &  4.0   &  6.5   &  1.5   &  7.2   &  1.0  &  2.9   &  0.1  &  14.1  &  1.0   &  -  &  -  &  -  &  -  &  135.5  \\
Central - high \pTt  &   0.9  &  73.3  &  2.5   &  15.7  &  1.9   &  5.5   &  2.0   &  3.4  &  2.4   &  1.3  &  0.9   &  0.8   &  -  &  -  &  -  &  -  &  11.3 \\
Forward - low \pTt   &  21.6  &  91.7  & 11.9   &  4.1   & 12.3   &  1.9   &  13.0  &  1.2  &  3.8   &  0.1  &  21.6  &  1.0   &  -  &  -  &  -  &  -  &  208.6 \\
Forward - high \pTt  &   1.3  &  71.9  &  3.6   &  16.2  &  3.2   &  6.4   &  3.3   &  3.9  &  2.5   &  0.9  &  1.3   &  0.8   &  -  &  -  &  -  &  -  &  16.1 \\
\VBF\ loose	     &   0.4  &  41.9  &  7.2   &  56.5  &  0.2   &  0.6   &  0.2   &  0.4  &  0.2   &  0.1  &  0.4   &  0.4   &  -  &  -  &  -  &  -  &  9.3 \\
\VBF\ tight          &   0.1  &  19.0  &  6.4   &  80.5  & $<0.1$ &  0.2   & $<0.1$ &  0.1  &  0.1   &  0.1  &  0.1   &  0.2   &  -  &  -  &  -  &  -  &  5.7 \\
\VH\ hadronic        &   0.2  &  45.9  &  0.1   &  3.2   &  3.0   & 30.3   &  3.1   & 18.8  &  0.7   &  1.3  &  0.2   &  0.5   &  -  &  -  &  -  &  -  &  3.2 \\
\VH\ \MET            & $<0.1$ &   2.3  & $<0.1$ &  0.3   &  1.3   & 36.9   &  3.0   & 51.0  &  1.8   &  9.5  & $<0.1$ & $<0.1$ &  -  &  -  &  -  &  -  &  1.1 \\
\VH\ one-lepton      & $<0.1$ &   0.5  & $<0.1$ &  0.2   &  4.8   & 89.8   &  0.6   &  6.3  &  1.0   &  3.3  & $<0.1$ & $<0.1$ &  -  &  -  &  -  &  -  &  1.7 \\
\VH\ dilepton        & $<0.1$ & $<0.1$ & $<0.1$ & $<0.1$ & $<0.1$ & $<0.1$ &  1.3   & 99.1  & $<0.1$ &  0.9  & $<0.1$ & $<0.1$ &  -  &  -  &  -  &  -  &  0.3 \\
\ttH\ hadronic       & $<0.1$ &   7.3  & $<0.1$ &  1.0   & $<0.1$ &  0.7   & $<0.1$ &  1.3  &  6.9   &  84.1 & $<0.1$ & $<0.1$ &  2.1  &  3.4  &  4.8  &  2.1  &  0.5 \\
\ttH\ leptonic       & $<0.1$ &   1.0  & $<0.1$ &  0.2   &  0.1   &  8.1   &  0.1   &  2.3  &  7.9   &  80.3 & $<0.1$ & $<0.1$ &  4.1  &  5.5  &  7.1  &  2.6  &  0.6 \\
\hline
 Total efficiency (\%)& 38.7  &  -     & 39.1   & -      & 33.3  & - & 33.8  & - & 30.2  & - & 38.7 & - & & & & & 38.5\%\\
\hline
 Events  &  \multicolumn{2}{c}{342.8}  &  \multicolumn{2}{c}{28.4}  &\multicolumn{2}{c}{10.7}
 & \multicolumn{2}{c}{6.4} & \multicolumn{2}{c}{1.8} & \multicolumn{2}{c}{3.6} & \multicolumn{2}{c}{$<0.1$}& \multicolumn{2}{c}{$<0.1$}& 393.8\\
\hline\hline
\end{tabular}
\end{center}
\end{table*}

The number of events observed in data in each category is reported in Table~\ref{tab:events_category_data}
separately for the 7~TeV and 8~TeV data.
The impact of the event categorization described in the previous sections on the uncertainty in
the combined signal strength is estimated on a representative signal plus MC background sample
generated under the SM hypothesis ($\mu=1$): the event categorization is found to provide a  
20\% reduction of the total uncertainty with respect to an inclusive analysis.

\begin{table}[!htb]
\footnotesize
\caption{Number of selected events in each event category for the 7 TeV and 8 TeV data and with  
a diphoton candidate invariant mass between 105~GeV and 160~\gev.}  
\label{tab:events_category_data}
\begin{center}
\begin{tabular}{lcc}
\hline\hline
      Category & ${\sqrt{s}=7}$~TeV & ${\sqrt{s}=8}$~TeV \\
\hline
Central - low  \pTt   &    4400  & 24080\\
Central - high \pTt  &       141  &     806 \\     
Forward - low  \pTt &   12131 & 66394 \\   
Forward - high \pTt &       429  &  2528 \\    
\VBF\ loose              &         58   &    411\\       
\VBF\ tight                &            7   &      67 \\       
\VH\ hadronic          &         34    &   185 \\      
\VH\ \MET                 &        14     &     35 \\       
\VH\ one-lepton       &          5     &     38 \\       
\VH\ dilepton            &          0     &       2 \\       
\ttH\ hadronic           &          3      &    15  \\       
\ttH\ leptonic             &          3      &      5 \\       
\hline\hline
\end{tabular}
\end{center}
\end{table}

\section{Signal and background models}
\label{sec:signalandbackgroundmodels}
The $\mgg$ distribution of the data in each category is fitted with the
sum of a signal model plus an analytic parameterization of the background.
The signal and background models are described in this section.

\subsection{Signal model}
\label{sec:signal_model}

The normalized distribution of \mgg\ for signal events in each category $c$
is described by a composite model $f_{\mathrm{S},c}$ resulting from
the sum of a Crystal 
Ball function $f_{\mathrm{CB},c}$~\cite{crystalball} (a Gaussian core with one exponential tail) and a small, wider Gaussian component $f_{\mathrm{GA},c}$.
The function $f_{\mathrm{CB},c}$ represents the core 
of well-reconstructed events, while the Gaussian component $f_{\mathrm{GA},c}$ 
is used to describe the outliers of the distribution. The signal model for a given event category and value of \mH\ can be written as:
\begin{multline}
f_{\mathrm{S},c} (\mgg,\mu_{\mathrm{CB},c},\sigma_{\mathrm{CB},c},\alpha_{\mathrm{CB},c}, n_\mathrm{CB}, \phi_{\mathrm{CB},c}, \mu_{\mathrm{GA},c},\sigma_{\mathrm{GA},c}) \\
= \phi_{\mathrm{CB},c}~ f_{\mathrm{CB},c} (\mgg,\mu_{\mathrm{CB},c},\sigma_{\mathrm{CB},c},\alpha_{\mathrm{CB},c}, n_\mathrm{CB}) \\
+ (1-\phi_{\mathrm{CB},c}) f_{\mathrm{GA},c} (\mgg,\mu_{\mathrm{GA},c},\sigma_{\mathrm{GA},c}),
\label{eq:signal_model}
\end{multline}
where $\mu_{\mathrm{CB},c}$, $\sigma_{\mathrm{CB},c}$ are the peak position and the width of the
Gaussian core of the Crystal Ball function 
\begin{widetext}
\begin{equation*}
  f_{\mathrm{CB},c}(\mgg,\mu_{\mathrm{CB},c},\sigma_{\mathrm{CB},c},\alpha_{\mathrm{CB},c}, n_\mathrm{CB}) = {\cal N}_c
\begin{cases}
  e^{-t^{2}/2} & t>-\alpha_{\mathrm{CB},c} \\
 \left( \frac{n_\mathrm{CB}}{|\alpha_{\mathrm{CB},c}|} \right)^{n_\mathrm{CB}}  e^{-|\alpha_{\mathrm{CB},c}|^{2}/2}  \left( \frac{n_\mathrm{CB}}{\alpha_{\mathrm{CB},c}} - \alpha_{\mathrm{CB},c} - t \right)^{-n_\mathrm{CB}} & t < -\alpha_{\mathrm{CB},c}
\end{cases},
\end{equation*}
\end{widetext}
where $t=(\mgg-\mu_{\mathrm{CB},c})/\sigma_{\mathrm{CB},c}$, ${\cal N}_c$ normalizes the distribution,
and $\mu_{\mathrm{GA},c}$, $\sigma_{\mathrm{GA},c}$ are the peak position and the
width of the Gaussian component of the model due to the outliers ($\mu_{\mathrm{CB},c}$ and $\mu_{\mathrm{GA},c}$
are fitted independently but both take on values close to \mH). 
The non-Gaussian tail of  $f_{\mathrm{CB},c}$ is parameterized by $\alpha_{\mathrm{CB},c}$ and $n_\mathrm{CB}$.
The fraction of the composite model due to the Crystal Ball component is described by $\phi_{\mathrm{CB},c}$.

Since the model parameters exhibit a smooth dependence on the values of \mH\ in the simulated signal
samples, the precision of the fit results is improved by assuming a polynomial dependence of the parameters on \mH.
The coefficients of the polynomials, except for $n_\mathrm{CB}$, which is fixed to a constant value
for all categories, are determined for each event category by a simultaneous fit to the relevant
sets of simulated signal mass peaks (Sec.~\ref{sec:dataset}) weighted by
their contributions to the signal yield expected in the SM.
For example, $\mu_{\mathrm{CB},c}$ is to a good approximation found to be equal to the test value of \mH. 
The model parameters extracted for ${\mH=125.4}$~GeV are inputs to the extended likelihood 
function described in Sec.~\ref{sec:statistic}.

The invariant mass resolutions $\sigma_{68}$, defined as half of the smallest $\mgg$-interval containing 
68\% of the signal events, for the 12 event categories are in the range 1.32--1.86 GeV (1.21--1.69 GeV) 
for the 8 TeV (7 TeV) data at ${\mH=\combmass}$~\GeV.
They are reported in Table~\ref{tab:signal_parameters}. 
The slightly smaller invariant mass resolution in the 7 TeV signal samples arises from a different
effective constant term in the energy resolution measured with \Zee\ events and from the 
lower pileup level in the 7~TeV data~\cite{calibration_paper}.

\begin{table}[!h]\footnotesize
  \caption{Effective signal mass resolutions $\sigma_{68}$ and $\sigma_{90}$
    for the 7~\TeV\ and 8~TeV data in each event category, where
    $\sigma_{68}$ ($\sigma_{90}$) is defined as half of
    the smallest interval expected to contain 68\% (90\%) of the signal events 
    ($N_\mathrm{S}$ in Table~\ref{tabefficiencies2012}) for a mass ${\mH=\combmass}$~\GeV. 
  }  
  \label{tab:signal_parameters}
  \begin{center}
    \begin{tabular}{lcccc}
      \hline\hline
      \multirow{2}{*}{Category}      &  \multicolumn{2}{c}{$\sqrt{s}$=7~TeV}       & \multicolumn{2}{c}{$\sqrt{s}$=8~TeV}     \\
                                     &  $\sigma_{68}$~[GeV]  & $\sigma_{90}$~[GeV]  & $\sigma_{68}$~[GeV]  & $\sigma_{90}$~[GeV] \\
      \hline
      Central - low  \pTt    & 1.36 & 2.32 & 1.47 & 2.50 \\
      Central - high  \pTt   & 1.21 & 2.04 & 1.32 & 2.21 \\
      Forward - low \pTt     & 1.69 & 3.03 & 1.86 & 3.31 \\
      Forward - high \pTt    & 1.48 & 2.59 & 1.64 & 2.88 \\
      \VBF\ loose            & 1.43 & 2.53 & 1.57 & 2.78 \\
      \VBF\ tight            & 1.37 & 2.39 & 1.47 & 2.61 \\
      \VH\ hadronic          & 1.35 & 2.32 & 1.45 & 2.57 \\
      \VH\ \MET              & 1.41 & 2.44 & 1.56 & 2.74 \\
      \VH\ one-lepton        & 1.48 & 2.55 & 1.61 & 2.80 \\
      \VH\ dilepton          & 1.45 & 2.59 & 1.59 & 2.76 \\
      \ttH\ hadronic         & 1.39 & 2.37 & 1.53 & 2.64 \\
      \ttH\ leptonic         & 1.42 & 2.45 & 1.56 & 2.69 \\
      \hline\hline
    \end{tabular}
  \end{center}
\end{table}

The \mgg\ distributions of simulated signal events generated with ${\mH=125}$~\gev\
at ${\sqrt{s}=8}$~TeV assigned to the categories with the best (Central - high \pTt) and worst 
mass resolution (Forward - low \pTt) are shown in Fig.~\ref{fig:signalinvmass_comp} together with the
signal models resulting from the simultaneous fits described above.
The signal resolution predicted by the MC simulation varies by less than 10\% over the full range of pileup
conditions in the analyzed data, as shown in Fig.~\ref{fig:signal_pv_pileup}. 
This figure also shows the predicted signal resolution obtained
using the two  primary vertex algorithms discussed in Sec.~\ref{sec:selection} 
compared with the ideal case in which the true vertex from the MC simulation is used.

\begin{figure}[!h]
  \centering
  \includegraphics[width=\columnwidth]{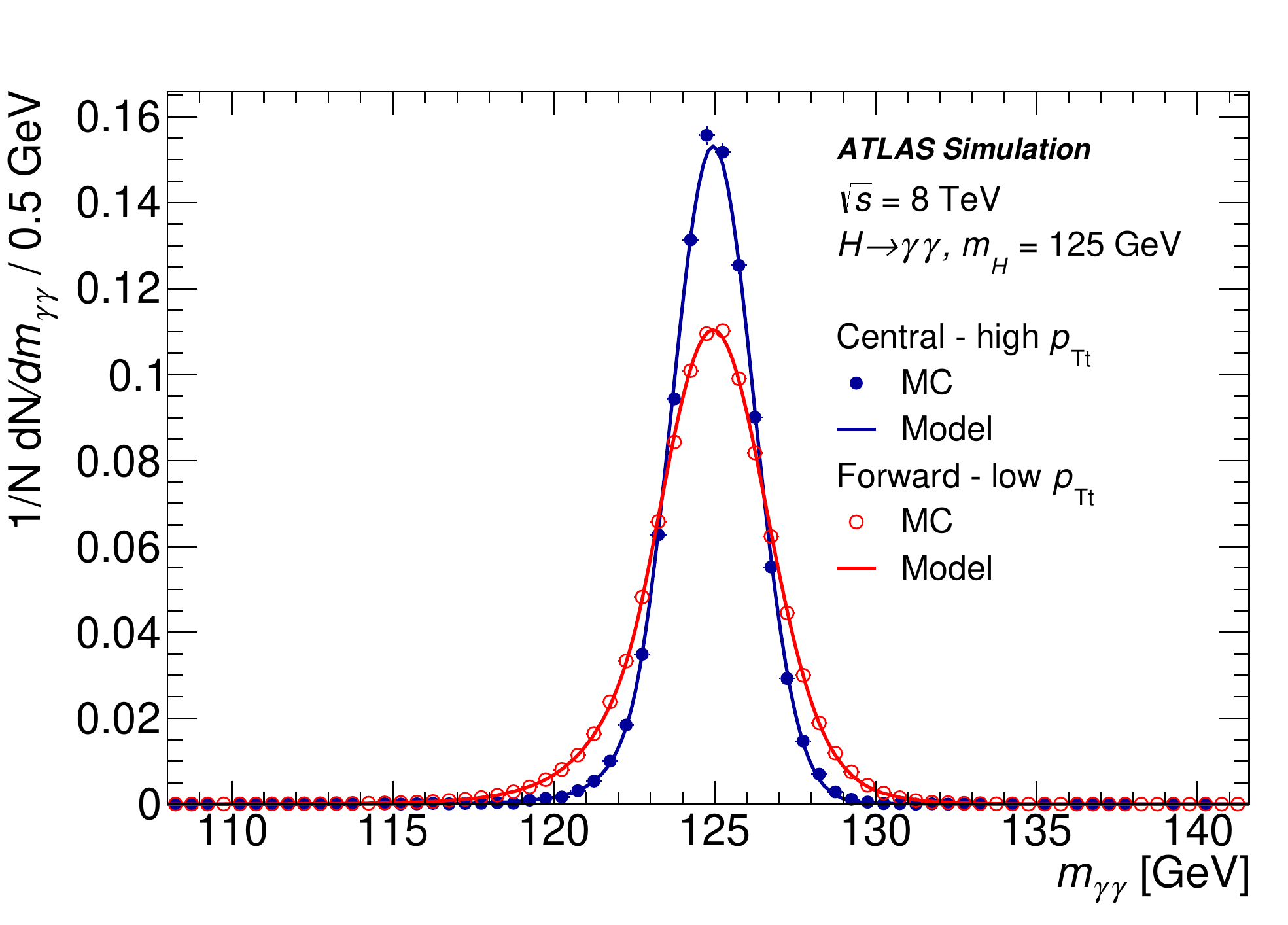}
  \caption{Distributions of diphoton invariant mass \mgg\ in a sample of Higgs boson events generated with ${\mH=125}$~\gev\
    at ${\sqrt{s}=8}$~TeV in the categories with the best resolution (Central - high \pTt, $\sigma_{68}=1.32$~GeV) and worst resolution
    (Forward - low \pTt, $\sigma_{68}=1.86$~GeV) together with the signal models resulting from the
    simultaneous fits described in the text.
  \label{fig:signalinvmass_comp}}
\end{figure}

\begin{figure}[!h]
  \centering
  \includegraphics[width=\columnwidth]{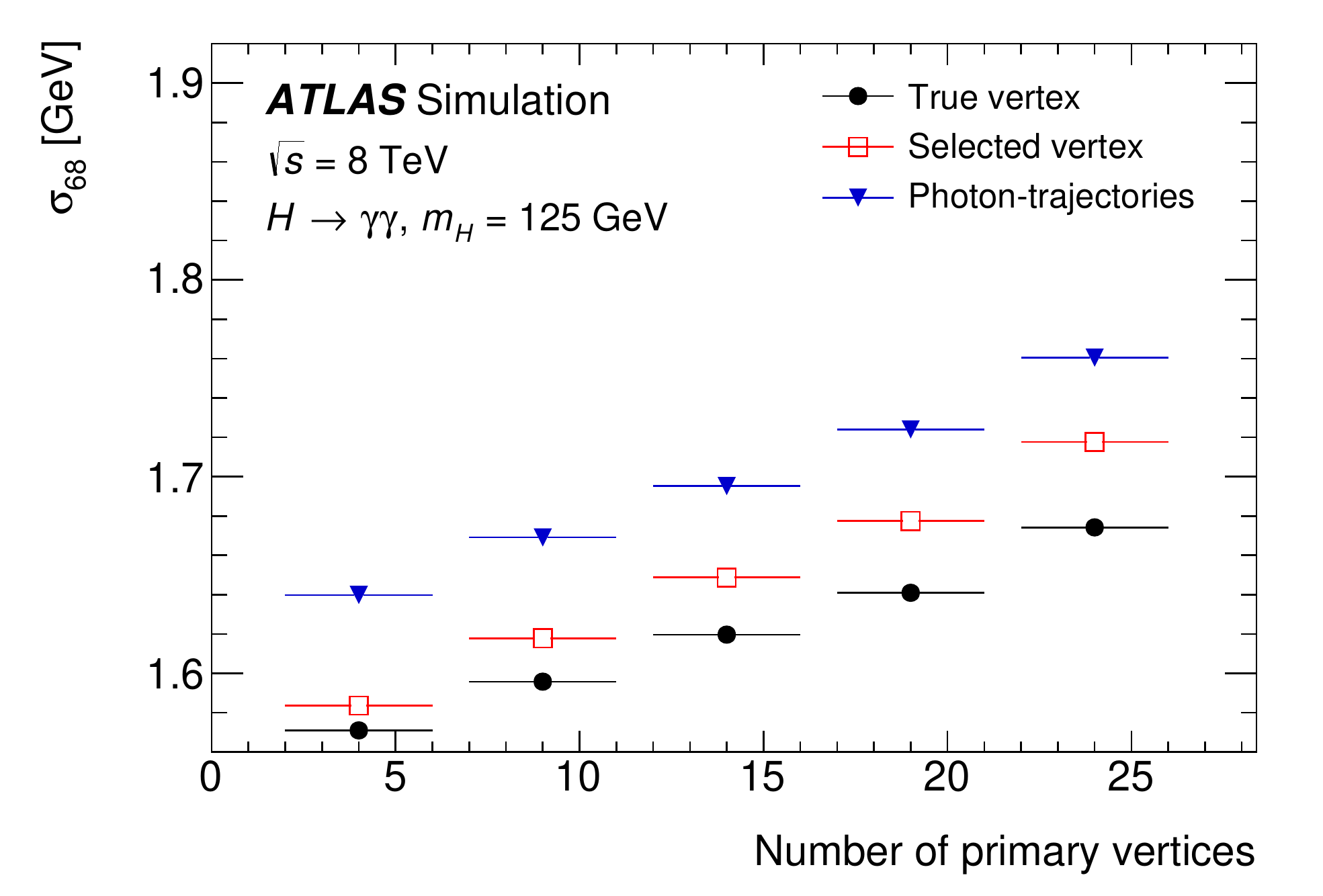}
  \caption{Signal invariant mass resolution $\sigma_{68}$ (defined in the text) as a function of the number of primary vertices per event 
    when using the true diphoton production vertex from the MC simulation (points), the production vertex reconstructed by the multivariate 
    algorithm described in Sec.~\ref{sec:selection} (open squares), and the production vertex reconstructed using only the
    photon trajectories (triangles). The events from different production processes are weighted according the SM cross sections
    and are required to fulfill the diphoton selection criteria (Sec.~\ref{sec:selection}) with
    no categorization applied.}
  \label{fig:signal_pv_pileup}
\end{figure}

\subsection{Background models}
\label{sec:background_model}
The background parameterizations are selected using MC samples or control
samples of data as described in the following.

For the four untagged, the two \VBF, the \VH\ hadronic and \MET\ categories,
the background parameterizations are tested with a mixture of $\gamma\gamma$,
$\gamma$--jet and jet--jet samples with the detector response simulated using
the simplified models mentioned in Sec.~\ref{sec:dataset}. 
The numbers of $\gamma\gamma$, $\gamma$--jet and \mbox{jet--jet} events in the selected diphoton
event sample are estimated by means of a double two-dimensional sideband method. The event fractions are fitted to
the distribution of the numbers of events in two bins of loose and tight photon identification criteria times two bins
of loose and tight photon isolation criteria, for each of the two photon candidates per event. The method relies on the
negligible correlation between these two variables for the jet background and that the sidebands (the regions where either the
photon identification or isolation is loose) are essentially populated by jets. The small signal contamination in the 
control regions is estimated using the MC simulation and accounted for.
The method
is cross-checked with alternative in situ techniques as described in Refs.~\cite{Aad:2011mh,diphoton2011}.
The number of events for each component in the selected diphoton events sample,
obtained independently in each bin of $\mgg$, is shown in Fig.~\ref{fig:back_dec}.
\begin{figure}[h]
 \centering
 \includegraphics[width=1.0\columnwidth]{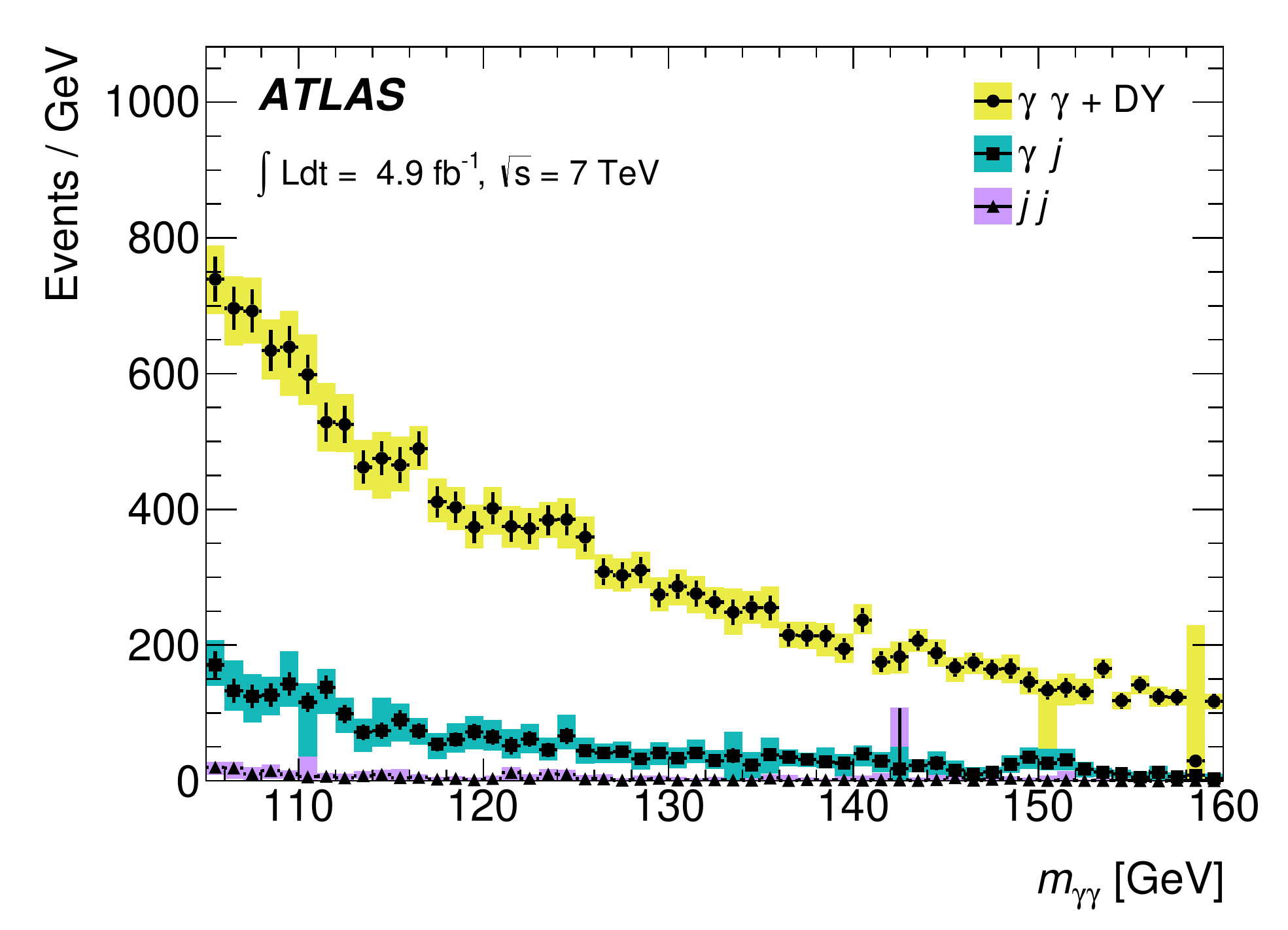}
 \includegraphics[width=1.0\columnwidth]{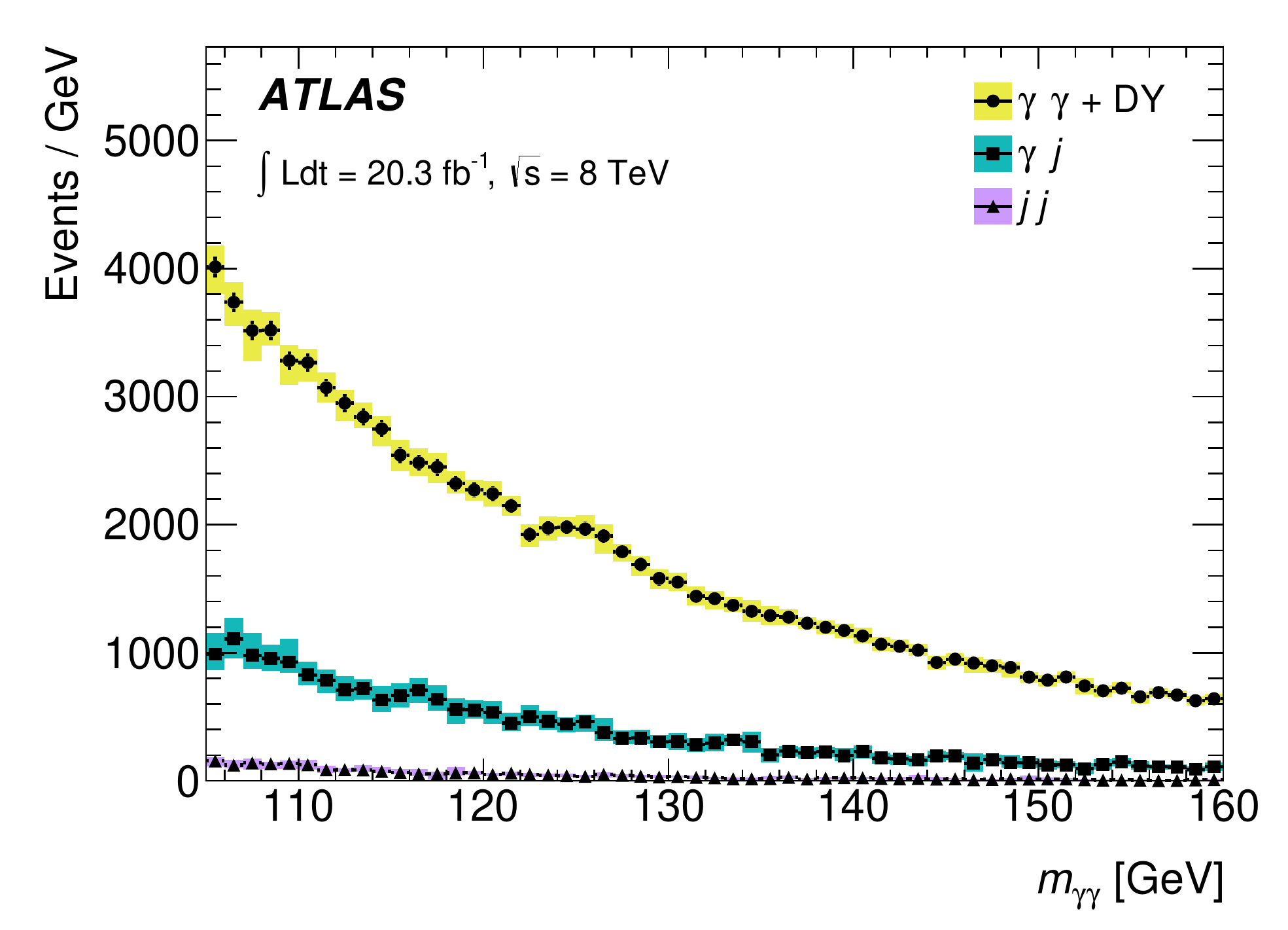} 
\caption{Cumulative components (jet--jet, $\gamma$--jet and $\gamma\gamma$) of the inclusive 
diphoton invariant mass spectrum, estimated using the double two-dimensional sideband
method as described in the text, in 7~TeV and 8~TeV data for 
all events passing the diphoton event selection. The $\gamma\gamma$ component also
includes a small $e^{+}e^{-}$ contribution from the Drell--Yan process.
The error bars on each point represent the
statistical uncertainty on the measurement while the colored bands represent the total
uncertainty.} 
 \label{fig:back_dec}
\end{figure}
The fractions of the three contributions, integrated over the \mgg\ spectrum, are found to be
84$\pm 8\%$ (77$\pm 3$\%), 15$\pm 8 \%$ (20$\pm 2$\%), and 1$\pm 1\%$ (3$\pm 1$\%) for the 7 TeV (8~TeV) data, respectively.
The MC components mentioned above are combined according to these
fractions and different background templates 
are derived for each category by applying the specific event selection of the category.
The combined background samples are then normalized to the numbers of events observed in
these categories (Table~\ref{tab:events_category_data}). Since this representative
background sample for each category contains many times more events than the corresponding data sample, 
the invariant mass distribution normalized to the data has negligible statistical fluctuations relative
to the statistical uncertainties that are taken from the data. 
Median expectations for quantities such as signal 
significance, signal amplitude, and their uncertainties 
are estimated using a single fit to the representative
background sample~\cite{stat}.
Other components that contribute less than 1\% of the total
background, such as Drell--Yan and $W\gamma$ and $Z\gamma$ production, 
are neglected. 
For the \VH\ \MET\  category, since the effect of the \MET\ cut on the
background shape is found to be negligible, it is not applied to the MC
events. 
The background samples for the \VH\ one-lepton category are obtained
from the MC $\gamma\gamma$ and $\gamma$--jet events 
introduced previously, where one jet is treated 
as a lepton for the category selection. 

An example of the diphoton invariant mass distributions in data and a
MC background sample is shown in  Fig.~\ref{fig:smeared_MC1}
for the Central - low \pTt\ category. For each category, the simulation describes the
distributions of the data sufficiently well (apart from the signal region ${\mgg\sim 125}$~GeV) 
to be used to select the parameterization of the background model
and to assess the corresponding systematic uncertainty on the signal yield.

\begin{figure}[!h]
  \centering  
  \includegraphics[width=\columnwidth]{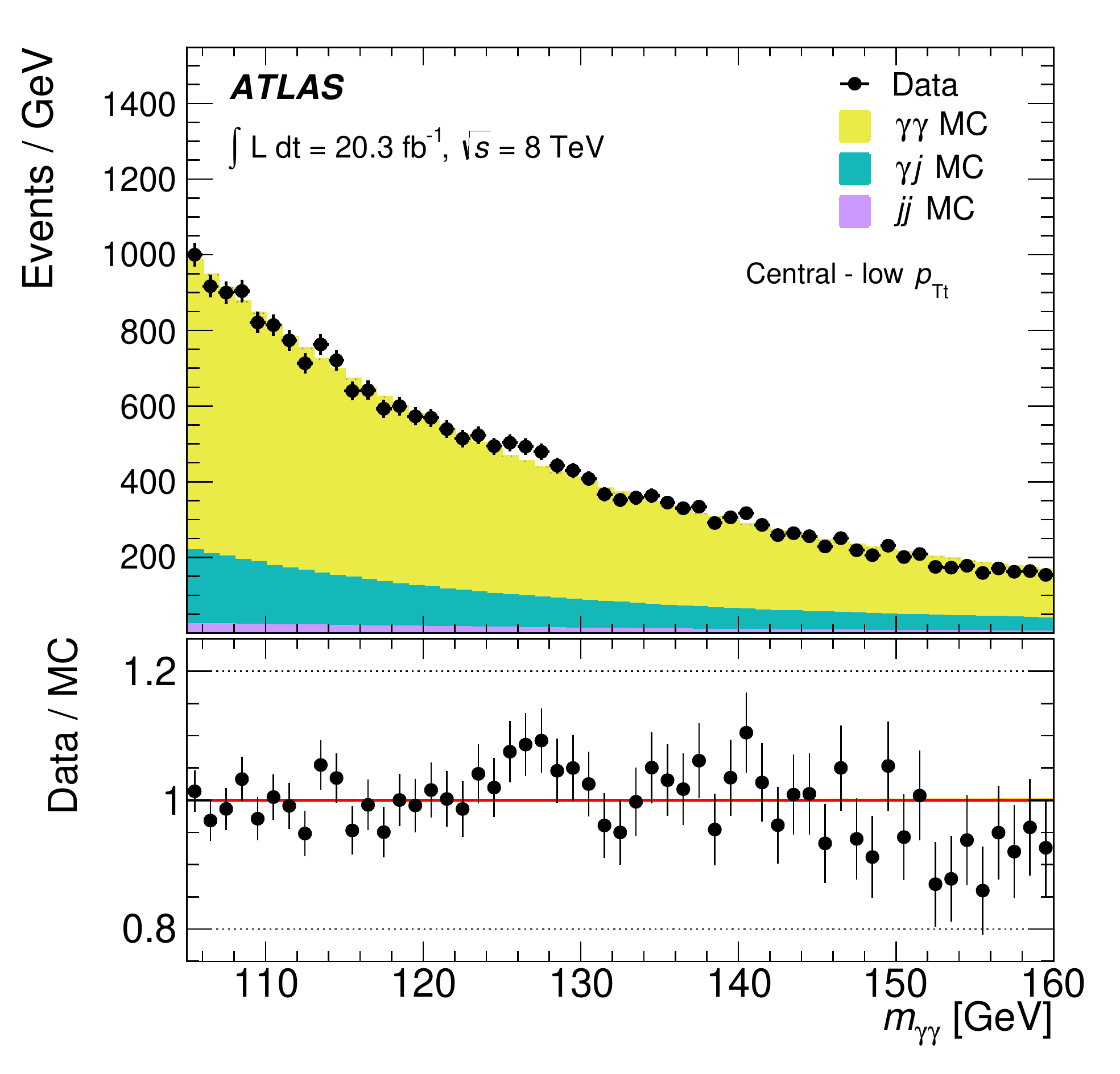}
  \caption{The distributions of diphoton invariant mass \mgg\ in the untagged Central - low \pTt\ category
    in data (points), and MC samples for the jet--jet, $\gamma$--jet and $\gamma\gamma$
    components of the continuum background (shaded cumulative histograms). The lower plot
    shows the ratio of data to MC simulation.
  }
\label{fig:smeared_MC1}
\end{figure}

A sample of fully simulated ${\Zboson\gamma\gamma}$ events is used for the \VH\ dilepton category since the 
contributions from \mbox{${\Zboson\gamma}$+jets} and \mbox{\Zboson+jets} events 
are estimated to be negligible after the event selection.
For the \ttH\ categories, the background parameterizations are tested on 
data control samples obtained 
by inverting photon identification criteria, isolation and the 
$b$-tagging, replacing the electron(s) with jet(s) and/or loosening the 
requirement on the number of jets. 

The selection of the parameterization for the background model
proceeds as follows. The distributions of \mgg\ from the samples described above are
fitted in the same 105--160~GeV range as the data with a signal 
at a given \mH\ (as described in Sec.~\ref{sec:signal_model}) plus
a background model. Since no signal is present in those background-only
samples, the resulting number of signal events from the fit $N_\mathrm{sp}(m_H)$
is
taken as an estimate of the bias in a particular background model
under test. For such a bias to be considered
acceptable, $N_\mathrm{sp}(m_H)$ has to be much smaller than the expected
signal rate or much smaller than the statistical uncertainty on the number of background
events in the fitted signal peak $\sigma_\mathrm{bkgd}(\mH)$, for cases where the number of expected signal events is
very small. 
The following criteria are adopted:
\begin{eqnarray}
& |N_\mathrm{sp}(\mH)|  < 10\% ~ N_\mathrm{S,exp}(\mH)  \nonumber \\
& \mathrm{or} \nonumber  \\
& |N_\mathrm{sp}(\mH) | < 20\% ~ \sigma_\mathrm{bkgd}(\mH) 
\label{eq:spurious}
\end{eqnarray}
for all \mH\ in the mass range 119--135~GeV.
The mass range was decided a priori to cover a region of approximately
five times the expected signal mass resolution on either side of the value of \mH\
measured by ATLAS in the ${\Hgg}$ channel~\cite{atlas-couplings-diboson}.
Here $N_\mathrm{S,exp}(\mH)$ is the number of signal events for a given value of \mH\ expected
to pass the $H\to\gamma\gamma$ selection.
For a given category, the parameterization with the smallest number of
free parameters satisfying the criteria in Eq.~(\ref{eq:spurious}) is chosen as
background model.

As an illustration of the procedure, the ratio $\mu_\mathrm{sp}(m_H)$  of 
$N_\mathrm{sp}(m_H)$ to the expected number of signal events is shown
in Fig.~\ref{fig:sp_MC1} for different candidate background models as functions of the test mass \mH\ for the 
Central - low \pTt\ category. The candidate parameterizations include exponentials
of first-, second- or third-order polynomials (exp1, exp2, exp3) and third-, fourth- or fifth-order Bernstein
polynomials~\cite{Bernstein} (bern3, bern4, bern5). The bands representing the criteria in Eq.~(\ref{eq:spurious}) 
are also shown. 
In this category exp1 and bern3 are excluded by the selection procedure and exp2 
is chosen since it has the fewest degrees of freedom of the parameterizations
that satsify the selection criteria.

\begin{figure}[!h]
  \centering
  \includegraphics[width=\columnwidth]{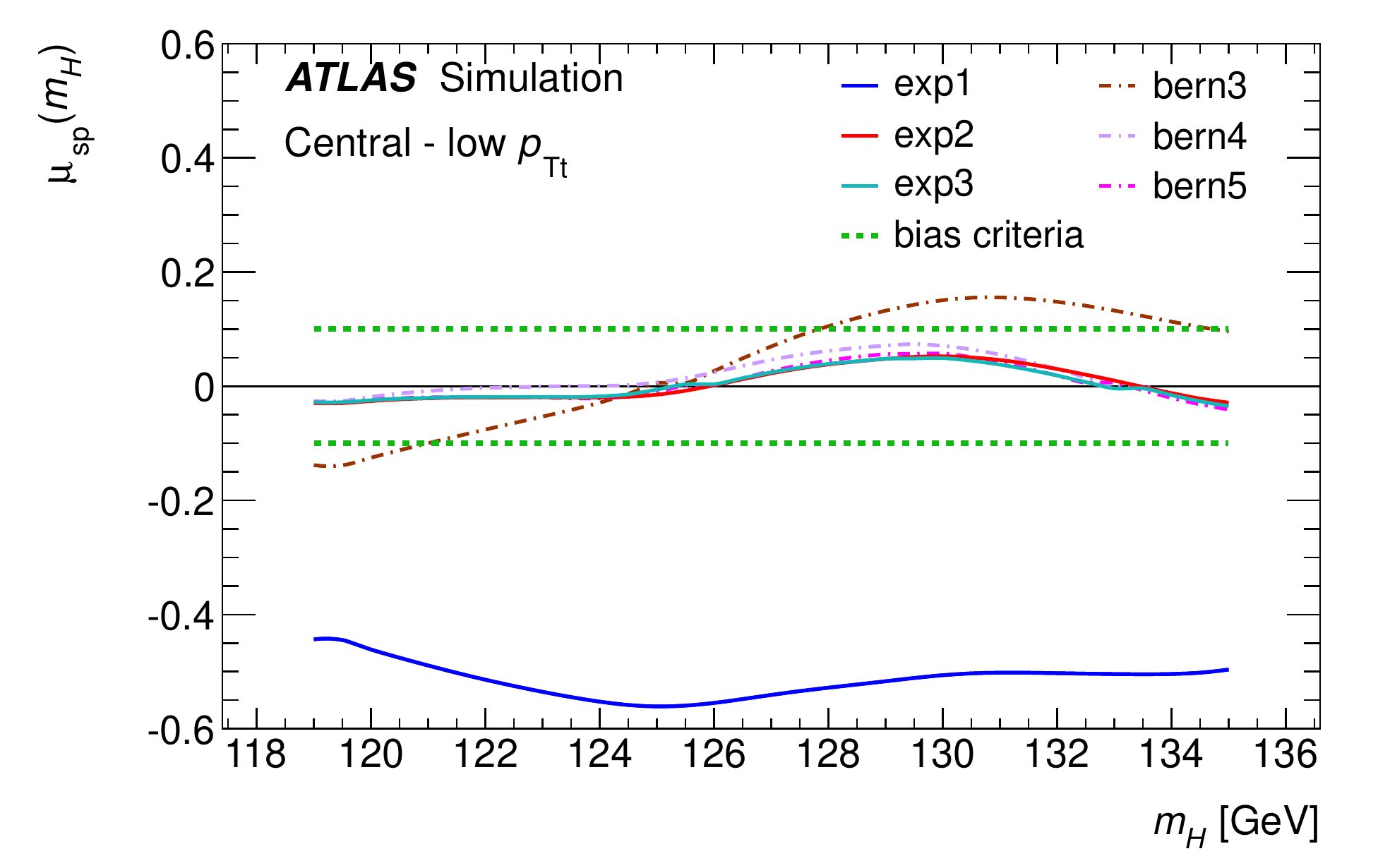}
  \caption{Ratio of the fitted number of signal events to the number expected for the SM $\mu_\mathrm{sp}(m_H)$ as a
function of the test mass \mH\ for the untagged Central - low \pTt\  category.
A single fit per value of \mH\ is performed on the represenative pure MC background sample described in the text 
with signal plus a variety of background parameterizations
(exp1, exp2, exp3 for the exponentials of first, second or third-order polynomials, respectively, and bern3, bern4, bern5 for
third, fourth and fifth-order Bernstein polynomials, respectively). 
The bias criteria in Eq.~(\ref{eq:spurious}) are indicated by the dashed lines.
}
\label{fig:sp_MC1}
\end{figure}

The largest $N_\mathrm{sp}(\mH)$ in the mass range 119--135~\GeV\ of a chosen parameterization, 
the \emph{spurious signal} $N_\mathrm{spur}$,
is assigned as the systematic uncertainty on the signal amplitude due to the
background modeling.
Table~\ref{tab:spurtab} summarizes the parameterizations used for the background model 
in each category described in Sec.~\ref{sec:categorisation} together with the 
derived uncertainties in terms of both spurious signal and its ratio
to the predicted number of signal events in each category ($\mu_\mathrm{spur}$).

\begin{table}[!htbp]
 \caption{List of the functions chosen to model the background distributions of \mgg\ 
and the associated systematic uncertainties on the signal amplitudes in terms of spurious
signal ($N_\mathrm{spur}$) and its ratio to the predicted number of signal events in each category 
($\mu_\mathrm{spur}$)
for the 12 categories and the 7 TeV and 8 TeV
datasets. Model exp1 (exp2) is the exponential of a first-order (second-order) polynomial.}
 \label{tab:spurtab}
 \begin{center}
 \begin{tabular}{lccccc}
 \hline \hline
  \multirow{2}{*}{Category} & \multirow{2}{*}{Model}  & \multicolumn{2}{c}{${\sqrt{s}=7}$~TeV} & \multicolumn{2}{c}{${\sqrt{s}=8}$~TeV} \bigstrut[t]\\ 
          &         &  $N_\mathrm{spur}$ & $\mu_\mathrm{spur}$  & $N_\mathrm{spur}$ & $\mu_\mathrm{spur}$ \bigstrut[b]\\
 \hline
 Central - low  \pTt  & exp2 &  1.1    &  0.041 & 6.7    & 0.050 \bigstrut[t] \\ 
 Central - high \pTt  & exp1 &  0.1    &  0.029 & 0.4    & 0.036 \\ 
 Forward - low  \pTt  & exp2 &  0.6    &  0.016 & 7.0    & 0.034 \\ 
 Forward - high \pTt  & exp2 &  0.3    &  0.088 & 1.2    & 0.073 \\ 
 \VBF\ loose          & exp1 &  0.2    &  0.091 & 1.3    & 0.14  \\ 
 \VBF\ tight          & exp1 &  $<0.1$ &  0.031 & 0.3    & 0.054 \\ 
 \VH\ hadronic        & exp1 &  0.1    &  0.14  & 0.5    & 0.14  \\ 
 \VH\ \MET            & exp1 &  0.1    &  0.18  & 0.1    & 0.11  \\ 
 \VH\ one-lepton      & exp1 &  $<0.1$ &  0.094 & 0.1    & 0.064 \\ 
 \VH\ dilepton        & exp1 &  $<0.1$ &  0.080 & $<0.1$ & 0.08 \\ 
 \ttH\ hadronic       & exp1 &  0.1    &  0.86  & 0.2    & 0.49  \\ 
 \ttH\ leptonic       & exp1 &  $<0.1$ &  0.10  & 0.2    & 0.28  \\ 

 \hline \hline
 \end{tabular}
 \end{center}
\end{table}

The numbers of measured background events $B_{90}$ within windows of invariant mass expected to contain 90\%\ of
the numbers of signal events predicted by the SM $S_{90}$ are listed in Table~\ref{tab:background_purity}
together with the expected signal purity $S_{90}/(S_{90}+B_{90})$ and signficance $S_{90}/\sqrt{S_{90}+B_{90}}$, for
each event category and the 7~TeV and 8~TeV datasets.

\begin{table}[!h]\footnotesize
\caption{Number of background events $B_{90}$ in the smallest interval expected to contain 90\%\ of 
the signal events $S_{90}$ (see $N_\mathrm{S}$ in Tables~\ref{tabefficiencies2011} and~\ref{tabefficiencies2012}), 
measured by fits to the data, and the expected purity $f_{90}\equiv S_{90}/(S_{90}+B_{90})$ and signal significance
$Z_{90}\equiv S_{90}/\sqrt{S_{90}+B_{90}}$ in each event category for the 7~\TeV\ and 8~TeV data.
}  
\label{tab:background_purity}
\begin{center}
\begin{tabular}{lcccccc}
\hline\hline
\multirow{2}{*}{Category}      & \multicolumn{3}{c}{${\sqrt{s}=7}$} & \multicolumn{3}{c}{${\sqrt{S}=8}$}  \\
                               & $B_{90}$ &  $f_{90}$ & $Z_{90}$ & $B_{90}$ & $f_{90}$ & $Z_{90}$  \\  
\hline
Central - low  \pTt  &  400  & 0.05 & 1.1  & 2400 & 0.05 & 2.4 \\
Central - high  \pTt &  11   & 0.14 & 0.52 & 68   & 0.13 & 1.2  \\
Forward - low \pTt   &  1400 & 0.02 & 0.94 & 8500 & 0.02 & 2.0  \\
Forward - high \pTt  &  47   & 0.05 & 0.38 & 280  & 0.05 & 0.84 \\
\VBF\ loose          &  6.6  & 0.18 & 0.52 & 44   & 0.16 & 1.2  \\
\VBF\ tight          &  0.48 & 0.64 & 0.75 & 6.7  & 0.44 & 1.5  \\
\VH\ hadronic        &  2.9  & 0.16 & 0.29 & 18   & 0.14 & 0.62 \\
\VH\ \MET            &  0.95 & 0.21 & 0.23 & 3.2  & 0.24 & 0.49 \\
\VH\ one-lepton      &  0.24 & 0.55 & 0.40 & 4.4  & 0.26 & 0.63 \\
\VH\ dilepton        &  0.00 & 1.0  & 0.20 & 0.27 & 0.46 & 0.32 \\
\ttH\ hadronic       &  0.21 & 0.22 & 0.11 & 1.8  & 0.20 & 0.30 \\
\ttH\ leptonic       &  0.11 & 0.46 & 0.21 & 0.53 & 0.50 & 0.51 \\
\hline\hline
\end{tabular}
\end{center}
\end{table}

\section{Systematic uncertainties}
\label{sec:systematics}
The various types of systematic uncertainties are presented in this section according
to the way they affect the determination of the signal strengths.
The theoretical and experimental uncertainties on the yields of diphoton events 
from Higgs boson decays
are discussed in Sec.~\ref{sec:unc_signal_yield}.
The systematic uncertainties affecting the event categorization due to
migrations of signal events from or to other categories
are presented in Sec.~\ref{sec:unc_migration}.
The systematic uncertainties 
related to the photon energy scale and
resolution are reported in Sec.~\ref{sec:unc_mass}.
The systematic uncertainties 
due to potential spurious signals
induced by systematic differences between the background parameterization and the background
component of the data are obtained with the technique described in Sec.~\ref{sec:background_model}
and reported in Table~\ref{tab:spurtab}.

\subsection{Uncertainties affecting the integrated signal yield} 
\label{sec:unc_signal_yield}

\subsubsection{Theoretical uncertainties}

The predicted total cross sections for the signal processes have uncertainties
due to missing higher-order terms in the perturbative calculations of QCD processes
that are estimated by varying the factorization and renormalization scales.
There are additional uncertainties related to the PDFs,
the strong coupling constant $\alpha_\mathrm{S}$, and the ${\hgg}$ branching ratio.
The uncertainties on the Higgs boson production cross sections 
are listed in Table~\ref{tab:syst_yield_theory} 
for ${\mH=\combmass}$~\GeV, separately for ${\sqrt{s}=7}$~\TeV\ and ${\sqrt{s}=8}$~\TeV.
The uncertainties estimated by varying the QCD scales affect the production processes independently, apart
from \WH\ and \ZH\ uncertainties, which are treated as fully correlated.
For the \tHbj\ and \WtH\ production processes, the scale uncertainties are obtained by varying the
renormalization and factorization scales by factors of 1/2 and 2 in the event generators (Sec.~\ref{sec:dataset}) and 
the PDF uncertainties are estimated by studying the impact of the variations within the CT10 PDF set. For the other
processes these uncertainties are taken from Ref.~\cite{lhcxs}.
The combined uncertainties on the effective luminosities for $gg$- and $qq$-initiated processes due to PDF
and $\alpha_\mathrm{S}$ uncertainties are independent but they affect the relevant 
processes coherently. Both of these sets of uncertainties affect the 7 TeV and the 8 TeV cross sections
coherently.
The impact of scale and PDF uncertainties on the kinematic acceptance for signal events 
is found to be negligible relative to the impact of the uncertainties on
the cross sections.
The uncertainty on the ${\hgg}$ branching ratio for ${\mH=\combmass}$~\GeV\ is
${\pm 5\%}$. 
These theoretical uncertainties, which vary only at the per mille level
within 1--2~GeV of ${\mH=\combmass}$~\GeV, are taken from Ref.~\cite{lhcxs}. 

\label{sec:yieldtheory}

\begin{table}[h!]
\footnotesize
\caption{Theoretical uncertainties $[\%]$ on cross sections for Higgs boson production processes
for ${\sqrt{s}=7}$~\TeV\ and ${\sqrt{s}=8}$~\TeV\ for  ${\mH = \combmass}$~GeV, as described in 
Sec.~\ref{sec:yieldtheory}. Except for the \tHbj\ and \WtH\ processes, the uncertainties are taken
from Ref.~\cite{lhcxs}.}
\label{tab:syst_yield_theory}
\begin{center}
\begin{tabular}{lcccc}
\hline\hline

\multirow{2}{*}{Process} & \multicolumn{2}{c}{QCD scale}  & \multicolumn{2}{c}{PDF+$\alpha_\mathrm{S}$}  \\
        & $\sqrt{s}=7$ \TeV  &$\sqrt{s}=8$ \TeV & $\sqrt{s}=7$ \TeV &$\sqrt{s}=8$ \TeV \\
\hline

\multirow{2}{*}{\ggH}       & $+7.1$       & $+7.2$  & $+7.6$ & $+7.5$  \\
                            & $-7.8$       & $-7.8$  & $-7.1$ & $-6.9$  \\

\multirow{2}{*}{\VBF}       & $+0.3$       & $+0.2$  & $+2.5$ & $+2.6$  \\
                            & $-0.3$       & $-0.2$  & $-2.1$ & $-2.8$  \\

\multirow{2}{*}{\WH}        & $+1.0$       & $+1.0$  & $+2.6$ & $+2.4$  \\
                            & $-1.0$       & $-1.0$  & $-2.6$ & $-2.4$  \\

\multirow{2}{*}{\ZH}        & $+2.9$       & $+3.1$  & $+2.6$ & $+2.5$  \\
                            & $-2.9$       & $-3.1$  & $-2.6$ & $-2.5$  \\

\multirow{2}{*}{\ttH}       & $+3.2$       & $+3.8$  & $+8.4$ & $+8.1$  \\
                            & $-9.3$       & $-9.3$  & $-8.4$ & $-8.1$  \\

\multirow{2}{*}{\bbH}       & $+10$        & $+10$   & $+6.2$ & $+6.1$  \\
                            & $-15$        & $-15$   & $-6.2$ & $-6.1$  \\

\multirow{2}{*}{\tHbj}      & $+7$       & $+6$  & $+4$ & $+4$  \\
		            & $-6$       & $-5$  & $-4$ & $-4$  \\

\multirow{2}{*}{\WtH}       & $+7$       & $+9$  & $+10$  & $+10$  \\
	                    & $-6$       & $-7$  & $-10$  & $-10$  \\

\hline\hline
\end{tabular}
\end{center}
\end{table}

\subsubsection{Sizes of MC samples}

\label{sec:MCstat}

The finite size of the MC signal samples may induce non-negligible
statistical uncertainties depending on the category and the production
process.
The impact of
these uncertainties on the individual signal strength parameters is
estimated for each event category by analyzing representative MC
datasets containing both Higgs boson signal and continuum background:
for the signal the sample size is fixed to the one expected in data by
the SM predictions, and for the background to the observed numbers of
events. The uncertainties that contribute more than 0.1\% in quadrature
to the total expected uncertainties are retained. 
The 14 uncertainties that contribute more than 0.1\% in quadrature to the total expected uncertainties 
are propagated, but their contribution is at the level of 1\% or less, which is much smaller than the 
expected statistical uncertainties on the individual signal strengths.

\subsubsection{Experimental uncertainties}
\label{sec:yieldexpt}

The expected signal yields are affected by the experimental systematic uncertainties listed below. 

\begin{enumerate}
\item{} The uncertainties on the integrated luminosities are 1.8\% for the 7 TeV data and 2.8\% 
for the 8 TeV data~\cite{lumi}. They are treated as uncorrelated.
\label{itm:lumi}

\item{} The trigger efficiencies in data are determined by combining the results from two different measurements.
The first measurement is performed
with photons in ${\Zll\gamma}$ events, where $\ell$ is an electron or a muon.
These events are collected with lepton-based triggers, making the photon candidates in these samples
unbiased with respect to the trigger.
The second measurement, based on the bootstrap technique described in Ref.~\cite{Aad:2012xs}, 
is performed on a background-corrected photon sample selected only by
a first-level trigger,  which has an efficiency of 100\% for signal-like photons in events
that pass the diphoton selection criteria (Sec.~\ref{sec:selection}).
Both measurements are dominated by statistical uncertainties.
The uncertainties on the trigger efficiencies based on these measurements
are estimated to be 0.2\% for both the 7~TeV and 8~TeV data and are fully uncorrelated. 
\label{itm:trigger}

\item{} The uncertainty on the photon identification efficiency for the 8~TeV data is derived from 
measurements performed with data using three different methods~\cite{PhotonID} that cover the full \ET\ spectrum 
relevant for this analysis.
In the first method, the efficiency is measured in a pure and unbiased sample of photons obtained by 
selecting radiative ${\Zll\gamma}$ decays 
without using the photon identification to
select the photon, and where $\ell$ is an electron or a muon. 
In the second method, the photon efficiency is measured using ${\Zee}$ data 
by extrapolating the properties of electron showers to photon showers using MC events~\cite{PhotonID}.
In the third method, the photon efficiency is determined from a data sample of isolated photon candidates from
prompt $\gamma$--jet production after subtracting the measured fraction of jet--jet background events.
The combined uncertainty on the photon identification efficiency in data relative to MC simulation ranges between 0.5\% and 2.0\% depending  
on the \ET\ and $\eta$ of the photon and on whether the photon is unconverted or converted and reconstructed
with one or two tracks.
For the 7~TeV data, more conservative
uncertainties,  ranging 
from 4\% to 7\%, are used 
because of the stronger 
correlation of the NN-based identification algorithm with the photon isolation, and because it 
relies more strongly on the correlations between the individual shower shape 
variables. 
Because these two effects complicate the 
measurement of the identification performance in data, conservative uncertainties, 
taken as the full difference between the efficiencies measured in data and the ones predicted by simulation, 
are used.
The uncertainties on the signal yield due to the uncertainty on the photon identification
efficiency are 8.4\% for the 7~TeV data and 1.0\% for the 8~TeV data and are treated as uncorrelated.
\label{itm:photonID}

\item{}  
The uncertainty on the isolation efficiency is conservatively
taken as the full size of the applied correction described in Sec.~\ref{photon_reco}. The effect on 
the signal yield varies among categories 
(depending on their photon \ET\ spectrum). These uncertainties, which range between 1.3\% and 2.3\%,   
are estimated with the 8~TeV dataset and are assumed to be
the same in the 7~TeV data but uncorrelated between the two datasets.
 
\label{itm:isolation}

\end{enumerate}

The estimated values of the experimental uncertainties for both 
datasets are summarized in Table~\ref{tab:syst_inclusive}. Larger uncertainties, also shown in the table,
on the photon identification and isolation selection efficiencies are assigned 
to the categories sensitive to \ttH\ and \VH\ production modes. The presence of large hadronic activity
(high jet multiplicity) in these events, which is partially correlated with the photon selection and
isolation efficiency, makes it difficult to measure the efficiencies precisely.
The impact of these additional systematic uncertainties is, however, negligible relative to the 
statistical uncertainties on the measurements of
$\mu_{\ttH}$, $\mu_{\ZH}$, and $\mu_{\WH}$. 

\begin{table}[h!]
\caption{Relative systematic uncertainties on the inclusive yields $[\%]$ for the 7 TeV and
8 TeV data. The numbers in parentheses refer to the uncertainties applied to events in the categories 
that are sensitive to \ttH\ and \VH\ production modes. 
The ranges of the category-dependent uncertainties due to the isolation efficiency are reported.
}
\label{tab:syst_inclusive}
\begin{center}
\begin{tabular}{lcc}
\hline\hline
Uncertainty source & ${\sqrt{s}=7}$~TeV & ${\sqrt{s}=8}$~ TeV \\
\hline
Luminosity & 1.8 & 2.8 \\
Trigger & 0.2 & 0.2 \\
Photon Id. & 8.4(9.3) & 1.0(4.1) \\
Isolation & 1.3--2.3(3.8) & 1.3--2.3(3.8) \\
\hline\hline
\end{tabular}
\end{center}
\end{table}

Finally, uncertainties on the signal yields due to the photon energy scale and primary
vertex selection are found to be negligible relative to the ones discussed above.

\subsection{Migration uncertainties}
\label{sec:unc_migration}

The impacts of theoretical and experimental uncertainties on
the predicted contributions from the various Higgs boson production processes
to each event category are summarized in the following.

\subsubsection{Theory uncertainties}
\label{sec:migrationTheory}

\begin{enumerate}

\item{} The uncertainty on the Higgs boson production cross section through gluon fusion in association with two or
more jets is estimated by applying an extension of the so-called Stewart--Tackmann method~\cite{PhysRevD.85.034011,STmethod} to
predictions made by the \textsc{mcfm}~\cite{MCFM} generator:  an
uncertainty of 20\% is assigned to the \ggH\ component in the \VBF\ loose, \VBF\ tight, and \VH\ hadronic categories. 
Since the \VBF\ categories make use of the azimuthal angle between the diphoton and dijet
systems, which is sensitive to the presence of a third jet, additional uncertainties are 
introduced for the \ggF\ contribution in these categories using a technique described in 
Ref.~\cite{lhcxs}. 
These uncertainties are found to be 25\%\ and 52\%\ for the VBF loose and
VBF tight categories, respectively.
\label{itm:jets}

\item{} The presence of additional hadronic activity from the underlying event (UE) 
may produce significant migrations of \ggH\ events to the \VBF\ and \ttH\ hadronic categories.  
The uncertainties on the UE modeling are conservatively estimated as the full change in signal migration in MC 
simulation with and without the UE.
The uncertainties are 5--6\%  of the 18--41\% component of \ggF\ in the \VBF\ categories, and 60\% of the 8--11\% \ggF\ contribution
in the \ttH\ hadronic category. In addition, the presence of the UE directly affects the \ttH\ yield in the \ttH\ hadronic and \ttH\ 
leptonic categories by 11\% and 3\%, respectively.
The differences between the uncertainties for the 7~TeV and 8~TeV data are small. 
Tables~\ref{tabefficiencies2011} and~\ref{tabefficiencies2012} show details of the nominal yields of the signal processes in the event categories.
The impacts of these uncertainties are small compared with the statistical uncertainties on the signal strengths
for these categories.
\label{itm:UE}

\item{ } The uncertainty on the modeling of the \pt\ spectrum of the Higgs boson for the \ggH\ process
can cause migrations of events between the low and the high \pTt\ categories. The size of the effect
has been checked using the 
\textsc{HRes2.1} prediction by varying the renormalization,
factorization, and two resummation scales. 
The uncertainties for the high-\pTt\ categories are estimated from 
the absolute values of the largest changes in the event categorization caused by the scale variations.
Events in the low-\pTt\ categories are assigned an uncertainty according 
to the Stewart--Tackmann procedure. The size of the effect varies among 
categories; it is as large as about $24\%$ 
in the high-\pTt\ categories.
\label{itm:HpT}

\item{} The \VBF\  selection uses angular variables $\Delta \phi_{jj}$ and $\eta^{*}$
that involve the two leading jets, as discussed in \mbox{Sec.~\ref{subsec:vbf_cat}}. 
The second jet in the 
generation of \ggH\ events by
\textsc{powheg-box}+\textsc{pythia8} predominantly comes from 
the parton shower generated by \textsc{pythia8}; therefore, the angular
correlation between the two jets is not well modeled. The uncertainty due to this
modeling is taken to be the difference in the event categorization caused by re-weighting
the events in the \textsc{powheg-box} sample
to reproduce the \pt\ spectrum of the Higgs boson predicted by \textsc{minlo hjj}~\cite{minlo_hjj}, 
which models the angular correlation between the first and second jet produced in gluon fusion to NLO accuracy. 
The mis-modeling of $\Delta \phi_{jj}$ ($\eta^{*}$) typically changes the 
number of \ggH\ events in the \VBF\ tight and \VBF\ loose categories by
at most 11.2\% (6.6\%) and 8.9\% (4.8\%), respectively.
\label{itm:dphijj} 

\item{} Additional uncertainties are estimated for production processes contributing
significantly to the \ttH\ categories due to acceptance changes observed when varying the renormalization and
factorization scales. 
The uncertainty on \ttH\ production itself is 2\% (1\%) in the \ttH\ leptonic (hadronic) category.
An uncertainty of 50\% is attributed to the \ggH\ contribution in the \ttH\ sensitive categories while 
an uncertainty of 4--8\% is attributed to the \WH, \tHbj\ and \WtH\ contributions to account for the sensitivity 
of the acceptance to scale changes.
The impact is independent for the three \ttH\ and \tH\ production processes, but coherent in the two \ttH\ event categories 
and for $\sqrt{s}$ = 7 TeV and 8 TeV.
In addition, the uncertainties on the \ggH, \VBF\, and \WH\
contributions to the \ttH\ categories are assumed to be
100\%\ to account for the uncertainty on the heavy flavor (HF)
fraction in these production processes. The overall impact of these large uncertainties on
$\mu_{\ttH}$ is about 10\% (and much less for the other signal strength measurements),
due to the small contributions from \ggH, \VBF\, and \WH\ production
to the \ttH\ categories (Tables~\ref{tabefficiencies2011} and~\ref{tabefficiencies2012}).
\label{itm:ttH}
 
\end{enumerate}

\subsubsection{Experimental uncertainties}
\label{sec:migrationexpt}

The following potential sources of signal migration between categories caused by experimental
effects are investigated.

\begin{enumerate} 

\item{} Uncertainties related to jet and \met~  reconstruction 
affect the predicted distributions
of signal events from the various production modes among the categories.
The effect of the uncertainty on the jet energy scale, jet energy resolution and jet vertex fraction
is estimated by varying individually each component of the
uncertainties~\cite{jet_reco}. The effect of the \met~energy scale and resolution uncertainty
is estimated by varying independently the uncertainty in the energy scale and resolution
of each type of physics object entering the calculation of \met\ as well as the uncertainty 
 on the scale and resolution of the soft term~\cite{met_perf}.
There are 20 and 5 uncorrelated components that account for the
jet- and \met-related uncertainties, respectively. 
Tables~\ref{tab:syst_jet} and~\ref{tab:syst_MET} show the impact of the jet
and \met\ uncertainties. 
To simplify the presentation of the results, categories and processes for which each
source of uncertainty has a similar impact are merged.
These uncertainties are fully correlated between the 7~TeV and 8~TeV datasets.
\label{itm:JESMES}

\item{} The impact of the uncertainty in the $b$-tagging efficiency on the migration of events
to and from the \ttH\ categories is decomposed into 10 (3) independent contributions in the 8 TeV (7 TeV) 
data analysis. 
The uncertainty on the \ttH\ yield in the \ttH\ categories from the uncertainty on the $b$-tagging efficiency ranges 
from 1 to 3\%. The uncertainties affecting other production processes that have the largest impact on the yield 
in the ttH categories are 20-30\% of the ggF component
in the hadronic category and 6-7\% to the WH contribution in the leptonic channel.
\label{itm:btag}

\item{} The total impact of the lepton reconstruction, identification and isolation
uncertainties on any of the selection efficiencies and event fractions of the signal production processes for
the event categories in Tables~\ref{tabefficiencies2011} and~\ref{tabefficiencies2012} is found to be below 1\%.
\label{itm:leptonall}

\end{enumerate}

\begin{table}
\caption{Relative uncertainties $[\%]$ on the Higgs boson signal yield in each category and for
each production process induced by the combined effects of the systematic uncertainties on the jet energy scale, 
jet energy resolution and jet vertex fraction. These uncertainties are approximately the same for the 7~TeV and
the 8~TeV data.}
\label{tab:syst_jet}
\begin{center}
\begin{tabular}{lcccc}
\hline\hline
Category  & \ggH & \VBF & \ttH  &  \WH+\ZH  \\
\hline
Central+Forward - low  \pTt   &  0.1 &  2.9  &  4.0 &  0.1  \\  
Central+Forward - high \pTt   &  1.1 &  4.5  &  3.5 &  1.4  \\  
\VBF\ loose                   &  12  &  4.4  &  7.6 &  13   \\  
\VBF\ tight                   &  13  &  9.1  &  6.3 &  17   \\  
\VH\ hadronic                 &  2.8 &  4.1  &  9.5 &  2.5  \\  
\VH\ \MET                     &  2.6 &  9.0  &  1.2 &  0.2  \\  
\VH\ one-lepton               &  4.9 &  6.2  &  2.8 &  0.5  \\  
\VH\ dilepton                 &  0   &  0    &  5.1 &  1.0  \\  
\ttH\ hadronic                &  11  &  21   &  7.3 &  22   \\  
\ttH\ leptonic                &  37  &  7.7  &  0.5 &  7.4  \\  
\hline\hline
\end{tabular}
\end{center}
\end{table}

\begin{table}
\caption{Relative uncertainties $[\%]$ on the Higgs boson signal yield in each category and for
each production process induced by systematic uncertainty on the \met\ energy scale and resolution.
The uncertainties, which are approximately the same for the 7~TeV and 8~TeV data, 
are obtained by summing in quadrature the impacts on the signal yield
of the variation of each component of the \met\ energy scale within its uncertainty.
}
\label{tab:syst_MET}
\begin{center}
\begin{tabular}{lcccc}
\hline\hline
Category  & \ggH +\VBF  &  \ttH  &  \WH  &  \ZH  \\
\hline
Untagged               &  0.0   &  0.2 &  0.1 &  0.2 \\ 
\VBF\ loose            &  0.0   &  1.0 &  0.1 &  0.2 \\ 
\VBF\ tight            &  0.0   &  2.7 &  1.1 &  0.0 \\ 
\VH\ hadronic          &  0.0   &  0.7 &  0.0 &  0.1 \\ 
\VH\ \MET              &  35    &  1.1 &  1.3 &  0.9 \\ 
\VH\ one-lepton        &  4.5   &  0.6 &  0.4 &  4.0 \\ 
\VH\ dilepton          &  0.0   &  2.0 &  0.0 &  0.1 \\ 
\ttH\ hadronic         &  0.0   &  0.0 &  0.0 &  0.0 \\ 
\ttH\ leptonic         &  1.9   &  0.1 &  1.0 &  3.0 \\ 
\hline\hline                                             
\end{tabular}
\end{center}
\end{table}

\subsection{Impact of diphoton resolution and mass scale uncertainties on the fitted signal yield}
\label{sec:unc_mass}
\subsubsection{Diphoton mass resolution}
\label{sec:unc_massres}

The precise determination of the uncertainty on the signal strengths due to the diphoton mass resolution 
is a key point in this analysis. It defines the range over which the 
signal model width is allowed to change in the fit, thus directly affecting the estimation of the number of signal events. 
The energy resolution and its uncertainty for photons are estimated by extrapolating from 
the ones for electrons. 
The electron energy resolution and its uncertainty
are measured in data using \Zee\ events 
that, however, can only provide constraints for electrons 
with $\pT\simeq 40$~GeV.  
The extrapolation from electrons to photons and to different energy ranges relies on an
accurate modeling of the resolution in the detector simulation. In the model used in this analysis, the total resolution 
is described in terms of four energy-dependent 
contributions~\cite{calibration_paper}: the asymptotic resolution 
at high energy, i.e.\ the constant term; the intrinsic sampling
fluctuations of the calorimeter; the effect of passive
material upstream of the calorimeter; and the electronic and pileup noise.
The effects on the various categories due to the the four
contributions to the uncertainty in the mass resolution are summarized in
Table~\ref{tab:syst_mres} for the 8 \TeV\ data: the typical relative uncertainty on the diphoton mass resolution 
obtained from the sum in quadrature of these contributions is 10--15\% for ${\mH\simeq 125}$~GeV.
The uncertainties for the 7 \TeV\ data
are very similar except for the reduced size of the pileup contribution, which ranges from 0.9\%  
to 1.4\%. 
These four contributions are uncorrelated while each contribution affects both of the
parametric width parameters $\sigma_\mathrm{CB}$ and $\sigma_\mathrm{GA}$ in the signal model (Sec.~\ref{sec:signal_model})
for all the categories and for both the 7~TeV and 8~TeV data coherently. 

\begin{table}
\caption{Systematic uncertainties on the diphoton mass resolution for the 8 \TeV\ data
$[\%]$ due to the four contributions described in the text.  For each category, the uncertainty is
estimated by using a simulation of the Higgs boson production process which makes the largest contribution
to the signal yield.}
\label{tab:syst_mres}
\begin{center}
\begin{tabular}{lcccc}
\hline\hline
\multirow{2}{*}{Category}    & Constant & Sampling & Material  & Noise \\
                             & term          & term       & modeling & term \\
\hline
Central - low  \pTt & 7.5      & 2.6       & 4.9        & 2.6  \\ 
Central - high \pTt & 9.6      & 5.6       & 6.2        & 1.7  \\ 
Forward - low  \pTt & 9.9      & 1.3       & 6.0        & 2.1  \\ 
Forward - high \pTt & 12       & 2.8       & 7.8        & 1.9  \\ 
\VBF\ loose         & 9.4      & 2.6       & 6.0        & 2.1  \\ 
\VBF\ tight         & 10       & 3.8       & 6.5        & 2.1  \\ 
\VH\ hadronic       & 11       & 4.0       & 7.2        & 1.6  \\ 
\VH\ \MET           & 11       & 3.6       & 7.4        & 1.7  \\ 
\VH\ one-lepton     & 9.8      & 2.8       & 6.3        & 2.1  \\ 
\VH\ dilepton       & 9.5      & 2.7       & 6.2        & 2.1  \\ 
\ttH\ hadronic      & 9.6      & 3.6       & 6.3        & 1.9  \\ 
\ttH\ leptonic      & 9.5      & 3.4       & 6.2        & 2.1  \\ 
\hline\hline
\end{tabular}
\end{center}
\end{table}

\subsubsection{Diphoton mass scale}
\label{sec:unc_escale}
The uncertainties
on the diphoton mass scale affect the position of the signal mass peak through variations of the peak of the Crystal Ball ($\mu_\mathrm{CB}$) 
and Gaussian ($\mu_\mathrm{GA}$) components of the signal model.
The dominant systematic uncertainties on the position of the mass peak arise from uncertainties on the photon energy
scale. These uncertainties, discussed in detail in Refs.~\cite{calibration_paper,Aad:2014aba}, are propagated to the diphoton 
mass distribution in the signal model for each of the 12 categories. 
The total uncertainty on the position of the mass peak from the photon
energy scale systematic uncertainties ranges from 0.18\% to 0.31\%
depending on the category. A second contribution, varying from 0.02\%
to 0.31\%, comes from the choice of the background model 
and is evaluated using the technique presented in Ref.~\cite{Aad:2014aba}.
Finally, the systematic uncertainty on the mass scale related to the reconstruction of the diphoton vertex is estimated to be 0.03\%\ for 
all the categories. 
As discussed in Sec.~\ref{sec:results}, 
the uncertainty on the diphoton mass scale is expected to flatten the dependence of $\mu$ as a function of \mH\ in the region
around the true value of \mH.

\section{Statistical procedure}
\label{sec:statistic}

The data are interpreted following the statistical procedure summarized in
Ref.~\cite{stat} and described in detail in Ref.~\cite{Aad:2012an}. An extended
likelihood function is built from the number of observed events and
analytic functions describing the distributions of \mgg\ in the range 105--160~\GeV\ 
for the signal (see Sec.~\ref{sec:signal_model}) and the background 
(see Sec.~\ref{sec:background_model}). 

The likelihood for a given category $c$ is a marked Poisson probability
distribution,
\begin{equation*}
\LL_c = \mathrm{Pois}(n_c|N_c(\vecth))\cdot\prod_{i=1}^{n_c}f_c(\mgg^i,\vecth)
\cdot G(\vecth),
\end{equation*}
where $n_c$ is the number of candidates, $N_c$ is the expected number of
candidates, $f_c(\mgg^i)$ is the value of the probability density
function (pdf) of the invariant mass distribution evaluated for each candidate
$i$, $\vecth$ are nuisance parameters and $G(\vecth)$ 
is a set of unit Gaussian constraints on certain of the nuisance parameters,
as described in the following.

The number of expected candidates is 
the sum of the hypothesized number of signal events
plus the fitted number of background candidates, $N_{\mathrm{bkg},c}$,
and the fitted spurious signal, $N_{\mathrm{spur},c}\cdot\thetaspurc$,
\begin{equation*}
N_c=\mu \cdot N_{\mathrm{S},c}(\vecth_c^\mathrm{yield},\vecth_c^\mathrm{migr},\mH) + N_{\mathrm{\mathrm{bkg}},c} 
+ N_{\mathrm{spur},c}\cdot\thetaspurc,
\end{equation*}
where $N_{\mathrm{S},c}(\vecth_c^\mathrm{yield},\vecth_c^\mathrm{migr},\mH)$ is the number of 
signal events predicted by the SM from all
production processes, $\vecth_c^\mathrm{yield}$ and $\vecth_c^\mathrm{migr}$ are the nuisance 
parameters that implement the systematic uncertainties affecting
the yields of the Higgs boson production (Sec.~\ref{sec:unc_signal_yield}) in 
and migration between the 12 categories (Sec.~\ref{sec:unc_migration}), respectively.
In more detail, the invariant mass distribution for each category 
has signal and background components,
\begin{multline*}
f_c(\mgg^i)= [(\mu \cdot N_{\mathrm{S},c} + N_{\mathrm{spur},c}\cdot\thetaspurc)\cdot f_{\mathrm{S},c}(\mgg^i,\vecth_{\mathrm{S},c}^\mathrm{shape})
+\\  N_{\mathrm{bkg},c} \cdot f_{\mathrm{bkg},c}(\mgg^i,\vecth_{\mathrm{bkg},c}^\mathrm{shape})]/N_c,
\end{multline*}
where $\vecth_{\mathrm{S},c}^\mathrm{shape}$ and $\vecth_{\mathrm{bkg},c}^\mathrm{shape}$ are nuisance parameters
associated with systematic uncertainties affecting the resolutions (Sec.~\ref{sec:unc_massres}) and positions (Sec.~\ref{sec:unc_escale}) 
of the invariant mass distributions of the signal $f_{\mathrm{S},c}$
(described in Sec.~\ref{sec:signal_model}) and 
background $f_{\mathrm{bkg},c}$ (described in Sec.~\ref{sec:background_model}), 
respectively. 

Apart from the spurious signal, systematic uncertainties are incorporated
into the likelihood by multiplying the relevant parameter of the statistical
model by a factor 
\begin{equation}
F_\mathrm{G}(\sigma,\theta)=(1+\sigma\cdot\theta)
\label{eq:gaussnuisance}
\end{equation} 
in the case of a Gaussian pdf for the
effect of an uncertainty of size $\sigma$ or, for cases where a
negative model parameter does not make physical sense (e.g.\ the
uncertainty on a measured integrated luminosity), 
\begin{equation}
F_\mathrm{LN}(\sigma,\theta)=e^{\sqrt{\ln(1+\sigma^2)} \theta}
\label{eq:lognormnuisance}
\end{equation} 
for a log-normal pdf. In both cases the corresponding
component of the constraint product $G(\theta)$ is a unit Gaussian centered at
zero for $\theta$. The systematic uncertainties affecting the yield and
mass resolution use the log-normal form while a Gaussian form
is used for all others. When two uncertainties are considered fully
correlated they share the same nuisance parameter $\theta$ with different
values of $\sigma$. Systematic uncertainties with partial correlations are
decomposed into their uncorrelated and fully correlated components before
being assigned to nuisance parameters.

The likelihood for the combined signal strength is the product of 24 likelihoods,
consisting of the 12 category likelihoods for each dataset (7~TeV and 8~TeV).
The combined signal strength and its uncertainty are determined with the
profile likelihood ratio test statistic 
\begin{equation}
\lambda(\mu) = -2\ln{\LL(\mu,\hat{\vecth}_\mu)\over 
                  \LL(\hat{\mu},\hat{\vecth})},
\label{eq:profileLR}
\end{equation}
where $\hat{\mu}$ and $\hat{\vecth}$ are the values of the combined
signal strength and nuisance parameters that unconditionally maximize the
likelihood while $\hat{\vecth}_\mu$ are the values of the nuisance parameters
that maximize the likelihood on the condition that $\mu$ is held fixed to a
given value. In the asymptotic approximation, which is valid for all the results
presented here, $\lambda(\mu)$ may be interpreted as a change in
$\chi^2$ with respect to the minimum~\cite{stat}
such that approximate confidence intervals are easily constructed. 

\begin{table}[th!]
\footnotesize
\caption{Summary of sources of systematic uncertainty $\sigma$, the number of nuisance parameters $N_\mathrm{NP}$ used to
implement them for the combination of the 7~TeV and 8~TeV data ($i$ is the index to each of the unique nuisance parameters $\theta$), the factor in the
likelihood function ${F_\mathrm{G}(\sigma,\theta)}$ or ${F_\mathrm{LN}(\sigma,\theta)}$ (defined in 
Eqs.~(\ref{eq:gaussnuisance}) and (\ref{eq:lognormnuisance})) 
that implements their impact on signal yields,
mass resolution and scale, and the spurious signals resulting from the background parameterization, and the section in which they are presented.
When acting on $N_\mathrm{S}^\mathrm{tot}$ the  uncertainty value is  the same for  all processes,
whereas  the uncertainty has a different value for each signal process
for  the case  denoted $N_\mathrm{S}^{p}$.} 
\label{tab:big_systematics}
\begin{center}
\begin{tabular}{cccccc}
\hline\hline
& & Syst. source & $N_\mathrm{NP}$  & Implementation &  Section \bigstrut \\
\hline\hline

\parbox[t]{1mm}{\multirow{7}{*}{\rotatebox[origin=c]{90}{Yield}}} &
\parbox[t]{1mm}{\multirow{3}{*}{\rotatebox[origin=c]{90}{Theory}}}
        & Scales          & 7 & $N_\mathrm{S}^{p} \, F_\mathrm{LN}(\sigma_i,\theta_i)$           &  \ref{sec:yieldtheory} \bigstrut \\
      & & PDF+$\alpha_\mathrm{S}$  & 2 & $N_\mathrm{S}^{p} \, F_\mathrm{LN}(\sigma_i,\theta_i)$           &  \ref{sec:yieldtheory} \bigstrut \\
      & & Br.~ratio & 1 & $N_\mathrm{S}^\mathrm{tot} \, F_\mathrm{LN}(\sigma_i,\theta_i)$  & \ref{sec:yieldtheory}  \bigstrut \\
  \cline{2-5}                  
  & \parbox[t]{1mm}{\multirow{4}{*}{\rotatebox[origin=c]{90}{Exp.}}} 
         &  Luminosity     & 2 & $N_\mathrm{S}^\mathrm{tot}  \, F_\mathrm{LN}(\sigma_i,\theta_i)$ & \ref{sec:yieldexpt}.\ref{itm:lumi}    \bigstrut   \\
   &     &   Trigger       & 2 & $N_\mathrm{S}^\mathrm{tot}  \, F_\mathrm{LN}(\sigma_i,\theta_i)$ & \ref{sec:yieldexpt}.\ref{itm:trigger} \bigstrut   \\
   &     &   Photon ID       & 2 & $N_\mathrm{S}^\mathrm{p} \, F_\mathrm{LN}(\sigma_i,\theta_i)$  & \ref{sec:yieldexpt}.\ref{itm:photonID} \bigstrut   \\
   &     &   Isolation     & 2 & $N_\mathrm{S}^\mathrm{p}   \, F_\mathrm{LN}(\sigma_i,\theta_i)$& \ref{sec:yieldexpt}.\ref{itm:isolation}  \bigstrut \\

 \hline\hline
 
 \multicolumn{2}{c}{MC} & MC stats. & 14 & $N_\mathrm{S}^{p}  \, F_\mathrm{G}(\sigma_i^p,\theta_i)$ &  \ref{sec:MCstat} \bigstrut \\
\hline\hline

   \parbox[t]{1mm}{\multirow{11}{*}{\rotatebox[origin=c]{90}{Migrations}}} &
   \parbox[t]{1mm}{\multirow{8}{*}{\rotatebox[origin=c]{90}{Theory}}} 

   & Jet-bin           & 2 &  $N_\mathrm{S}^{\ggH} \, F_\mathrm{LN}(\sigma_i^{\ggH},\theta_i^{\ggH}) $ & \ref{sec:migrationTheory}.\ref{itm:jets} \bigstrut \\
 & & UE+PS             & 1 &  $N_\mathrm{S}^{p} \, F_\mathrm{G}(\sigma_i^{p},\theta_i)$       & \ref{sec:migrationTheory}.\ref{itm:UE}   \bigstrut \\ 
 & & Higgs $\pT$       & 1 &  $N_\mathrm{S}^{\ggH} \, F_\mathrm{G}(\sigma_i^{\ggH},\theta_i^{\ggH})$  & \ref{sec:migrationTheory}.\ref{itm:HpT}\bigstrut   \\ 
 & & $\Delta\phi_{jj}$  & 1 &  $N_\mathrm{S}^{\ggH} \, F_\mathrm{LN}(\sigma_i^{\ggH},\theta_i^{\ggH})$  & \ref{sec:migrationTheory}.\ref{itm:dphijj}\bigstrut  \\
 & &     $\eta^{*}$     & 1 &  $N_\mathrm{S}^{\ggH} \, F_\mathrm{LN}(\sigma_i^{\ggH},\theta_i^{\ggH})$  & \ref{sec:migrationTheory}.\ref{itm:dphijj} \bigstrut \\
 & & \ttH\ model   & 2 &   $N_\mathrm{S}^{\ttH}\,{F_\mathrm{LN}(\sigma_i^{\ttH},\theta_i^{\ttH})}$ & \ref{sec:migrationTheory}.\ref{itm:ttH}  \bigstrut  \\
 & & HF content        & 1 &   $N_\mathrm{S}^{p} \, F_\mathrm{LN}(\sigma_i^{p},\theta_i)$      & \ref{sec:migrationTheory}.\ref{itm:ttH}  \bigstrut \\
 & & Scale (\ttH\ cat.) & 4 &   $N_\mathrm{S}^{p}\,{F_\mathrm{LN}(\sigma_i^{\ttH},\theta_i^{\ttH})}$ & \ref{sec:migrationTheory}.\ref{itm:ttH} \bigstrut   \\
  \cline{2-5}
 & \parbox[t]{1mm}{\multirow{3}{*}{\rotatebox[origin=c]{90}{Exp.}}} 
    & Jet reco.   & 20 & $N_\mathrm{S}^{p}  \, F_\mathrm{G}(\sigma_i^{p},\theta_i)$  &  \ref{sec:migrationexpt}.\ref{itm:JESMES}  \bigstrut \\
 &  & \MET\       & 5  & $N_\mathrm{S}^{p}  \, F_\mathrm{G}(\sigma_i^{p},\theta_i)$  &  \ref{sec:migrationexpt}.\ref{itm:JESMES}  \bigstrut  \\
 &  & $b$-tagging & 13 & $N_\mathrm{S}^{p} \, F_\mathrm{G}(\sigma_i^{p},\theta_i)$   &  \ref{sec:migrationexpt}.\ref{itm:btag}   \bigstrut  \\
 &  & Lepton  ID+isol. &  2 & $N_\mathrm{S}^{p} \, F_\mathrm{G}(\sigma_i^{p},\theta_i)$   &  \ref{sec:migrationexpt}.\ref{itm:leptonall} \bigstrut \\
 &  & Lepton isolation &  2 & $N_\mathrm{S}^{p} \, F_\mathrm{G}(\sigma_i^{p},\theta_i)$   &  \ref{sec:migrationexpt}.\ref{itm:leptonall} \bigstrut \\
 
 \hline\hline
  \parbox[t]{1mm}{\multirow{4}{*}{\rotatebox[origin=c]{90}{Mass}}} 
        &       &    \multirow{2}{*}{Resolution}        & \multirow{2}{*}{4}  & $\sigma_\mathrm{CB}  \, F_\mathrm{LN}(\sigma_i,\theta_i)$ & \multirow{2}{*}{\ref{sec:unc_massres}} \bigstrut \\
        &       &                                       &                     & $\sigma_\mathrm{GA} \, F_\mathrm{LN}(\sigma_i,\theta_i)$  & \bigstrut \\
   \cline{2-5}
        &       &   \multirow{2}{*}{Scale}              & \multirow{2}{*}{43} & $\mu_\mathrm{CB} \, F_\mathrm{G}(\sigma_i,\theta_i)$    &  \multirow{2}{*}{\ref{sec:unc_escale}} \bigstrut \\
        &       &                                       &                     & $\mu_\mathrm{GA}\, F_\mathrm{G}(\sigma_i,\theta_i)$     &                                      \bigstrut   \\
 \hline\hline
 Back.   & & Spurious  signal                          & 12 &  $N_{\mathrm{spur},c}\, \theta_{\mathrm{spur},c}$                  & \ref{sec:background_model} \bigstrut \\
\hline\hline
\end{tabular}
\end{center}
\end{table}

A summary of the different sources of systematic uncertainty, the
number of associated nuisance parameters and the functional forms
used as constraints is reported in Table~\ref{tab:big_systematics}.
As can be seen in Table~\ref{tab:big_systematics} there are 146
constrained nuisance parameters associated with systematic uncertainties.
Twelve of these are associated with the
spurious signal in each of the 12 event categories. There are 49 unconstrained
nuisance parameters that describe the normalizations and shapes of the fitted
backgrounds in the 12 categories for the 7 TeV and 8 TeV data. 
As at least 2 events are needed to constrain the slope of the exponential background model, 
the categories with low expected yields are assumed to have the same shape parameters for the 7~TeV and the 8~TeV data.
The \VH\ \met, one-lepton, and dilepton categories are defined to have low yield since
the probabilities to observe 2 events in the 7~TeV data are less than 1\% based on the numbers
of events observed in the corresponding 8 TeV data categories. 

To test the
signal strengths of individual production processes or groups of them, the 
hypothesized number of signal events
and invariant mass distribution are decomposed into individual contributions,
\begin{equation}
\mu N_{\mathrm{S},c} \rightarrow \sum_p\mu_p N_{p,c},
\label{eqn:muprocess}
\end{equation}
where $\mu_p$ is the hypothesized signal strength for production process 
$p\in$\{\ggF,\VBF,\ZH,\WH,\ttH,\bbH,\tH\} and $N_{p,c}$ is the number of signal
events predicted by the SM in category $c$ for production process $p$
(the nuisance parameters are not shown in Eq.~(\ref{eqn:muprocess}), but they follow the
decomposition). In several of the results in the
next section some of the signal strengths are required to have the same value, such as for
the measurement of the combined signal strength where all seven are set equal.
For the measurements of individual signal strengths and signal strength
ratios, $\mu_{\bbH}$ and $\mu_{\tH}$ are held constant at 1, thus treating them
effectively as backgrounds.

The total uncertainty ${^{+\delta\mu_+}_{-\delta\mu_-}}$
at the 68\% confidence level (CL) 
of a measured signal strength $\mu_\mathrm{X}$ with best fit value $\hat{\mu}_\mathrm{X}$ is 
estimated by finding the points where ${\Lambda(\hat{\mu}_\mathrm{X}+\delta\mu_+)=\Lambda(\hat{\mu}_\mathrm{X}-\delta\mu_-)=1}$. 
The statistical component of the total uncertainty is
estimated by fixing all the 146 constrained nuisance parameters associated with
systematic uncertainties summarised in Table~\ref{tab:big_systematics}
to their maximum likelihood values and finding the new
points where ${\Lambda_\mathrm{stat.}(\mu_\mathrm{X}) = 1}$. The total systematic uncertainty
is given by the quadratic difference between the total and statistical uncertainties.
The separate contributions of the total experimental
and total theoretical uncertainties are estimated by finding the points where
${\Lambda_\mathrm{stat.\oplus expt.}(\mu_\mathrm{X})=1}$ and ${\Lambda_\mathrm{stat.\oplus theory}(\mu_\mathrm{X})=1}$,
respectively,
when fixing the 123 (23) constrained nuisance parameters
associated with experimental (theoretical) uncertainty
to their maximum likelihood values, and 
subtracting the resulting uncertainties in quadrature from the total uncertainty. For cases 
where the confidence intervals are approximately symmetric around the best fit value of $\mu_\mathrm{X}$, the 
positive and negative uncertainty contributions are reported as a single value ${\pm\delta\mu}$.

\section{Results}
\label{sec:results}

The observed diphoton invariant mass distribution for the sum of the 7 \TeV\ and 8 \TeV\ data is shown in Fig.~\ref{fig:invmass}
and in \mbox{Fig.~\ref{fig:invmass_cat}}  for the  sums of categories most sensitive to different
production modes. In all cases, for illustration purposes, each event is weighted according to the expected signal-to-background ratio $S_{90}/B_{90}$ for
the relevant category  and center-of-mass energy. 
The results of signal plus background fits to these spectra with \mH\ set to ${\combmass}$~\GeV\
are shown together with the separate signal and background components. Both the signal plus background and background-only curves reported here are 
obtained from the sum of the individual curves in each category weighted in the same way as the data points.

\begin{figure}[!htb]
  \centering

	\includegraphics[width=1.0\columnwidth]{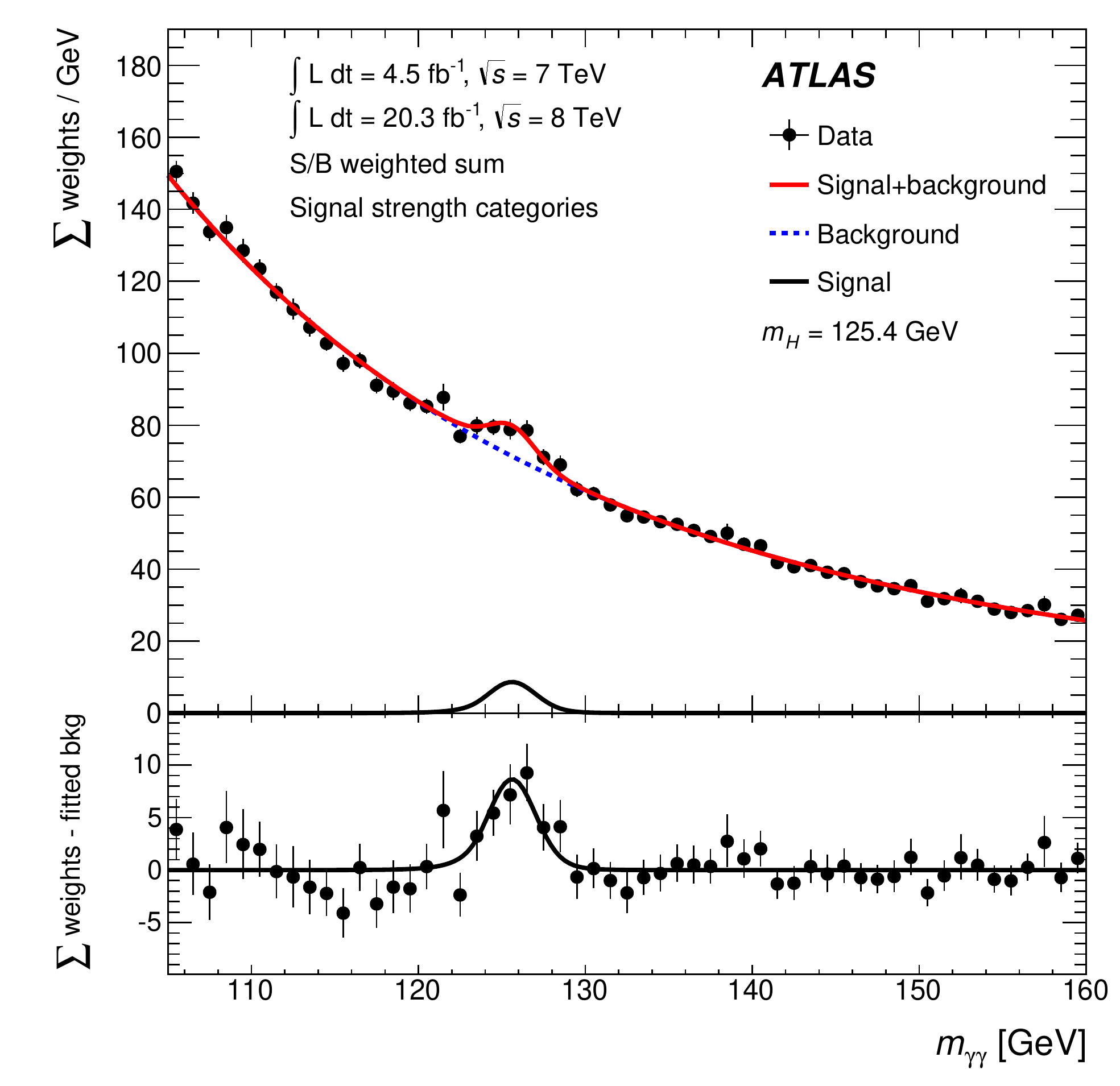}
	 \label{invmass_inclusive}

  \caption{Diphoton invariant mass \mgg\ spectrum observed in the sum of the 7~TeV and 8 TeV data. Each event is weighted by the signal-to-background
    ratio in the dataset and category it belongs to. The errors bars represent 68\% confidence intervals of the weighted sums.
The solid red curve shows the fitted signal plus background model when the Higgs boson 
    mass is fixed at ${\combmass}$~\GeV. The  background component of the fit is shown  with  the 
    dotted blue curve. The signal component of the fit is shown with the solid black curve.
    Both the signal plus background and background-only curves reported here are 
obtained from the sum of the individual curves in each category weighted by their
signal-to-background ratio.
    The bottom plot shows the data relative to the background component of the fitted model.} 
  \label{fig:invmass}
\end{figure}

\begin{figure*}[!htbp]
  \centering
	\subfigure[] {  \includegraphics[width=1.0\columnwidth]{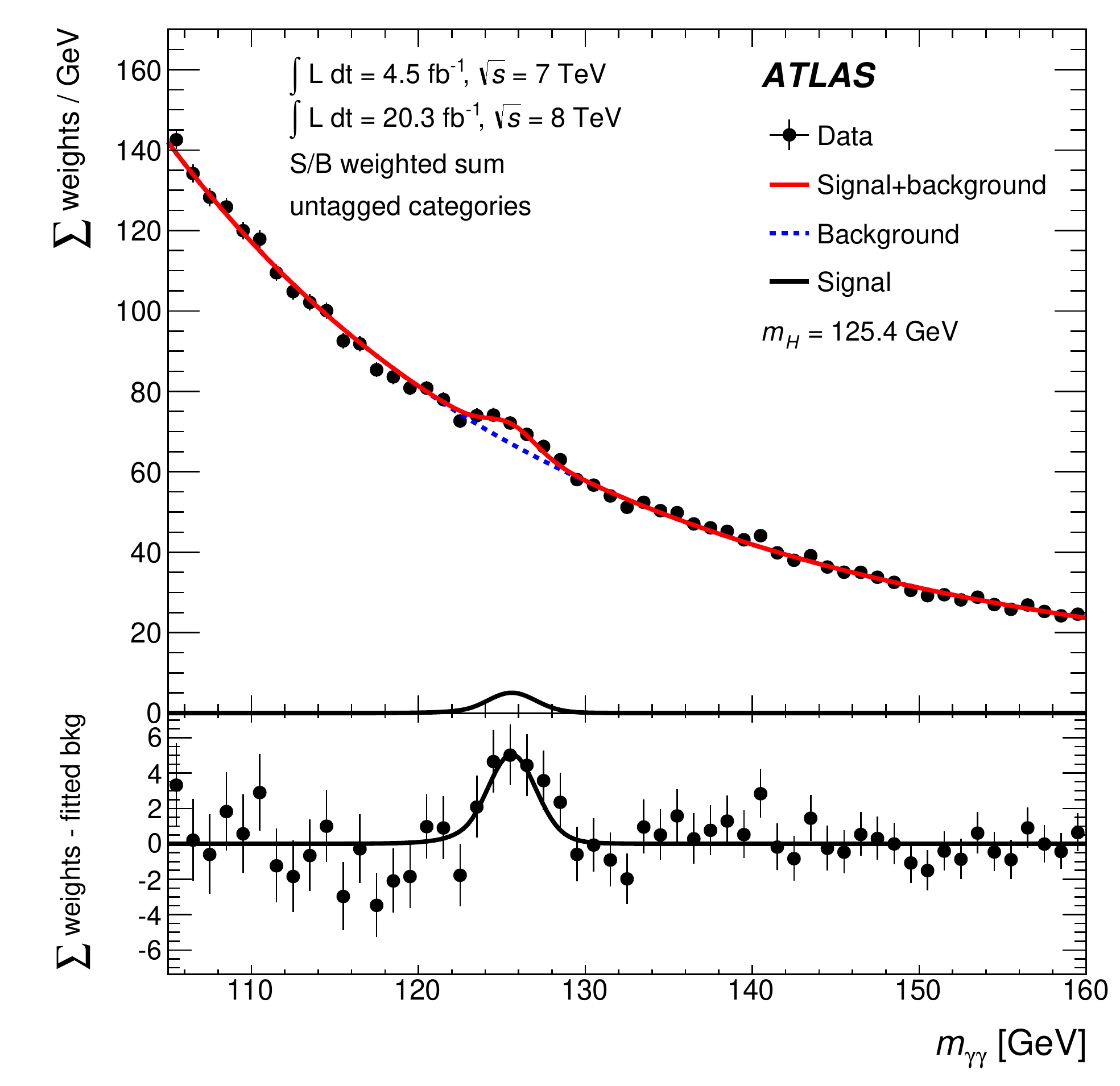} \label{invmass_untagged}}
	\subfigure[] {  \includegraphics[width=1.0\columnwidth]{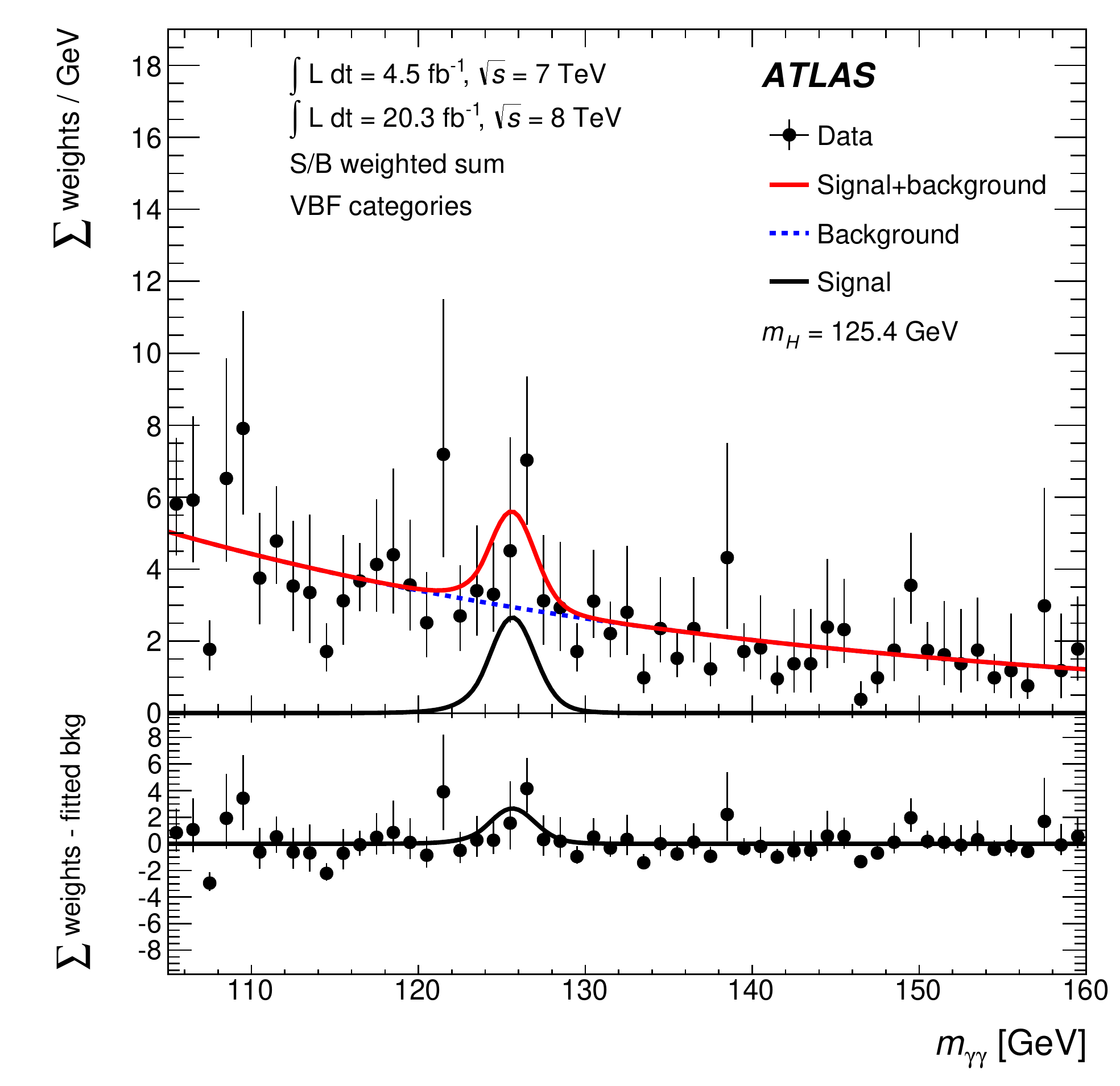} \label{invmass_VBF}}
	\subfigure[] {  \includegraphics[width=1.0\columnwidth]{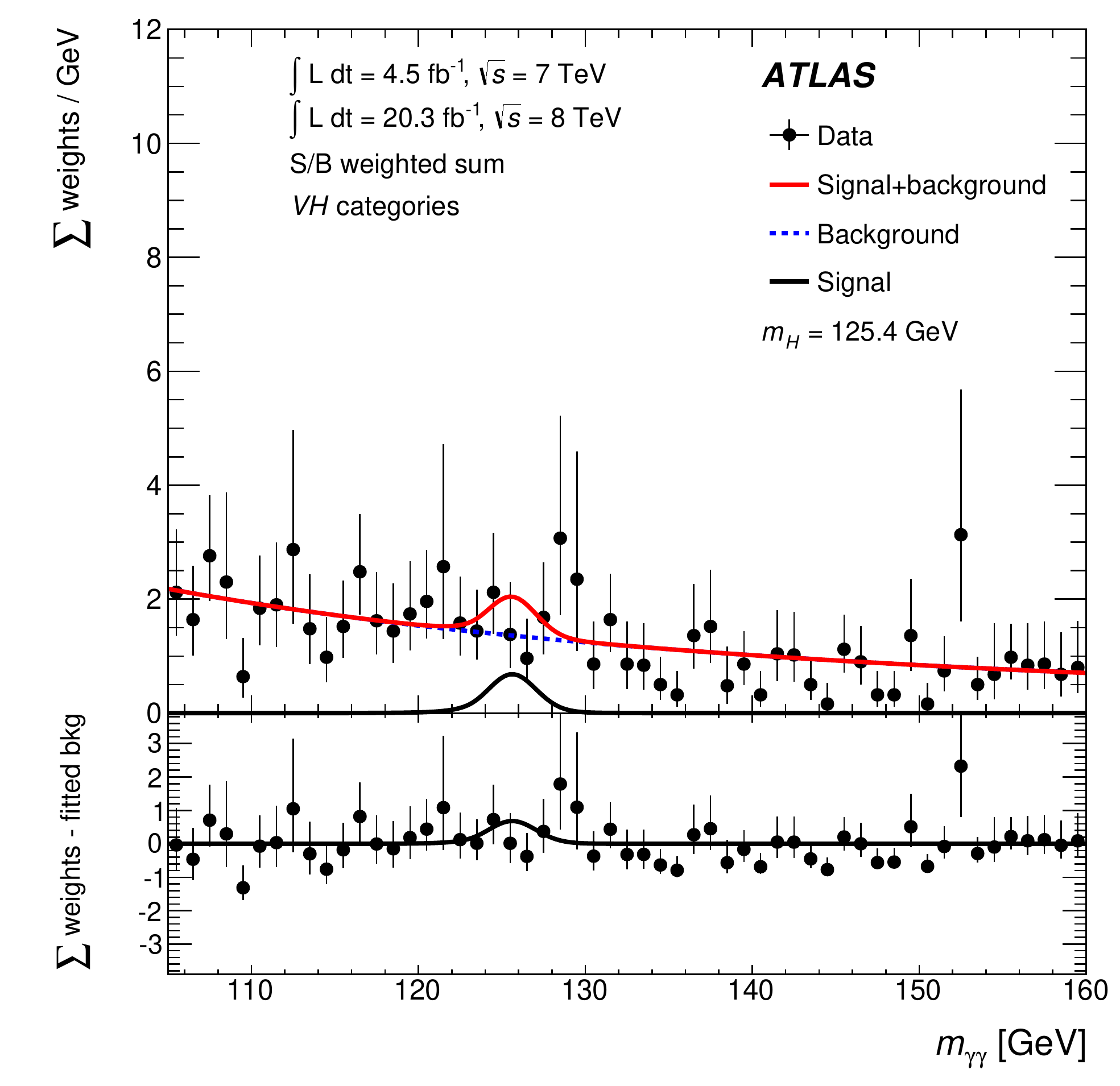}\label{invmass_VH}}
	\subfigure[] {  \includegraphics[width=1.0\columnwidth]{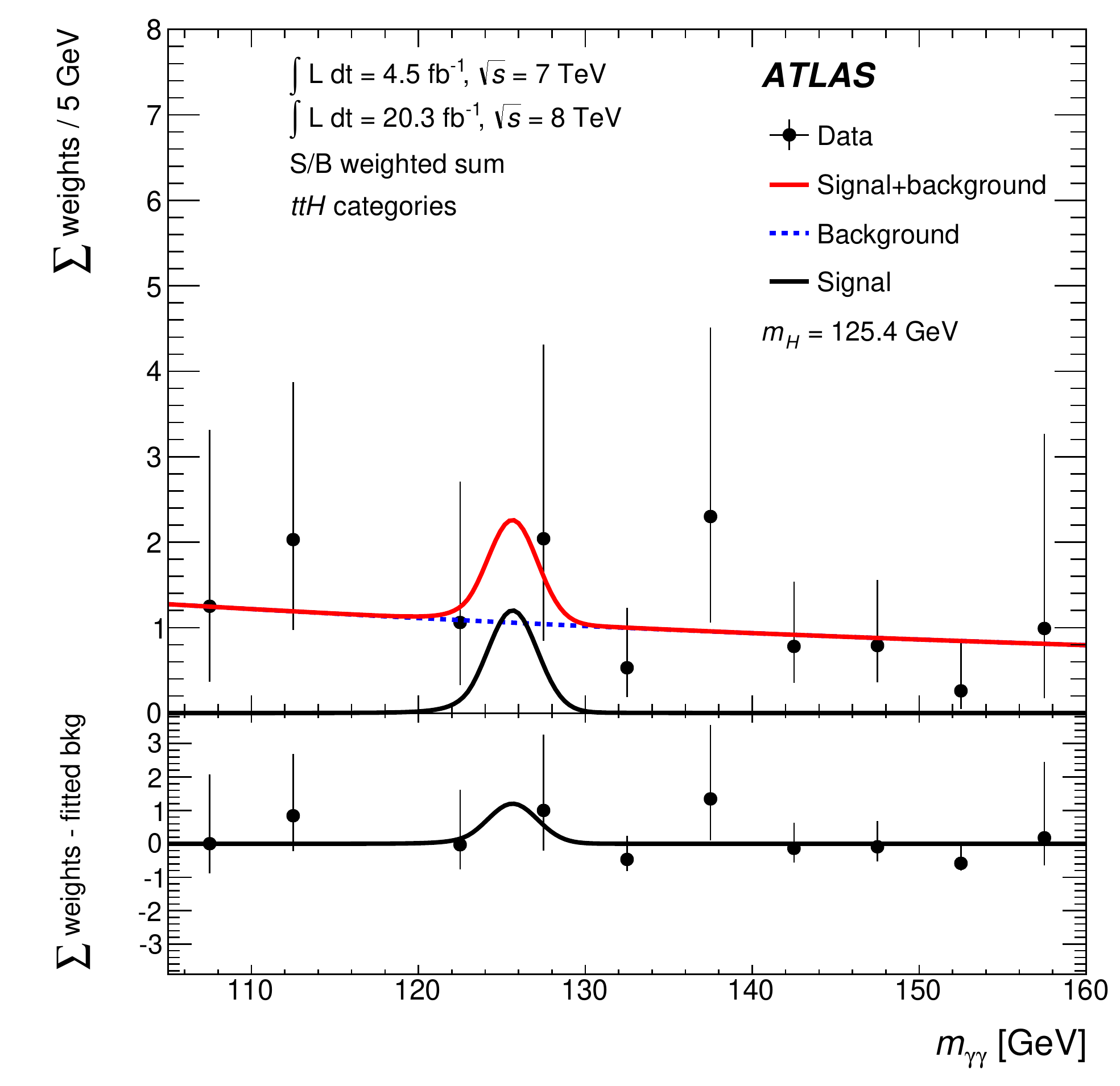}\label{invmass_ttH}}

  \caption{Diphoton invariant mass spectra observed in the 7 TeV and 8 TeV data in four groups
    of categories: (a) untagged categories, which are dominated by ggF, (b) \VBF\ categories, (c) \VH\ and (d) \ttH\ 
    categories.
    In each plot the contribution from the different categories in each group is weighted according to the $S/B$ ratio in each category.
    The errors bars represent 68\% confidence intervals of the weighted sums.
    The solid red line shows the fitted signal plus background model when the Higgs boson mass is fixed at ${\mH=\combmass}$~\GeV.
    The  background component of each fit is shown with a dotted blue line.  
    Both the signal plus background and background-only curves reported here are 
    obtained from the sum of the individual curves in each category weighted by their
    signal-to-background ratio.
    The bottom plot in each figure shows the data relative to the background component of the fitted model.} 
  \label{fig:invmass_cat}
\end{figure*}

The signal strengths are measured with the extended likelihood analysis described in Sec.~\ref{sec:statistic}.
The profile of the negative log-likelihood ratio ${\lambda(\mu)}$ (Eq.~(\ref{eq:profileLR}))
of the combined signal strength $\mu$ for ${\mH=\combmass}$~GeV is shown in Fig.~\ref{fig:nllscanmulti_mu}.
\begin{figure}[!htbp]
  \centering
  \includegraphics[width=1.0\columnwidth]{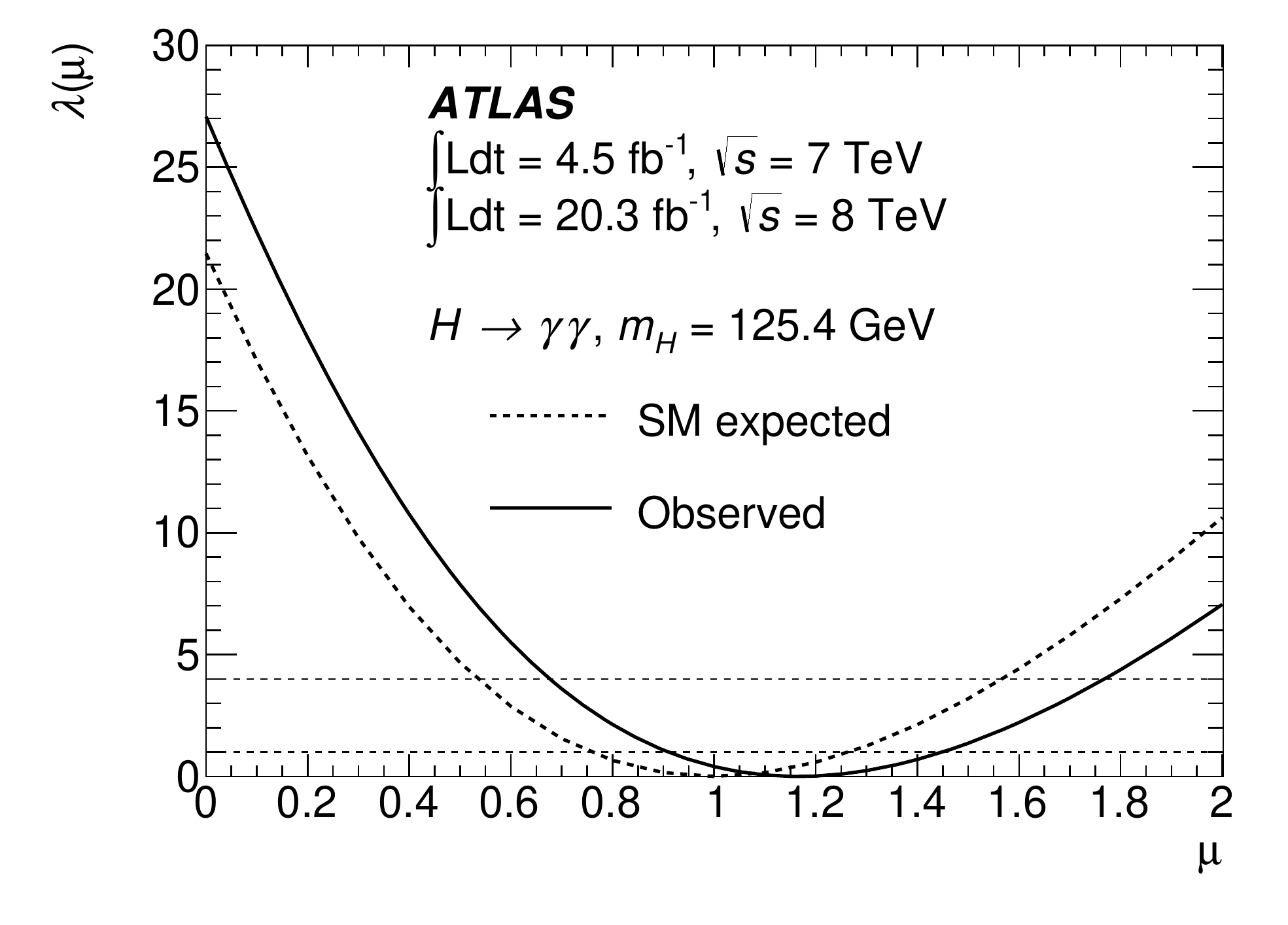} 
  \caption{The profile of the negative log-likelihood ratio ${\lambda(\mu)}$ of the combined signal strength
           $\mu$ for ${\mH=\combmass}$~GeV. The observed result is shown by the solid curve, the expectation for the
           SM by the dashed curve. The intersections of the solid and dashed curves with the horizontal dashed line 
           at ${\lambda(\mu)=1}$ indicate the 68\% confidence intervals of the observed and expected results, respectively.
  }
  \label{fig:nllscanmulti_mu}
\end{figure}
The local significance $Z$ of the observed combined excess of events, given by ${\sqrt{\lambda(0)}}$, is
${5.2\,\sigma}$ (${4.6\,\sigma}$ expected). 
The best fit value of $\mu$, determined by the minimum of ${\lambda(\mu)}$, is found to be 
\begin{align*}
\combmu \\
\combmucompactSS,
\end{align*}
corresponding to a ${0.7\,\sigma}$
compatibility with the SM prediction (${\mu=1}$). 
Figure~\ref{fig:muvsmh} shows the best fit value of $\mu$ as a function of \mH\ when mass scale
systematic uncertainties are included in or excluded from the fit. 
The figure illustrates that when the mass scale systematic
uncertainties are taken into account, the mass region compatible with the peak position
is broadened. Only a slight dependence of $\mu$ on 
\mH\ in the region compatible with the \combmassname\ ${\mH=\combmass\pm 0.4}$~\GeV\ is seen. This is also
a consequence of the small variation of the cross section times branching ratio versus \mH\ in the
same region (about 2\%/GeV).

\begin{figure}[!htb]
\centering
\includegraphics[width=\columnwidth]{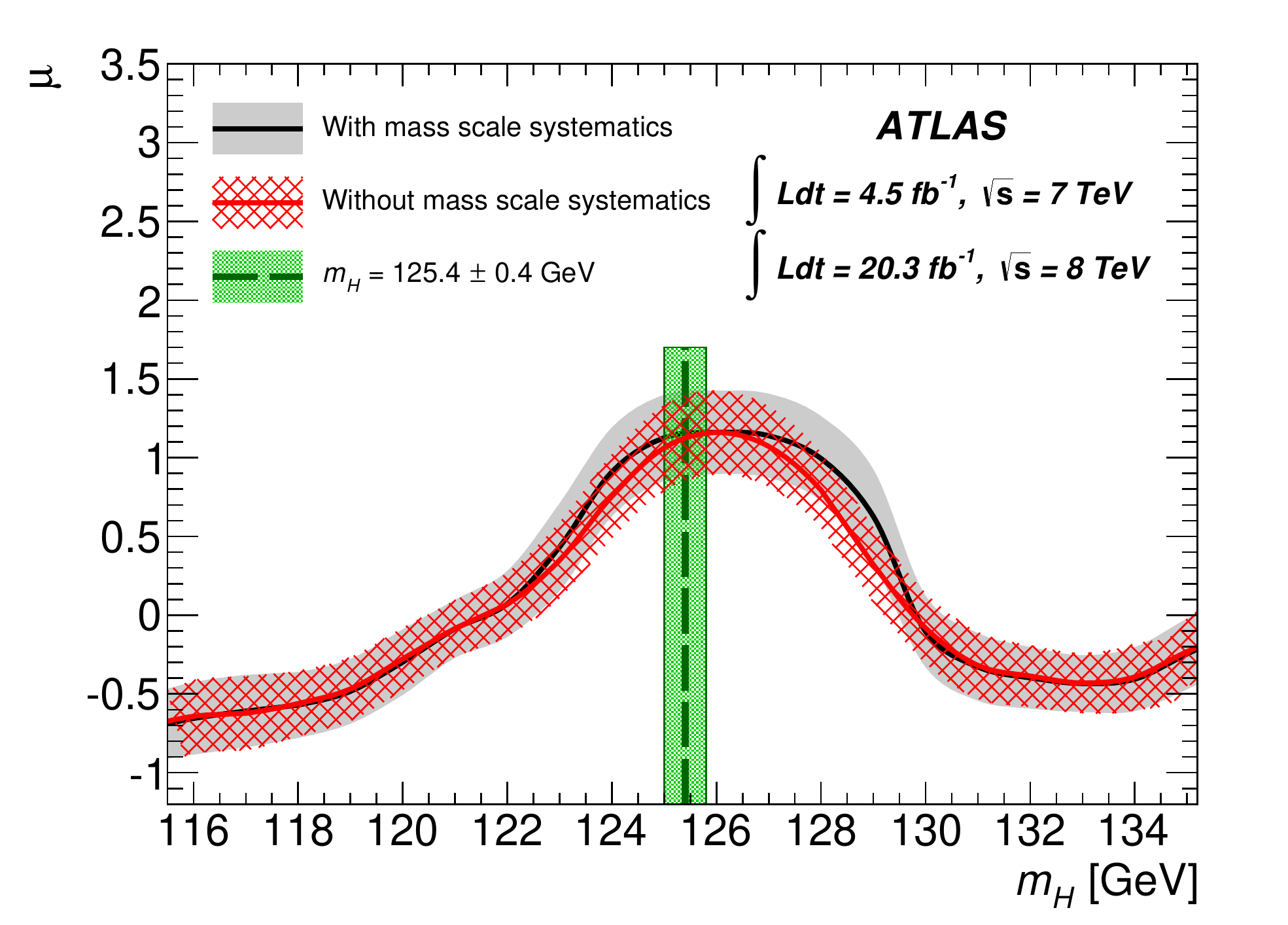}
\caption{The combined signal strength parameter $\mu$ versus \mH\ with mass scale systematic uncertainties
included (black curve) and excluded (red curve). The uncertainties on the measured $\mu$ are shown as gray
(red) bands with the mass scale systematic uncertainties included (excluded). The vertical dotted line and shaded
band indicate the value ${\mH=\combmass\pm 0.4}$~\GeV.}
\label{fig:muvsmh}
\end{figure}

The signal strengths measured in the individual event categories are shown in Fig.~\ref{fig:mu_allcats}. The signal strengths
measured in the four production mode--based groups of categories described in Sec.~\ref{sec:categorisation} are presented 
in Fig.~\ref{fig:mu_groupcat}. All of these individual and grouped signal strengths are compatible with the combined signal strength.

\begin{figure}[!htb]
\centering
\includegraphics[width=\columnwidth]{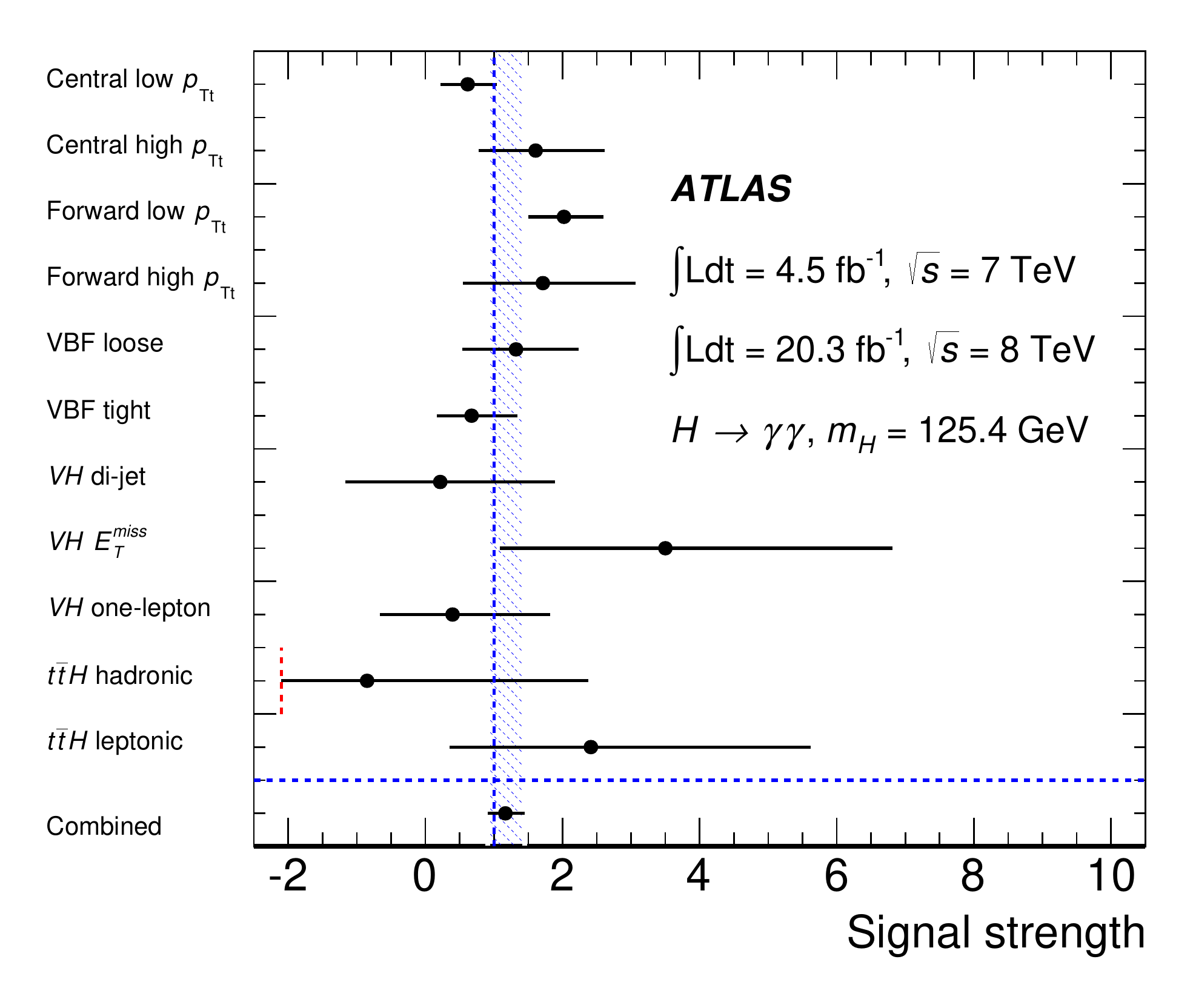}
\caption{The signal strength for a Higgs boson of mass ${\mH = \combmass}$~\gev\ decaying via \hgg\ 
as measured in the individual analysis categories, and the combined signal strength, for the combination of the 7~TeV and 8~\TeV\ data.
The vertical hatched band indicates the 68\%\ confidence interval of the combined signal strength. 
The vertical dashed line at signal strength 1 indicates the SM expectation.
The vertical dashed red line indicates the limit below which 
the fitted signal plus background mass distribution for the  \ttH\ hadronic category
becomes negative for some mass in the fit range. The \VH\ dilepton category is not shown because with
only two events in the combined sample, the fit results are not meaningful.
}
\label{fig:mu_allcats}
\end{figure}

\begin{figure}[!htb]
\centering
\includegraphics[width=\columnwidth]{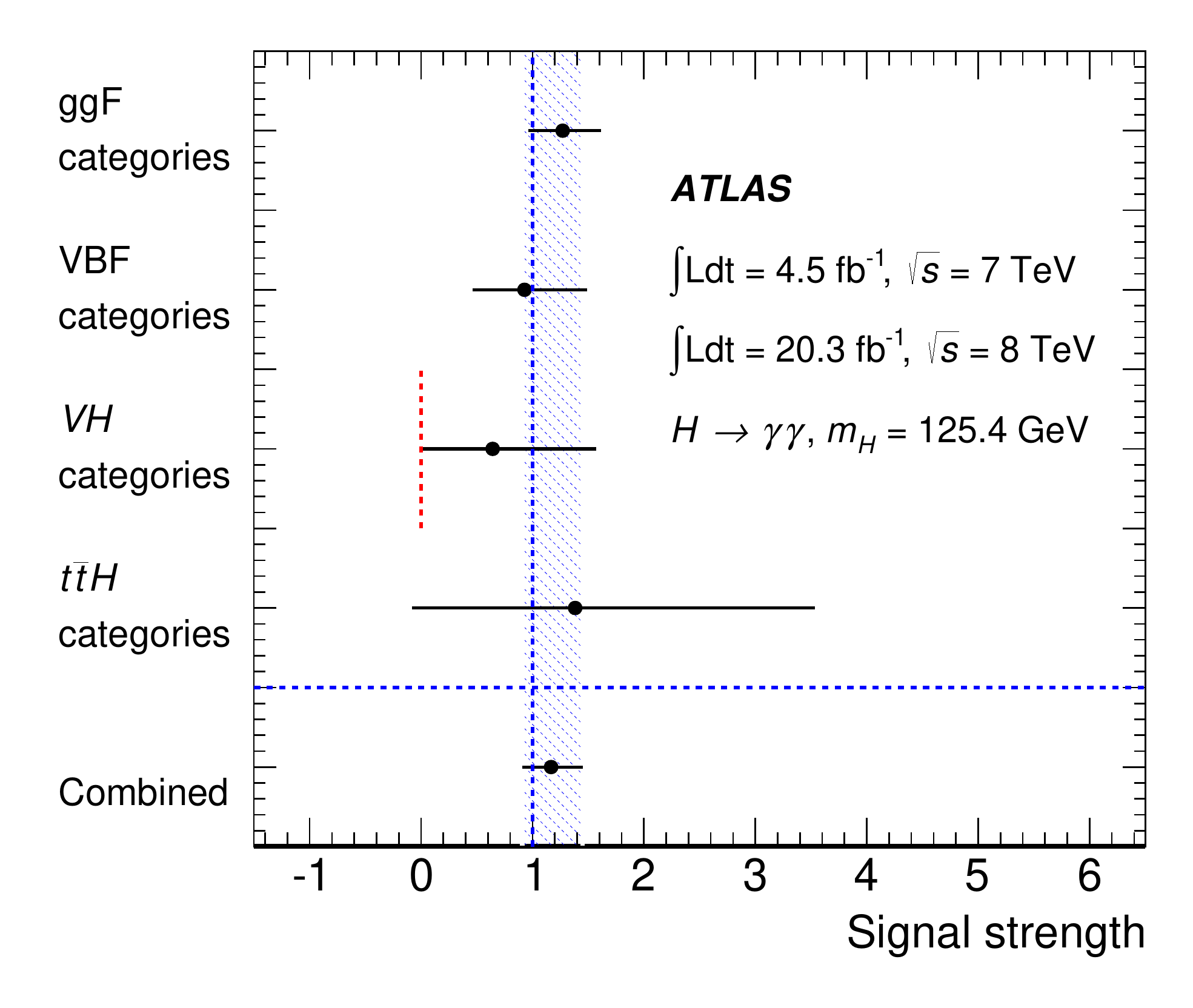}
\caption{The signal strength for a Higgs boson of mass ${\mH = \combmass}$~\gev\ decaying via \hgg\
as measured in groups of categories sensitive to individual production modes, and the combined signal strength, for the combination of the
7~TeV and 8~\TeV\ data. The vertical hatched band indicates the 68\%\ confidence interval of the combined signal strength.
The vertical dashed line at signal strength 1 indicates the SM expectation.
The vertical dashed red line indicates the limit below which
the fitted signal plus background mass distribution for the combination of the \VH\ categories
becomes negative for some mass in the fit range.  
}
\label{fig:mu_groupcat}
\end{figure}

The impacts of the main sources of systematic uncertainty
presented in Sec.~\ref{sec:systematics} on the combined 
signal strength parameter measurement
are presented in Table~\ref{tab:syst_combined}.
They are determined from the difference in quadrature between the nominal uncertainty 
and change in the 68\% CL range on $\mu$ when the corresponding nuisance parameters are fixed to their best fit values.
The sums of the squares of the theoretical uncertainties linked to the QCD scales, PDFs, and ${\hgg}$ 
branching ratio account for approximately 50\% of the square of the total systematic uncertainty. 
The dominant experimental uncertainty is from the photon energy resolution, 
which represents approximately 30\% of the total systematic uncertainty (as above in terms of its contribution to the square of the total 
systematic uncertainty). 
 In the fit to extract the signal strengths, the post-fit values of
 the most relevant nuisance parameters (those apart from the ones of the background model), do not show significant
 deviations from their pre-fit input values.

\begin{table}
  \caption{
    Main systematic uncertainties $\sigma_\mu^\mathrm{syst.}$ on the combined signal strength parameter $\mu$. 
    The values for each group of uncertainties are determined by subtracting in quadrature from the total uncertainty
    the change in the 68\% CL range on $\mu$ when the corresponding nuisance parameters are fixed to their best fit values.
    The experimental uncertainty on the yield does not include the luminosity contribution, which is accounted for
    separately.
  }
  \label{tab:syst_combined}
  \begin{center}
    \begin{tabular}{lc}
      \hline\hline
      Uncertainty group & $\sigma_\mu^\mathrm{syst.}$ \\
      \hline\hline

	Theory (yield)& 0.09 \\
	 Experimental (yield)  & 0.02 \\	
	 Luminosity & 0.03 \\
	
	\hline
	
  	MC statistics & $< 0.01$ \\
	
	\hline
	
	Theory (migrations) & 0.03 \\
	Experimental (migrations) & 0.02 \\

	\hline	

	Resolution & 0.07 \\
	Mass scale & 0.02 \\
	
	\hline
	
	Background shape & 0.02 \\

      \hline \hline
    \end{tabular}
  \end{center}
\end{table}

The compatibility of the combined signal strength presented in this article with the one published in 
Ref.~\cite{atlas-couplings-diboson}, ${\pubcombmu}$, is investigated using a jackknife resampling
technique~\cite{jk1,jk2} in which variances and covariances of observables are
estimated with a series of sub-samples of the observations.
The datasets used in the two analyses are highly correlated: 142681 events are selected in Ref.~\cite{atlas-couplings-diboson}, 
111791 events are selected in the current analysis, and 104407 events are selected in both analyses.
The significance of the 0.4 difference between the combined signal strengths, 
including the effect of the 74\%\ 
correlation between the two measurements, is calculated by applying the jackknife technique
to the union of the two datasets and 
is found to be ${2.3\,\sigma}$. 
An uncertainty of ${0.1\,\sigma}$ on the compatibility between the two measurements 
is estimated by varying the 
size of the jackknife sub-samples.
The decrease
 in the observed signal significance (${5.2\,\sigma}$) with respect to the one published in Ref.~\cite{atlas-couplings-diboson}
(${7.4\,\sigma}$) is related to the reduction of the measured signal strength 
 according to the asymptotic 
formula ${Z = \mu / \sigma_{\mu}^{\mathrm{stat}}}$, where $\sigma_{\mu}^{\mathrm{stat}}$ is the statistical component of the uncertainty on $\mu$.
In other words, the observed reductions of the significance and signal strength are consistent with each other
and consistent with a statistical fluctuation at the level of ${\sim 2.3\,\sigma}$.

As can be seen in Figs.~\ref{fig:mu_allcats} and \ref{fig:mu_groupcat}, 
the observed signal strengths of the tagged categories, which are dominated by production processes 
other than ggF, tend to be lower than the signal strengths measured with the untagged categories,
which are dominated by ggF production. 
This tendency, combined with the optimized sensitivity of this analysis to production processes
other than ggF, results in a lower combined signal strength than those measured using alternative analyses
of the same dataset (or where the datasets are largely overlapping) that are inclusive with respect
to the production process.
The compatibility of the combined signal strength obtained in this analysis with the signal strength 
${\mu = 1.29 \pm 0.30}$
obtained in the mass measurement analysis quoted in Ref.~\cite{Aad:2014aba} for the diphoton channel
(where the diphoton events are sorted into categories that depend only on the properties of the photons)
is evaluated with the same resampling technique described above and found to be within one standard
deviation. A measurement of the fiducial cross section of Higgs boson production
in the ${\hgg}$ decay channel with the ATLAS detector
is performed in Ref.~\cite{ggfiducial_xs}. In order to make that analysis more model-independent, there is
no use of production process--related event categories. The signal strength of the measured fiducial cross section,
using only the 8~TeV data, is approximately $1.4$
and found to be compatible with the combined
signal strength measured here within ${1.2\,\sigma}$ (using again the jackknife resampling technique).

\begin{figure}[!htb]
\centering
\includegraphics[width=\columnwidth]{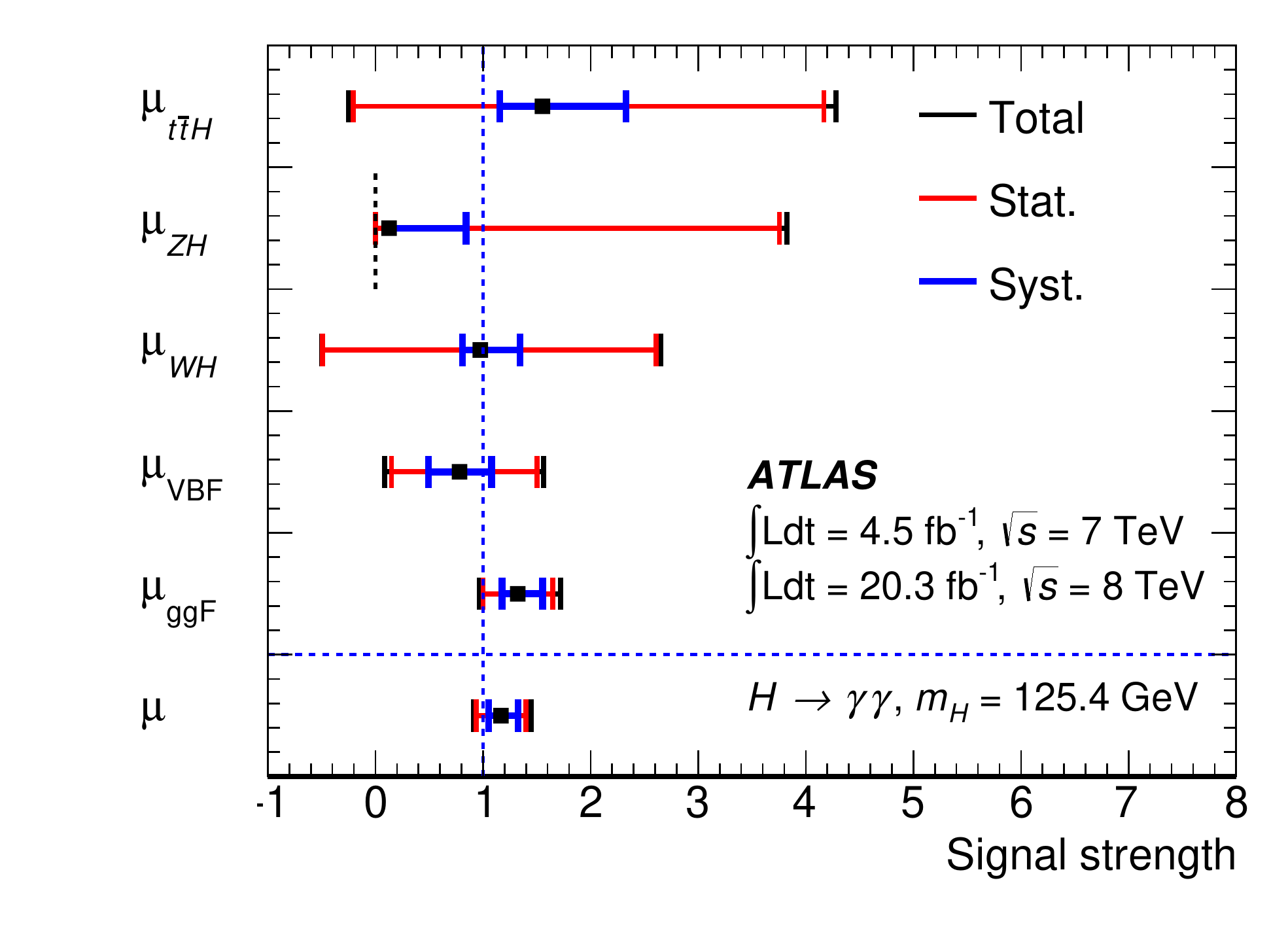}
\caption{Measured signal strengths, for a Higgs boson of mass ${\mH = \combmass}$~\gev\ decaying via \hgg,
  of the different Higgs boson production modes 
  and the combined signal strength $\mu$ obtained with the combination of the 7~TeV and 8~\TeV\ data.
  The vertical dashed line at ${\mu=1}$ indicates the SM expectation.
  The vertical dashed line at the left end of the $\mu_{\ZH}$ result indicates the limit below which
the fitted signal plus background mass distribution 
becomes negative for some mass in the fit range.}
\label{fig:5mu_summ}
\end{figure}

In addition to the combined signal strength, the signal strengths of the primary production processes are determined
by exploiting the sensitivities of the analysis categories to specific production processes, and found to be 
(see also Fig.~\ref{fig:5mu_summ}):
\begin{align*}
\mugghS \\
\mugghSS, \\
\muvbfS \\
\muvbfSS, \\
\muwhS \\
\muwhSS, \\
\muzhS \\
\muzhSS, \\
\mutthS \\
\mutthSS.
\end{align*}
In this measurement, both $\mu_{\tH}$ and $\mu_{\bbH}$ are fixed to the SM expectations ($\mu_{\tH}$=1 and $\mu_{\bbH}$=1).
The correlation between the fitted values of $\mu_{\ggH}$ and $\mu_{\VBF}$ has been studied by still
fixing both $\mu_{\tH}$ and $\mu_{\bbH}$ to 1 and profiling\footnote{Profiling here means maximizing the likelihood
with respect to all parameters apart from the parameters of interest $\mu_{\ggF}$ and $\mu_{\VBF}$.} 
the remaining signal strengths $\mu_{\ZH}$, $\mu_{\WH}$, and $\mu_{\ttH}$. The best-fit values of $\mu_{\ggH}$ and $\mu_{\VBF}$ and
the 68\% and 95\% CL contours are shown in Fig.~\ref{fig:ggh_vbf}.

\begin{figure}[!htb]
\centering
\includegraphics[width=\columnwidth]{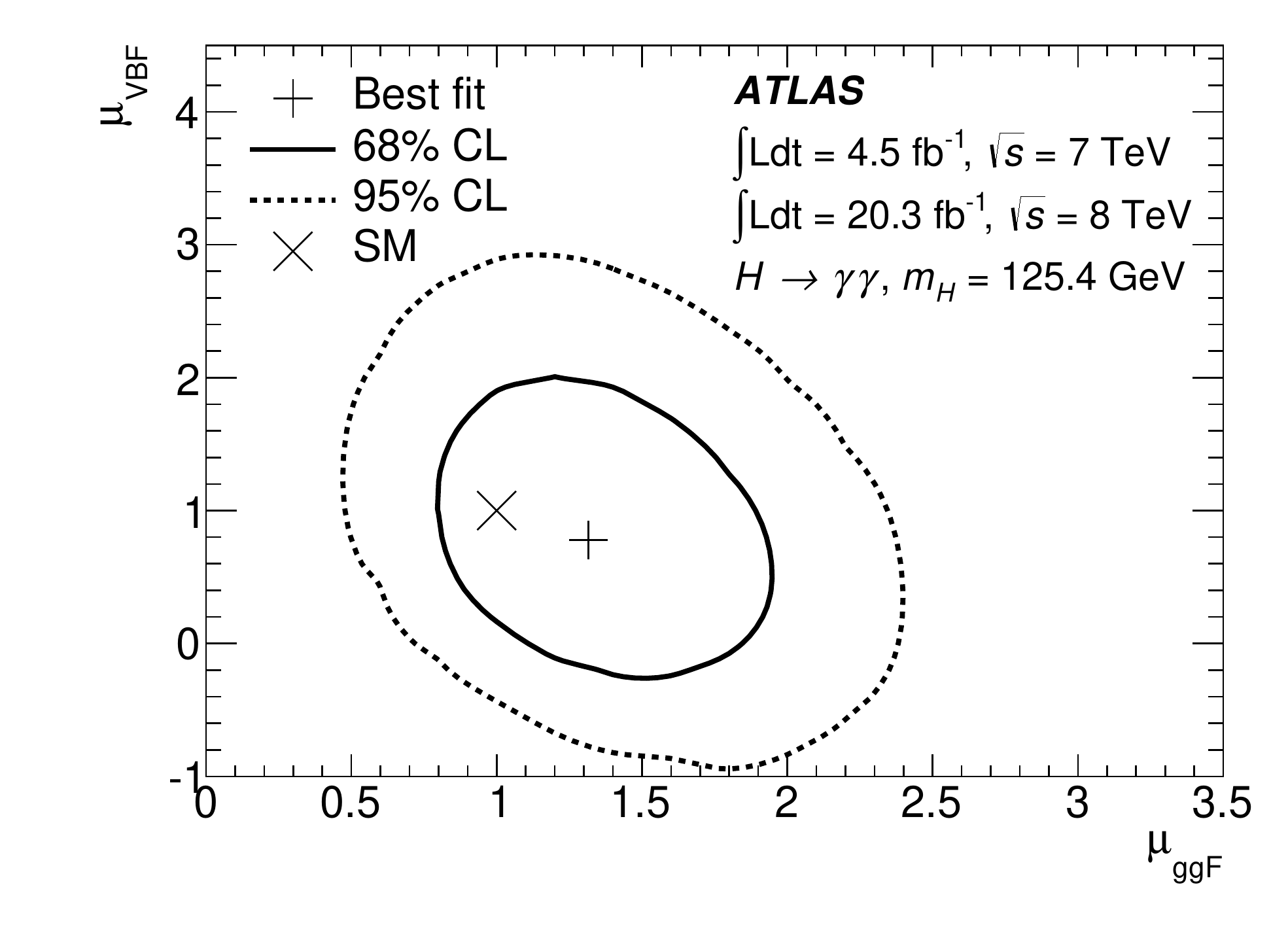}
\caption{The two-dimensional best-fit value of ($\mu_{\VBF}$, $\mu_{\ggH}$) for a Higgs boson of mass ${\mH = \combmass}$~\gev\
decaying via \hgg\ when fixing both $\mu_{\tH}$ and $\mu_{\bbH}$ to 1 and profiling
all the other signal strength parameters. The 68\% and 95\% CL contours are shown with the solid and dashed lines, respectively. 
The result is obtained for ${\mH = \combmass}$~\gev\ and the combination of the 7~TeV and 8~\TeV\ data.}
\label{fig:ggh_vbf}
\end{figure}
Compared with the measured \ttH\ signal strength parameter \mbox{\muttHpub} in Ref.~\cite{ttHpaper}, 
$\mu_{\ttH}$ measured in this analysis profits 
from the contribution of \ttH\ events in other categories such as \VH\ \MET\ and \VH\ one-lepton.
In addition, in this measurement the other contributions to the signal strength are profiled, whereas they 
are fixed at the SM predictions in Ref.~\cite{ttHpaper}.

As mentioned in the introduction, in order to test the production through VBF and associated 
production with a \Wboson\ or \Zboson\ boson or a $t\bar{t}$ pair,
independently of the ${\hgg}$ branching ratio, the ratios $\mu_{\VBF}$/$\mu_{\ggH}$, $\mu_{\VH}$/$\mu_{\ggH}$, 
and $\mu_{\ttH}$/$\mu_{\ggH}$ are fitted separately by fixing $\mu_{\tH}$ and $\mu_{\bbH}$ to 1 and profiling the remaining signal
strengths.  The measured ratios
\begin{align*}
\mu_{\VBF}/\mu_{\ggH} &= \muVBFomuggF, \\
\mu_{\VH}/\mu_{\ggH}  &=  \muVHomuggF, \\
\mu_{\ttH}/\mu_{\ggH} &= \muttHomuggF, 
\end{align*}
although not significantly different from zero, are consistent with the SM predictions of 1.0. 
Likelihood scans of these ratios are presented in Fig.~\ref{fig:compact_likelihood}. 
The result for $\mu_{\VBF}$/$\mu_{\ggH}$ is consistent with  ${\mu_{\VBF+\VH}/\mu_{\ggH+\ttH}=1.1^{+0.9}_{-0.5}}$
reported by ATLAS with the same data in Ref.~\cite{atlas-couplings-diboson}, although they are not directly comparable.

\begin{figure}[!htb]
\centering
\includegraphics[width=\columnwidth]{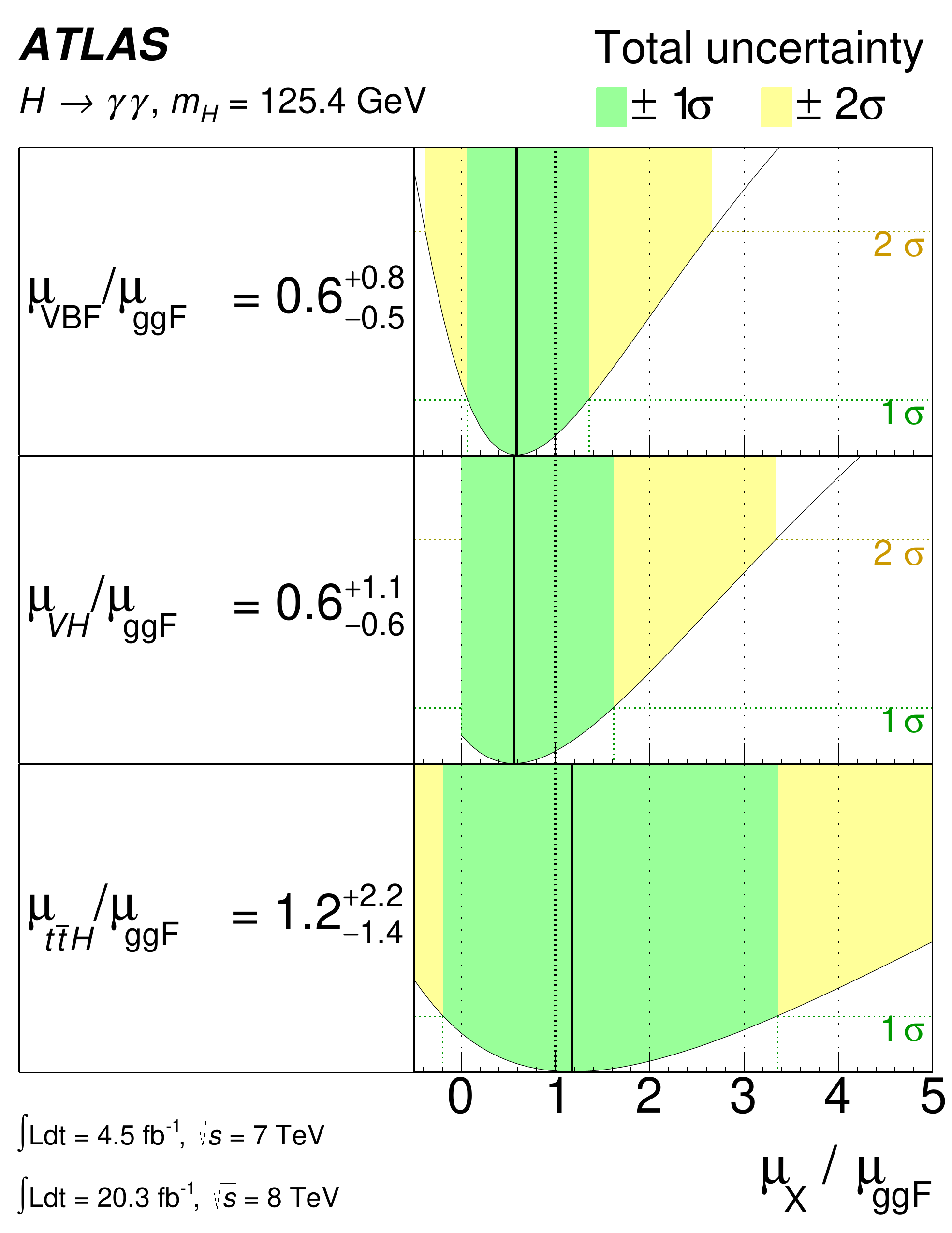}
\caption{Measurements of the $\mu_{\VBF}/\mu_{\ggH}$, $\mu_{\VH}/\mu_{\ggH}$  and $\mu_{\ttH}/\mu_{\ggH}$
ratios and their total errors
for a Higgs boson mass ${\mH=\combmass}$~\GeV. 
For a more complete illustration, the log-likelihood curves from 
which the total uncertainties are extracted are also shown:
the best fit values are represented by the solid vertical lines, with the total ${\pm 1 \sigma}$ and
${\pm 2 \sigma}$ uncertainties indicated by the dark- and light-shaded band, respectively.
The likelihood curve and uncertainty bands for $\mu_{\VH}/\mu_{\ggH}$ stop at zero
because below this the hypothesized signal plus background mass distribution
in the \VH\ dilepton channel becomes negative (unphysical) for some mass in the fit range.
}
\label{fig:compact_likelihood}
\end{figure}

%\clearpage

\phantom{nohting}

\phantom{nohting}

\section{Conclusion}
\label{sec:conclusion}

A refined measurement of Higgs boson signal strengths in the ${\hgg}$ decay channel is
performed using the proton-proton collision data recorded by the ATLAS experiment
at the CERN Large Hadron Collider at center-of-mass energies of ${\sqrt{s}=7}$~\TeV\ and ${\sqrt{s}=8}$~\TeV\
corresponding to a total integrated luminosity of 25 \ifb (the LHC Run 1 dataset).
The results are based on improved calibrations for photons, electrons and muons and on improved analysis 
techniques with respect to the previously published analysis of the same dataset. 
The strength of the signal relative to the SM expectation, measured at the combined 
ATLAS Higgs boson mass
${\mH=\combmass}$~GeV 
is found to be 
\begin{equation*}
\combmucompactS.
\end{equation*}
The compatibility with the SM prediction of ${\mu=1}$ corresponds to ${0.7\,\sigma}$. 
Signal strengths of the main 
production modes 
are measured separately by exploiting
event categories that are designed to be sensitive to particular production modes. They are
found to be
\begin{align*}
\mugghT, \\
\muvbfT, \\
\muwhT, \\
\muzhT, \\
\mutthT,
\end{align*}
where the statistical, systematic and theoretical uncertainties are combined. The total
uncertainty of both the combined and the five individual signal strength parameters presented above 
is dominated by the statistical uncertainty.
These are the first results obtained by ATLAS
in the diphoton final state for these five production mechanisms simultaneously. 
No significant deviations from the SM expectations are observed. More data are
needed to establish evidence for Higgs boson production in the ${\hgg}$ decay channel via the
\VBF, \WH, \ZH, and \ttH\ production mechanisms individually.
These results supersede the previous ones and represent the new reference for the
signal strengths of Higgs boson production in the ${\hgg}$ decay channel measured
by ATLAS with the LHC Run 1 data.

\section*{Acknowledgments}

We thank CERN for the very successful operation of the LHC, as well as the
support staff from our institutions without whom ATLAS could not be
operated efficiently.

We acknowledge the support of ANPCyT, Argentina; YerPhI, Armenia; ARC,
Australia; BMWF and FWF, Austria; ANAS, Azerbaijan; SSTC, Belarus; CNPq and FAPESP,
Brazil; NSERC, NRC and CFI, Canada; CERN; CONICYT, Chile; CAS, MOST and NSFC,
China; COLCIENCIAS, Colombia; MSMT CR, MPO CR and VSC CR, Czech Republic;
DNRF, DNSRC and Lundbeck Foundation, Denmark; EPLANET, ERC and NSRF, European Union;
IN2P3-CNRS, CEA-DSM/IRFU, France; GNSF, Georgia; BMBF, DFG, HGF, MPG and AvH
Foundation, Germany; GSRT and NSRF, Greece; ISF, MINERVA, GIF, I-CORE and Benoziyo Center,
Israel; INFN, Italy; MEXT and JSPS, Japan; CNRST, Morocco; FOM and NWO,
Netherlands; BRF and RCN, Norway; MNiSW and NCN, Poland; GRICES and FCT, Portugal; MNE/IFA, Romania; MES of Russia and ROSATOM, Russian Federation; JINR; MSTD,
Serbia; MSSR, Slovakia; ARRS and MIZ\v{S}, Slovenia; DST/NRF, South Africa;
MINECO, Spain; SRC and Wallenberg Foundation, Sweden; SER, SNSF and Cantons of
Bern and Geneva, Switzerland; NSC, Taiwan; TAEK, Turkey; STFC, the Royal
Society and Leverhulme Trust, United Kingdom; DOE and NSF, United States of
America.

The crucial computing support from all WLCG partners is acknowledged
gratefully, in particular from CERN and the ATLAS Tier-1 facilities at
TRIUMF (Canada), NDGF (Denmark, Norway, Sweden), CC-IN2P3 (France),
KIT/GridKA (Germany), INFN-CNAF (Italy), NL-T1 (Netherlands), PIC (Spain),
ASGC (Taiwan), RAL (UK) and BNL (USA) and in the Tier-2 facilities
worldwide.

\printfigures

\bibliographystyle{apsrev4-1}

\begin{thebibliography}{112}%
\makeatletter
\providecommand \@ifxundefined [1]{%
 \@ifx{#1\undefined}
}%
\providecommand \@ifnum [1]{%
 \ifnum #1\expandafter \@firstoftwo
 \else \expandafter \@secondoftwo
 \fi
}%
\providecommand \@ifx [1]{%
 \ifx #1\expandafter \@firstoftwo
 \else \expandafter \@secondoftwo
 \fi
}%
\providecommand \natexlab [1]{#1}%
\providecommand \enquote  [1]{``#1''}%
\providecommand \bibnamefont  [1]{#1}%
\providecommand \bibfnamefont [1]{#1}%
\providecommand \citenamefont [1]{#1}%
\providecommand \href@noop [0]{\@secondoftwo}%
\providecommand \href [0]{\begingroup \@sanitize@url \@href}%
\providecommand \@href[1]{\@@startlink{#1}\@@href}%
\providecommand \@@href[1]{\endgroup#1\@@endlink}%
\providecommand \@sanitize@url [0]{\catcode `\\12\catcode `\$12\catcode
  `\&12\catcode `\#12\catcode `\^12\catcode `\_12\catcode `\%12\relax}%
\providecommand \@@startlink[1]{}%
\providecommand \@@endlink[0]{}%
\providecommand \url  [0]{\begingroup\@sanitize@url \@url }%
\providecommand \@url [1]{\endgroup\@href {#1}{\urlprefix }}%
\providecommand \urlprefix  [0]{URL }%
\providecommand \Eprint [0]{\href }%
\providecommand \doibase [0]{http://dx.doi.org/}%
\providecommand \selectlanguage [0]{\@gobble}%
\providecommand \bibinfo  [0]{\@secondoftwo}%
\providecommand \bibfield  [0]{\@secondoftwo}%
\providecommand \translation [1]{[#1]}%
\providecommand \BibitemOpen [0]{}%
\providecommand \bibitemStop [0]{}%
\providecommand \bibitemNoStop [0]{.\EOS\space}%
\providecommand \EOS [0]{\spacefactor3000\relax}%
\providecommand \BibitemShut  [1]{\csname bibitem#1\endcsname}%
\let\auto@bib@innerbib\@empty
\bibitem [{\citenamefont {{ATLAS
  Collaboration}}(2012{\natexlab{a}})}]{ObservationATLAS}%
  \BibitemOpen
  \bibfield  {author} {\bibinfo {author} {\bibnamefont {{ATLAS
  Collaboration}}},\ }\href {\doibase 10.1016/j.physletb.2012.08.020}
  {\bibfield  {journal} {\bibinfo  {journal} {Phys.\ Lett.\ B}\ }\textbf
  {\bibinfo {volume} {716}},\ \bibinfo {pages} {1} (\bibinfo {year}
  {2012}{\natexlab{a}})},\ \Eprint {http://arxiv.org/abs/1207.7214}
  {arXiv:1207.7214 [hep-ex]} \BibitemShut {NoStop}%
\bibitem [{\citenamefont {{CMS Collaboration}}(2012)}]{ObservationCMS}%
  \BibitemOpen
  \bibfield  {author} {\bibinfo {author} {\bibnamefont {{CMS Collaboration}}},\
  }\href {\doibase 10.1016/j.physletb.2012.08.021} {\bibfield  {journal}
  {\bibinfo  {journal} {Phys.\ Lett.\ B}\ }\textbf {\bibinfo {volume} {716}},\
  \bibinfo {pages} {30} (\bibinfo {year} {2012})},\ \Eprint
  {http://arxiv.org/abs/1207.7235} {arXiv:1207.7235 [hep-ex]} \BibitemShut
  {NoStop}%
\bibitem [{\citenamefont {Englert}\ and\ \citenamefont
  {Brout}(1964)}]{Englert:1964et}%
  \BibitemOpen
  \bibfield  {author} {\bibinfo {author} {\bibfnamefont {F.}~\bibnamefont
  {Englert}}\ and\ \bibinfo {author} {\bibfnamefont {R.}~\bibnamefont
  {Brout}},\ }\href {\doibase 10.1103/PhysRevLett.13.321} {\bibfield  {journal}
  {\bibinfo  {journal} {Phys.\ Rev.\ Lett.}\ }\textbf {\bibinfo {volume}
  {13}},\ \bibinfo {pages} {321} (\bibinfo {year} {1964})}\BibitemShut
  {NoStop}%
\bibitem [{\citenamefont {Higgs}(1964{\natexlab{a}})}]{Higgs:1964ia}%
  \BibitemOpen
  \bibfield  {author} {\bibinfo {author} {\bibfnamefont {P.~W.}\ \bibnamefont
  {Higgs}},\ }\href {\doibase 10.1016/0031-9163(64)91136-9} {\bibfield
  {journal} {\bibinfo  {journal} {Phys.\ Lett.}\ }\textbf {\bibinfo {volume}
  {12}},\ \bibinfo {pages} {132} (\bibinfo {year}
  {1964}{\natexlab{a}})}\BibitemShut {NoStop}%
\bibitem [{\citenamefont {Higgs}(1964{\natexlab{b}})}]{Higgs:1964pj}%
  \BibitemOpen
  \bibfield  {author} {\bibinfo {author} {\bibfnamefont {P.~W.}\ \bibnamefont
  {Higgs}},\ }\href {\doibase 10.1103/PhysRevLett.13.508} {\bibfield  {journal}
  {\bibinfo  {journal} {Phys.\ Rev.\ Lett.}\ }\textbf {\bibinfo {volume}
  {13}},\ \bibinfo {pages} {508} (\bibinfo {year}
  {1964}{\natexlab{b}})}\BibitemShut {NoStop}%
\bibitem [{\citenamefont {Guralnik}\ \emph {et~al.}(1964)\citenamefont
  {Guralnik}, \citenamefont {Hagen},\ and\ \citenamefont
  {Kibble}}]{Guralnik:1964eu}%
  \BibitemOpen
  \bibfield  {author} {\bibinfo {author} {\bibfnamefont {G.}~\bibnamefont
  {Guralnik}}, \bibinfo {author} {\bibfnamefont {C.}~\bibnamefont {Hagen}}, \
  and\ \bibinfo {author} {\bibfnamefont {T.}~\bibnamefont {Kibble}},\ }\href
  {\doibase 10.1103/PhysRevLett.13.585} {\bibfield  {journal} {\bibinfo
  {journal} {Phys.\ Rev.\ Lett.}\ }\textbf {\bibinfo {volume} {13}},\ \bibinfo
  {pages} {585} (\bibinfo {year} {1964})}\BibitemShut {NoStop}%
\bibitem [{\citenamefont {Higgs}(1966)}]{Higgs:1966ev}%
  \BibitemOpen
  \bibfield  {author} {\bibinfo {author} {\bibfnamefont {P.~W.}\ \bibnamefont
  {Higgs}},\ }\href {\doibase 10.1103/PhysRev.145.1156} {\bibfield  {journal}
  {\bibinfo  {journal} {Phys.\ Rev.}\ }\textbf {\bibinfo {volume} {145}},\
  \bibinfo {pages} {1156} (\bibinfo {year} {1966})}\BibitemShut {NoStop}%
\bibitem [{\citenamefont {Kibble}(1967)}]{Kibble:1967sv}%
  \BibitemOpen
  \bibfield  {author} {\bibinfo {author} {\bibfnamefont {T.}~\bibnamefont
  {Kibble}},\ }\href {\doibase 10.1103/PhysRev.155.1554} {\bibfield  {journal}
  {\bibinfo  {journal} {Phys.\ Rev.}\ }\textbf {\bibinfo {volume} {155}},\
  \bibinfo {pages} {1554} (\bibinfo {year} {1967})}\BibitemShut {NoStop}%
\bibitem [{\citenamefont {{ATLAS
  Collaboration}}(2014{\natexlab{a}})}]{Aad:2014aba}%
  \BibitemOpen
  \bibfield  {author} {\bibinfo {author} {\bibnamefont {{ATLAS
  Collaboration}}},\ }\href@noop {} {\bibfield  {journal} {\bibinfo  {journal}
  {CERN-PH-EP-2014-122}\ } (\bibinfo {year} {2014}{\natexlab{a}})},\ \bibinfo
  {note} {submitted to Phys.\ Rev.\ D.},\ \Eprint
  {http://arxiv.org/abs/1406.3827} {arXiv:1406.3827 [hep-ex]} \BibitemShut
  {NoStop}%
\bibitem [{\citenamefont {{CMS Collaboration}}(2014)}]{ref:cmsmass}%
  \BibitemOpen
  \bibfield  {author} {\bibinfo {author} {\bibnamefont {{CMS Collaboration}}},\
  }\href {http://cdsweb.cern.ch/record/1728249} {\bibfield  {journal} {\bibinfo
   {journal} {CMS-PAS-HIG-14-009}\ } (\bibinfo {year} {2014})},\ \bibinfo
  {note}
  {\href{http://cdsweb.cern.ch/record/1728249}{http://cdsweb.cern.ch/record/1728249}}\BibitemShut
  {NoStop}%
\bibitem [{\citenamefont {{ATLAS
  Collaboration}}(2013{\natexlab{a}})}]{atlas-spin}%
  \BibitemOpen
  \bibfield  {author} {\bibinfo {author} {\bibnamefont {{ATLAS
  Collaboration}}},\ }\href {\doibase 10.1016/j.physletb.2013.08.026}
  {\bibfield  {journal} {\bibinfo  {journal} {Phys.\ Lett.\ B}\ }\textbf
  {\bibinfo {volume} {726}},\ \bibinfo {pages} {120} (\bibinfo {year}
  {2013}{\natexlab{a}})},\ \Eprint {http://arxiv.org/abs/1307.1432}
  {arXiv:1307.1432 [hep-ex]} \BibitemShut {NoStop}%
\bibitem [{\citenamefont {{CMS Collaboration}}(2013)}]{cms-spin}%
  \BibitemOpen
  \bibfield  {author} {\bibinfo {author} {\bibnamefont {{CMS Collaboration}}},\
  }\href {\doibase 10.1103/PhysRevLett.110.081803} {\bibfield  {journal}
  {\bibinfo  {journal} {Phys.\ Rev.\ Lett.}\ }\textbf {\bibinfo {volume}
  {110}},\ \bibinfo {pages} {081803} (\bibinfo {year} {2013})},\ \Eprint
  {http://arxiv.org/abs/1212.6639} {arXiv:1212.6639 [hep-ex]} \BibitemShut
  {NoStop}%
\bibitem [{\citenamefont {{ATLAS
  Collaboration}}(2013{\natexlab{b}})}]{atlas-couplings-diboson}%
  \BibitemOpen
  \bibfield  {author} {\bibinfo {author} {\bibnamefont {{ATLAS
  Collaboration}}},\ }\href {\doibase 10.1016/j.physletb.2013.08.010}
  {\bibfield  {journal} {\bibinfo  {journal} {Phys.\ Lett.\ B}\ }\textbf
  {\bibinfo {volume} {726}},\ \bibinfo {pages} {88} (\bibinfo {year}
  {2013}{\natexlab{b}})},\ \Eprint {http://arxiv.org/abs/1307.1427}
  {arXiv:1307.1427 [hep-ex]} \BibitemShut {NoStop}%
\bibitem [{\citenamefont {{CMS
  Collaboration}}(2014{\natexlab{b}})}]{newCMSHgg}%                                                                                                                
  \BibitemOpen
  \bibfield  {author} {\bibinfo {author} {\bibnamefont {{CMS Collaboration}}},\
  }\href@noop {} {\bibfield  {journal} {\bibinfo  {journal} {CERN-PH-EP-2014-117}\ }
  (\bibinfo {year} {2014}{\natexlab{b}})},\ \bibinfo {note} {submitted to Eur.\
  Phys.\ J.\ C},\ \Eprint {http://arxiv.org/abs/1407.0558} {arXiv:1407.0558
  [hep-ex]} \BibitemShut {NoStop}%                 
\bibitem [{\citenamefont {{ATLAS
  Collaboration}}(2014{\natexlab{b}})}]{calibration_paper}%
  \BibitemOpen
  \bibfield  {author} {\bibinfo {author} {\bibnamefont {{ATLAS
  Collaboration}}},\ }\href@noop {} {\bibfield  {journal} {\bibinfo  {journal}
  {CERN-PH-EP-2014-153}\ } (\bibinfo {year} {2014}{\natexlab{b}})},\ \bibinfo {note}
  {submitted to Eur.\ Phys.\ J.\ C},\ \Eprint {http://arxiv.org/abs/1407.5063}
  {arXiv:1407.5063 [hep-ex]} \BibitemShut {NoStop}%
\bibitem [{\citenamefont {{ATLAS Collaboration}}(2008)}]{ATLAS_detector}%
  \BibitemOpen
  \bibfield  {author} {\bibinfo {author} {\bibnamefont {{ATLAS
  Collaboration}}},\ }\href {\doibase 10.1088/1748-0221/3/08/S08003} {\bibfield
   {journal} {\bibinfo  {journal} {JINST}\ }\textbf {\bibinfo {volume} {3}},\
  \bibinfo {pages} {S08003} (\bibinfo {year} {2008})}\BibitemShut {NoStop}%
\bibitem [{\citenamefont {{ATLAS
  Collaboration}}(2012{\natexlab{b}})}]{atlas-trigger}%
  \BibitemOpen
  \bibfield  {author} {\bibinfo {author} {\bibnamefont {{ATLAS
  Collaboration}}},\ }\href {http://cdsweb.cern.ch/record/1450089} {\bibfield
  {journal} {\bibinfo  {journal} {ATLAS-CONF-2012-048}\ } (\bibinfo {year}
  {2012}{\natexlab{b}})},\ \bibinfo {note}
  {\href{http://cdsweb.cern.ch/record/1450089}{http://cdsweb.cern.ch/record/1450089}}\BibitemShut
  {NoStop}%
\bibitem [{\citenamefont {{ATLAS
  Collaboration}}(2012{\natexlab{c}})}]{atlasmctunes}%
  \BibitemOpen
  \bibfield  {author} {\bibinfo {author} {\bibnamefont {{ATLAS
  Collaboration}}},\ }\href {http://cdsweb.cern.ch/record/1474107} {\bibfield
  {journal} {\bibinfo  {journal} {ATL-PHYS-PUB-2012-003}\ } (\bibinfo {year}
  {2012}{\natexlab{c}})},\ \bibinfo {note}
  {\href{http://cdsweb.cern.ch/record/1474107}{http://cdsweb.cern.ch/record/1474107}}\BibitemShut
  {NoStop}%
\bibitem [{\citenamefont {{T. Sj\"{o}strand, S. Mrenna, P.
  Skands}}(2008)}]{pythia8}%
  \BibitemOpen
  \bibfield  {author} {\bibinfo {author} {\bibnamefont {{T. Sj\"{o}strand, S.
  Mrenna, P. Skands}}},\ }\href {\doibase 10.1016/j.cpc.2008.01.036} {\bibfield
   {journal} {\bibinfo  {journal} {Comput.\ Phys.\ Commun.}\ }\textbf {\bibinfo
  {volume} {178}},\ \bibinfo {pages} {852} (\bibinfo {year} {2008})},\ \Eprint
  {http://arxiv.org/abs/0710.3820} {arXiv:0710.3820 [hep-ph]} \BibitemShut
  {NoStop}%
\bibitem [{\citenamefont {Heinemeyer}\ \emph {et~al.}(2013)\citenamefont
  {Heinemeyer} \emph {et~al.}}]{lhcxs}%
  \BibitemOpen
  \bibfield  {author} {\bibinfo {author} {\bibfnamefont {S.}~\bibnamefont
  {Heinemeyer}} \emph {et~al.} (\bibinfo {collaboration} {LHC Higgs Cross
  Section Working Group}),\ }\href {\doibase 10.5170/CERN-2013-004} {\
  (\bibinfo {year} {2013}),\ 10.5170/CERN-2013-004},\ \Eprint
  {http://arxiv.org/abs/1307.1347} {arXiv:1307.1347 [hep-ph]} \BibitemShut
  {NoStop}%
\bibitem [{\citenamefont {Nason}(2004)}]{Nason:2004rx}%
  \BibitemOpen
  \bibfield  {author} {\bibinfo {author} {\bibfnamefont {P.}~\bibnamefont
  {Nason}},\ }\href {\doibase 10.1088/1126-6708/2004/11/040} {\bibfield
  {journal} {\bibinfo  {journal} {J.~High Energy Phys.}\ ,\ \bibinfo {pages}
  {040}} (\bibinfo {year} {2004})},\ \Eprint
  {http://arxiv.org/abs/hep-ph/0409146} {arXiv:hep-ph/0409146 [hep-ph]}
  \BibitemShut {NoStop}%
\bibitem [{\citenamefont {Frixione}\ \emph {et~al.}(2007)\citenamefont
  {Frixione}, \citenamefont {Nason},\ and\ \citenamefont
  {Oleari}}]{Frixione:2007vw}%
  \BibitemOpen
  \bibfield  {author} {\bibinfo {author} {\bibfnamefont {S.}~\bibnamefont
  {Frixione}}, \bibinfo {author} {\bibfnamefont {P.}~\bibnamefont {Nason}}, \
  and\ \bibinfo {author} {\bibfnamefont {C.}~\bibnamefont {Oleari}},\ }\href
  {\doibase 10.1088/1126-6708/2007/11/070} {\bibfield  {journal} {\bibinfo
  {journal} {J.~High Energy Phys.}\ ,\ \bibinfo {pages} {070}} (\bibinfo {year}
  {2007})},\ \Eprint {http://arxiv.org/abs/0709.2092} {arXiv:0709.2092
  [hep-ph]} \BibitemShut {NoStop}%
\bibitem [{\citenamefont {Alioli}\ \emph {et~al.}(2010)\citenamefont {Alioli},
  \citenamefont {Nason}, \citenamefont {Oleari},\ and\ \citenamefont
  {Re}}]{powhegbox}%
  \BibitemOpen
  \bibfield  {author} {\bibinfo {author} {\bibfnamefont {S.}~\bibnamefont
  {Alioli}}, \bibinfo {author} {\bibfnamefont {P.}~\bibnamefont {Nason}},
  \bibinfo {author} {\bibfnamefont {C.}~\bibnamefont {Oleari}}, \ and\ \bibinfo
  {author} {\bibfnamefont {E.}~\bibnamefont {Re}},\ }\href {\doibase
  10.1007/JHEP06(2010)043} {\bibfield  {journal} {\bibinfo  {journal} {J.~High
  Energy Phys.}\ ,\ \bibinfo {pages} {043}} (\bibinfo {year} {2010})},\ \Eprint
  {http://arxiv.org/abs/1002.2581} {arXiv:1002.2581 [hep-ph]} \BibitemShut
  {NoStop}%
\bibitem [{\citenamefont {{S. Alioli, P. Nason, C. Oleari and E.
  Re}}(2009)}]{powheg_ggF}%
  \BibitemOpen
  \bibfield  {author} {\bibinfo {author} {\bibnamefont {{S. Alioli, P. Nason,
  C. Oleari and E. Re}}},\ }\href {\doibase 10.1088/1126-6708/2009/04/002}
  {\bibfield  {journal} {\bibinfo  {journal} {J.~High Energy Phys.}\ ,\
  \bibinfo {pages} {002}} (\bibinfo {year} {2009})},\ \Eprint
  {http://arxiv.org/abs/0812.0578} {arXiv:0812.0578 [hep-ph]} \BibitemShut
  {NoStop}%
\bibitem [{\citenamefont {Bagnaschi}\ \emph {et~al.}(2012)\citenamefont
  {Bagnaschi}, \citenamefont {Degrassi}, \citenamefont {Slavich},\ and\
  \citenamefont {Vicini}}]{Bagnaschi:2011tu}%
  \BibitemOpen
  \bibfield  {author} {\bibinfo {author} {\bibfnamefont {E.}~\bibnamefont
  {Bagnaschi}}, \bibinfo {author} {\bibfnamefont {G.}~\bibnamefont {Degrassi}},
  \bibinfo {author} {\bibfnamefont {P.}~\bibnamefont {Slavich}}, \ and\
  \bibinfo {author} {\bibfnamefont {A.}~\bibnamefont {Vicini}},\ }\href
  {\doibase 10.1007/JHEP02(2012)088} {\bibfield  {journal} {\bibinfo  {journal}
  {J.~High Energy Phys.}\ ,\ \bibinfo {pages} {088}} (\bibinfo {year}
  {2012})},\ \Eprint {http://arxiv.org/abs/1111.2854} {arXiv:1111.2854
  [hep-ph]} \BibitemShut {NoStop}%
\bibitem [{\citenamefont {Djouadi}\ \emph {et~al.}(1991)\citenamefont
  {Djouadi}, \citenamefont {Spira},\ and\ \citenamefont
  {Zerwas}}]{Djouadi:1991tka}%
  \BibitemOpen
  \bibfield  {author} {\bibinfo {author} {\bibfnamefont {A.}~\bibnamefont
  {Djouadi}}, \bibinfo {author} {\bibfnamefont {M.}~\bibnamefont {Spira}}, \
  and\ \bibinfo {author} {\bibfnamefont {P.}~\bibnamefont {Zerwas}},\ }\href
  {\doibase 10.1016/0370-2693(91)90375-Z} {\bibfield  {journal} {\bibinfo
  {journal} {Phys.\ Lett.\ B}\ }\textbf {\bibinfo {volume} {264}},\ \bibinfo
  {pages} {440} (\bibinfo {year} {1991})}\BibitemShut {NoStop}%
\bibitem [{\citenamefont {Dawson}(1991)}]{Dawson:1990zj}%
  \BibitemOpen
  \bibfield  {author} {\bibinfo {author} {\bibfnamefont {S.}~\bibnamefont
  {Dawson}},\ }\href {\doibase 10.1016/0550-3213(91)90061-2} {\bibfield
  {journal} {\bibinfo  {journal} {Nucl.\ Phys.\ B}\ }\textbf {\bibinfo {volume}
  {359}},\ \bibinfo {pages} {283} (\bibinfo {year} {1991})}\BibitemShut
  {NoStop}%
\bibitem [{\citenamefont {Spira}\ \emph {et~al.}(1995)\citenamefont {Spira},
  \citenamefont {Djouadi}, \citenamefont {Graudenz},\ and\ \citenamefont
  {Zerwas}}]{Spira:1995rr}%
  \BibitemOpen
  \bibfield  {author} {\bibinfo {author} {\bibfnamefont {M.}~\bibnamefont
  {Spira}}, \bibinfo {author} {\bibfnamefont {A.}~\bibnamefont {Djouadi}},
  \bibinfo {author} {\bibfnamefont {D.}~\bibnamefont {Graudenz}}, \ and\
  \bibinfo {author} {\bibfnamefont {P.}~\bibnamefont {Zerwas}},\ }\href
  {\doibase 10.1016/0550-3213(95)00379-7} {\bibfield  {journal} {\bibinfo
  {journal} {Nucl.\ Phys.\ B}\ }\textbf {\bibinfo {volume} {453}},\ \bibinfo
  {pages} {17} (\bibinfo {year} {1995})},\ \Eprint
  {http://arxiv.org/abs/hep-ph/9504378} {arXiv:hep-ph/9504378 [hep-ph]}
  \BibitemShut {NoStop}%
\bibitem [{\citenamefont {Harlander}\ and\ \citenamefont
  {Kilgore}(2002)}]{Harlander:2002wh}%
  \BibitemOpen
  \bibfield  {author} {\bibinfo {author} {\bibfnamefont {R.~V.}\ \bibnamefont
  {Harlander}}\ and\ \bibinfo {author} {\bibfnamefont {W.~B.}\ \bibnamefont
  {Kilgore}},\ }\href {\doibase 10.1103/PhysRevLett.88.201801} {\bibfield
  {journal} {\bibinfo  {journal} {Phys.\ Rev.\ Lett.}\ }\textbf {\bibinfo
  {volume} {88}},\ \bibinfo {pages} {201801} (\bibinfo {year} {2002})},\
  \Eprint {http://arxiv.org/abs/hep-ph/0201206} {arXiv:hep-ph/0201206 [hep-ph]}
  \BibitemShut {NoStop}%
\bibitem [{\citenamefont {Anastasiou}\ and\ \citenamefont
  {Melnikov}(2002)}]{Anastasiou:2002yz}%
  \BibitemOpen
  \bibfield  {author} {\bibinfo {author} {\bibfnamefont {C.}~\bibnamefont
  {Anastasiou}}\ and\ \bibinfo {author} {\bibfnamefont {K.}~\bibnamefont
  {Melnikov}},\ }\href {\doibase 10.1016/S0550-3213(02)00837-4} {\bibfield
  {journal} {\bibinfo  {journal} {Nucl.\ Phys.\ B}\ }\textbf {\bibinfo {volume}
  {646}},\ \bibinfo {pages} {220} (\bibinfo {year} {2002})},\ \Eprint
  {http://arxiv.org/abs/hep-ph/0207004} {arXiv:hep-ph/0207004 [hep-ph]}
  \BibitemShut {NoStop}%
\bibitem [{\citenamefont {Ravindran}\ \emph {et~al.}(2003)\citenamefont
  {Ravindran}, \citenamefont {Smith},\ and\ \citenamefont {van
  Neerven}}]{Ravindran:2003um}%
  \BibitemOpen
  \bibfield  {author} {\bibinfo {author} {\bibfnamefont {V.}~\bibnamefont
  {Ravindran}}, \bibinfo {author} {\bibfnamefont {J.}~\bibnamefont {Smith}}, \
  and\ \bibinfo {author} {\bibfnamefont {W.~L.}\ \bibnamefont {van Neerven}},\
  }\href {\doibase 10.1016/S0550-3213(03)00457-7} {\bibfield  {journal}
  {\bibinfo  {journal} {Nucl.\ Phys.}\ }\textbf {\bibinfo {volume} {665}},\
  \bibinfo {pages} {325} (\bibinfo {year} {2003})},\ \Eprint
  {http://arxiv.org/abs/hep-ph/0302135} {arXiv:hep-ph/0302135 [hep-ph]}
  \BibitemShut {NoStop}%
\bibitem [{\citenamefont {Aglietti}\ \emph {et~al.}(2004)\citenamefont
  {Aglietti}, \citenamefont {Bonciani}, \citenamefont {Degrassi},\ and\
  \citenamefont {Vicini}}]{Aglietti:2004nj}%
  \BibitemOpen
  \bibfield  {author} {\bibinfo {author} {\bibfnamefont {U.}~\bibnamefont
  {Aglietti}}, \bibinfo {author} {\bibfnamefont {R.}~\bibnamefont {Bonciani}},
  \bibinfo {author} {\bibfnamefont {G.}~\bibnamefont {Degrassi}}, \ and\
  \bibinfo {author} {\bibfnamefont {A.}~\bibnamefont {Vicini}},\ }\href
  {\doibase 10.1016/j.physletb.2004.06.063} {\bibfield  {journal} {\bibinfo
  {journal} {Phys.\ Lett.\ B}\ }\textbf {\bibinfo {volume} {595}},\ \bibinfo
  {pages} {432} (\bibinfo {year} {2004})},\ \Eprint
  {http://arxiv.org/abs/hep-ph/0404071} {arXiv:hep-ph/0404071 [hep-ph]}
  \BibitemShut {NoStop}%
\bibitem [{\citenamefont {Actis}\ \emph {et~al.}(2008)\citenamefont {Actis},
  \citenamefont {Passarino}, \citenamefont {Sturm},\ and\ \citenamefont
  {Uccirati}}]{Actis:2008ug}%
  \BibitemOpen
  \bibfield  {author} {\bibinfo {author} {\bibfnamefont {S.}~\bibnamefont
  {Actis}}, \bibinfo {author} {\bibfnamefont {G.}~\bibnamefont {Passarino}},
  \bibinfo {author} {\bibfnamefont {C.}~\bibnamefont {Sturm}}, \ and\ \bibinfo
  {author} {\bibfnamefont {S.}~\bibnamefont {Uccirati}},\ }\href {\doibase
  10.1016/j.physletb.2008.10.018} {\bibfield  {journal} {\bibinfo  {journal}
  {Phys.\ Lett.\ B}\ }\textbf {\bibinfo {volume} {670}},\ \bibinfo {pages} {12}
  (\bibinfo {year} {2008})},\ \Eprint {http://arxiv.org/abs/0809.1301}
  {arXiv:0809.1301 [hep-ph]} \BibitemShut {NoStop}%
\bibitem [{\citenamefont {Dixon}\ \emph {et~al.}()\citenamefont {Dixon},
  \citenamefont {Li},\ and\ \citenamefont {Hoeche}}]{Dixon_pc}%
  \BibitemOpen
  \bibfield  {author} {\bibinfo {author} {\bibfnamefont {L.}~\bibnamefont
  {Dixon}}, \bibinfo {author} {\bibfnamefont {Y.}~\bibnamefont {Li}}, \ and\
  \bibinfo {author} {\bibfnamefont {S.}~\bibnamefont {Hoeche}},\ }\href@noop {}
  {\ }\bibinfo {note} {{Private Communication}}\BibitemShut {NoStop}%
\bibitem [{\citenamefont {Dixon}\ and\ \citenamefont
  {Siu}(2003)}]{dixon_interference}%
  \BibitemOpen
  \bibfield  {author} {\bibinfo {author} {\bibfnamefont {L.~J.}\ \bibnamefont
  {Dixon}}\ and\ \bibinfo {author} {\bibfnamefont {M.~S.}\ \bibnamefont
  {Siu}},\ }\href {\doibase 10.1103/PhysRevLett.90.252001} {\bibfield
  {journal} {\bibinfo  {journal} {Phys.\ Rev.\ Lett.}\ }\textbf {\bibinfo
  {volume} {90}},\ \bibinfo {pages} {252001} (\bibinfo {year} {2003})},\
  \Eprint {http://arxiv.org/abs/hep-ph/0302233} {arXiv:hep-ph/0302233 [hep-ph]}
  \BibitemShut {NoStop}%
\bibitem [{\citenamefont {Dixon}\ and\ \citenamefont
  {Li}(2013)}]{Dixon:2013haa}%
  \BibitemOpen
  \bibfield  {author} {\bibinfo {author} {\bibfnamefont {L.~J.}\ \bibnamefont
  {Dixon}}\ and\ \bibinfo {author} {\bibfnamefont {Y.}~\bibnamefont {Li}},\
  }\href {\doibase 10.1103/PhysRevLett.111.111802} {\bibfield  {journal}
  {\bibinfo  {journal} {Phys.\ Rev.\ Lett.}\ }\textbf {\bibinfo {volume}
  {111}},\ \bibinfo {pages} {111802} (\bibinfo {year} {2013})},\ \Eprint
  {http://arxiv.org/abs/1305.3854} {arXiv:1305.3854 [hep-ph]} \BibitemShut
  {NoStop}%
\bibitem [{\citenamefont {{P. Nason and C. Oleari}}(2010)}]{powheg_VBF}%
  \BibitemOpen
  \bibfield  {author} {\bibinfo {author} {\bibnamefont {{P. Nason and C.
  Oleari}}},\ }\href {\doibase 10.1007/JHEP02(2010)037} {\bibfield  {journal}
  {\bibinfo  {journal} {J.~High Energy Phys.}\ ,\ \bibinfo {pages} {037}}
  (\bibinfo {year} {2010})},\ \Eprint {http://arxiv.org/abs/0911.5299}
  {arXiv:0911.5299 [hep-ph]} \BibitemShut {NoStop}%
\bibitem [{\citenamefont {Ciccolini}\ \emph
  {et~al.}(2007{\natexlab{a}})\citenamefont {Ciccolini}, \citenamefont
  {Denner},\ and\ \citenamefont {Dittmaier}}]{Ciccolini:2007jr}%
  \BibitemOpen
  \bibfield  {author} {\bibinfo {author} {\bibfnamefont {M.}~\bibnamefont
  {Ciccolini}}, \bibinfo {author} {\bibfnamefont {A.}~\bibnamefont {Denner}}, \
  and\ \bibinfo {author} {\bibfnamefont {S.}~\bibnamefont {Dittmaier}},\ }\href
  {\doibase 10.1103/PhysRevLett.99.161803} {\bibfield  {journal} {\bibinfo
  {journal} {Phys.\ Rev.\ Lett.}\ }\textbf {\bibinfo {volume} {99}},\ \bibinfo
  {pages} {161803} (\bibinfo {year} {2007}{\natexlab{a}})},\ \Eprint
  {http://arxiv.org/abs/0707.0381} {arXiv:0707.0381 [hep-ph]} \BibitemShut
  {NoStop}%
\bibitem [{\citenamefont {Ciccolini}\ \emph
  {et~al.}(2008{\natexlab{a}})\citenamefont {Ciccolini}, \citenamefont
  {Denner},\ and\ \citenamefont {Dittmaier}}]{Ciccolini:2007ec}%
  \BibitemOpen
  \bibfield  {author} {\bibinfo {author} {\bibfnamefont {M.}~\bibnamefont
  {Ciccolini}}, \bibinfo {author} {\bibfnamefont {A.}~\bibnamefont {Denner}}, \
  and\ \bibinfo {author} {\bibfnamefont {S.}~\bibnamefont {Dittmaier}},\ }\href
  {\doibase 10.1103/PhysRevD.77.013002} {\bibfield  {journal} {\bibinfo
  {journal} {Phys.\ Rev.\ D}\ }\textbf {\bibinfo {volume} {77}},\ \bibinfo
  {pages} {013002} (\bibinfo {year} {2008}{\natexlab{a}})},\ \Eprint
  {http://arxiv.org/abs/0710.4749} {arXiv:0710.4749 [hep-ph]} \BibitemShut
  {NoStop}%
\bibitem [{\citenamefont {Arnold}\ \emph {et~al.}(2009)\citenamefont {Arnold}
  \emph {et~al.}}]{Arnold:2008rz}%
  \BibitemOpen
  \bibfield  {author} {\bibinfo {author} {\bibfnamefont {K.}~\bibnamefont
  {Arnold}} \emph {et~al.},\ }\href {\doibase 10.1016/j.cpc.2009.03.006}
  {\bibfield  {journal} {\bibinfo  {journal} {Comput.\ Phys.\ Commun.}\
  }\textbf {\bibinfo {volume} {180}},\ \bibinfo {pages} {1661} (\bibinfo {year}
  {2009})},\ \Eprint {http://arxiv.org/abs/0811.4559} {arXiv:0811.4559
  [hep-ph]} \BibitemShut {NoStop}%
\bibitem [{\citenamefont {Bolzoni}\ \emph {et~al.}(2010)\citenamefont
  {Bolzoni}, \citenamefont {Maltoni}, \citenamefont {Moch},\ and\ \citenamefont
  {Zaro}}]{Bolzoni:2010xr}%
  \BibitemOpen
  \bibfield  {author} {\bibinfo {author} {\bibfnamefont {P.}~\bibnamefont
  {Bolzoni}}, \bibinfo {author} {\bibfnamefont {F.}~\bibnamefont {Maltoni}},
  \bibinfo {author} {\bibfnamefont {S.-O.}\ \bibnamefont {Moch}}, \ and\
  \bibinfo {author} {\bibfnamefont {M.}~\bibnamefont {Zaro}},\ }\href {\doibase
  10.1103/PhysRevLett.105.011801} {\bibfield  {journal} {\bibinfo  {journal}
  {Phys.\ Rev.\ Lett.}\ }\textbf {\bibinfo {volume} {105}},\ \bibinfo {pages}
  {011801} (\bibinfo {year} {2010})},\ \Eprint {http://arxiv.org/abs/1003.4451}
  {arXiv:1003.4451 [hep-ph]} \BibitemShut {NoStop}%
\bibitem [{\citenamefont {Brein}\ \emph {et~al.}(2004)\citenamefont {Brein},
  \citenamefont {Djouadi},\ and\ \citenamefont {Harlander}}]{Brein:2003wg}%
  \BibitemOpen
  \bibfield  {author} {\bibinfo {author} {\bibfnamefont {O.}~\bibnamefont
  {Brein}}, \bibinfo {author} {\bibfnamefont {A.}~\bibnamefont {Djouadi}}, \
  and\ \bibinfo {author} {\bibfnamefont {R.}~\bibnamefont {Harlander}},\ }\href
  {\doibase 10.1016/j.physletb.2003.10.112} {\bibfield  {journal} {\bibinfo
  {journal} {Phys.\ Lett.\ B}\ }\textbf {\bibinfo {volume} {579}},\ \bibinfo
  {pages} {149} (\bibinfo {year} {2004})},\ \Eprint
  {http://arxiv.org/abs/hep-ph/0307206} {arXiv:hep-ph/0307206 [hep-ph]}
  \BibitemShut {NoStop}%
\bibitem [{\citenamefont {Ciccolini}\ \emph {et~al.}(2003)\citenamefont
  {Ciccolini}, \citenamefont {Dittmaier},\ and\ \citenamefont
  {{Kr\"amer}}}]{Ciccolini:2003jy}%
  \BibitemOpen
  \bibfield  {author} {\bibinfo {author} {\bibfnamefont {M.}~\bibnamefont
  {Ciccolini}}, \bibinfo {author} {\bibfnamefont {S.}~\bibnamefont
  {Dittmaier}}, \ and\ \bibinfo {author} {\bibfnamefont {M.}~\bibnamefont
  {{Kr\"amer}}},\ }\href {\doibase 10.1103/PhysRevD.68.073003} {\bibfield
  {journal} {\bibinfo  {journal} {Phys.\ Rev.\ D}\ }\textbf {\bibinfo {volume}
  {68}},\ \bibinfo {pages} {073003} (\bibinfo {year} {2003})},\ \Eprint
  {http://arxiv.org/abs/hep-ph/0306234} {arXiv:hep-ph/0306234 [hep-ph]}
  \BibitemShut {NoStop}%
\bibitem [{\citenamefont {Bevilacqua}\ \emph {et~al.}(2013)\citenamefont
  {Bevilacqua} \emph {et~al.}}]{HelacNLO}%
  \BibitemOpen
  \bibfield  {author} {\bibinfo {author} {\bibfnamefont {G.}~\bibnamefont
  {Bevilacqua}} \emph {et~al.},\ }\href {\doibase 10.1016/j.cpc.2012.10.033}
  {\bibfield  {journal} {\bibinfo  {journal} {Comput.Phys.Commun.}\ }\textbf
  {\bibinfo {volume} {184}},\ \bibinfo {pages} {986} (\bibinfo {year}
  {2013})},\ \Eprint {http://arxiv.org/abs/1110.1499} {arXiv:1110.1499
  [hep-ph]} \BibitemShut {NoStop}%
\bibitem [{\citenamefont {Beenakker}\ \emph {et~al.}(2001)\citenamefont
  {Beenakker} \emph {et~al.}}]{Beenakker:2001rj}%
  \BibitemOpen
  \bibfield  {author} {\bibinfo {author} {\bibfnamefont {W.}~\bibnamefont
  {Beenakker}} \emph {et~al.},\ }\href {\doibase 10.1103/PhysRevLett.87.201805}
  {\bibfield  {journal} {\bibinfo  {journal} {Phys.\ Rev.\ Lett.}\ }\textbf
  {\bibinfo {volume} {87}},\ \bibinfo {pages} {201805} (\bibinfo {year}
  {2001})},\ \Eprint {http://arxiv.org/abs/hep-ph/0107081}
  {arXiv:hep-ph/0107081 [hep-ph]} \BibitemShut {NoStop}%
\bibitem [{\citenamefont {Beenakker}\ \emph {et~al.}(2003)\citenamefont
  {Beenakker} \emph {et~al.}}]{Beenakker:2002nc}%
  \BibitemOpen
  \bibfield  {author} {\bibinfo {author} {\bibfnamefont {W.}~\bibnamefont
  {Beenakker}} \emph {et~al.},\ }\href {\doibase 10.1016/S0550-3213(03)00044-0}
  {\bibfield  {journal} {\bibinfo  {journal} {Nucl.\ Phys.\ B}\ }\textbf
  {\bibinfo {volume} {653}},\ \bibinfo {pages} {151} (\bibinfo {year}
  {2003})},\ \Eprint {http://arxiv.org/abs/hep-ph/0211352}
  {arXiv:hep-ph/0211352 [hep-ph]} \BibitemShut {NoStop}%
\bibitem [{\citenamefont {Dawson}\ \emph
  {et~al.}(2003{\natexlab{a}})\citenamefont {Dawson}, \citenamefont {Orr},
  \citenamefont {Reina},\ and\ \citenamefont {Wackeroth}}]{Dawson:2002tg}%
  \BibitemOpen
  \bibfield  {author} {\bibinfo {author} {\bibfnamefont {S.}~\bibnamefont
  {Dawson}}, \bibinfo {author} {\bibfnamefont {L.}~\bibnamefont {Orr}},
  \bibinfo {author} {\bibfnamefont {L.}~\bibnamefont {Reina}}, \ and\ \bibinfo
  {author} {\bibfnamefont {D.}~\bibnamefont {Wackeroth}},\ }\href {\doibase
  10.1103/PhysRevD.67.071503} {\bibfield  {journal} {\bibinfo  {journal}
  {Phys.\ Rev.\ D}\ }\textbf {\bibinfo {volume} {67}},\ \bibinfo {pages}
  {071503} (\bibinfo {year} {2003}{\natexlab{a}})},\ \Eprint
  {http://arxiv.org/abs/hep-ph/0211438} {arXiv:hep-ph/0211438 [hep-ph]}
  \BibitemShut {NoStop}%
\bibitem [{\citenamefont {Dawson}\ \emph
  {et~al.}(2003{\natexlab{b}})\citenamefont {Dawson}, \citenamefont {Jackson},
  \citenamefont {Orr}, \citenamefont {Reina},\ and\ \citenamefont
  {Wackeroth}}]{Dawson:2003zu}%
  \BibitemOpen
  \bibfield  {author} {\bibinfo {author} {\bibfnamefont {S.}~\bibnamefont
  {Dawson}}, \bibinfo {author} {\bibfnamefont {C.}~\bibnamefont {Jackson}},
  \bibinfo {author} {\bibfnamefont {L.}~\bibnamefont {Orr}}, \bibinfo {author}
  {\bibfnamefont {L.}~\bibnamefont {Reina}}, \ and\ \bibinfo {author}
  {\bibfnamefont {D.}~\bibnamefont {Wackeroth}},\ }\href {\doibase
  10.1103/PhysRevD.68.034022} {\bibfield  {journal} {\bibinfo  {journal}
  {Phys.\ Rev.\ D}\ }\textbf {\bibinfo {volume} {68}},\ \bibinfo {pages}
  {034022} (\bibinfo {year} {2003}{\natexlab{b}})},\ \Eprint
  {http://arxiv.org/abs/hep-ph/0305087} {arXiv:hep-ph/0305087 [hep-ph]}
  \BibitemShut {NoStop}%
\bibitem [{\citenamefont {Maltoni}\ and\ \citenamefont
  {Stelzer}(2003)}]{Maltoni:2002qb}%
  \BibitemOpen
  \bibfield  {author} {\bibinfo {author} {\bibfnamefont {F.}~\bibnamefont
  {Maltoni}}\ and\ \bibinfo {author} {\bibfnamefont {T.}~\bibnamefont
  {Stelzer}},\ }\href {\doibase 10.1088/1126-6708/2003/02/027} {\bibfield
  {journal} {\bibinfo  {journal} {J.~High Energy Phys.}\ }\textbf {\bibinfo
  {volume} {02}},\ \bibinfo {pages} {027} (\bibinfo {year} {2003})},\ \Eprint
  {http://arxiv.org/abs/hep-ph/0208156} {arXiv:hep-ph/0208156 [hep-ph]}
  \BibitemShut {NoStop}%
\bibitem [{\citenamefont {Maltoni}\ \emph {et~al.}(2001)\citenamefont
  {Maltoni}, \citenamefont {Paul}, \citenamefont {Stelzer},\ and\ \citenamefont
  {Willenbrock}}]{Maltoni:2001hu}%
  \BibitemOpen
  \bibfield  {author} {\bibinfo {author} {\bibfnamefont {F.}~\bibnamefont
  {Maltoni}}, \bibinfo {author} {\bibfnamefont {K.}~\bibnamefont {Paul}},
  \bibinfo {author} {\bibfnamefont {T.}~\bibnamefont {Stelzer}}, \ and\
  \bibinfo {author} {\bibfnamefont {S.}~\bibnamefont {Willenbrock}},\ }\href
  {\doibase 10.1103/PhysRevD.64.094023} {\bibfield  {journal} {\bibinfo
  {journal} {Phys.\ Rev.\ D}\ }\textbf {\bibinfo {volume} {64}},\ \bibinfo
  {pages} {094023} (\bibinfo {year} {2001})},\ \Eprint
  {http://arxiv.org/abs/hep-ph/0106293} {arXiv:hep-ph/0106293 [hep-ph]}
  \BibitemShut {NoStop}%
\bibitem [{\citenamefont {Barger}\ \emph {et~al.}(2010)\citenamefont {Barger},
  \citenamefont {McCaskey},\ and\ \citenamefont {Shaughnessy}}]{Barger:2009ky}%
  \BibitemOpen
  \bibfield  {author} {\bibinfo {author} {\bibfnamefont {V.}~\bibnamefont
  {Barger}}, \bibinfo {author} {\bibfnamefont {M.}~\bibnamefont {McCaskey}}, \
  and\ \bibinfo {author} {\bibfnamefont {G.}~\bibnamefont {Shaughnessy}},\
  }\href {\doibase 10.1103/PhysRevD.81.034020} {\bibfield  {journal} {\bibinfo
  {journal} {Phys.\ Rev.\ D}\ }\textbf {\bibinfo {volume} {81}},\ \bibinfo
  {pages} {034020} (\bibinfo {year} {2010})},\ \Eprint
  {http://arxiv.org/abs/0911.1556} {arXiv:0911.1556 [hep-ph]} \BibitemShut
  {NoStop}%
\bibitem [{\citenamefont {Farina}\ \emph {et~al.}(2013)\citenamefont {Farina},
  \citenamefont {Grojean}, \citenamefont {Maltoni}, \citenamefont {Salvioni},\
  and\ \citenamefont {Thamm}}]{Farina:2012xp}%
  \BibitemOpen
  \bibfield  {author} {\bibinfo {author} {\bibfnamefont {M.}~\bibnamefont
  {Farina}}, \bibinfo {author} {\bibfnamefont {C.}~\bibnamefont {Grojean}},
  \bibinfo {author} {\bibfnamefont {F.}~\bibnamefont {Maltoni}}, \bibinfo
  {author} {\bibfnamefont {E.}~\bibnamefont {Salvioni}}, \ and\ \bibinfo
  {author} {\bibfnamefont {A.}~\bibnamefont {Thamm}},\ }\href {\doibase
  10.1007/JHEP05(2013)022} {\bibfield  {journal} {\bibinfo  {journal} {J.~High
  Energy Phys.}\ }\textbf {\bibinfo {volume} {05}},\ \bibinfo {pages} {022}
  (\bibinfo {year} {2013})},\ \Eprint {http://arxiv.org/abs/1211.3736}
  {arXiv:1211.3736 [hep-ph]} \BibitemShut {NoStop}%
\bibitem [{\citenamefont {Biswas}\ \emph {et~al.}(2013)\citenamefont {Biswas},
  \citenamefont {Gabrielli},\ and\ \citenamefont {Mele}}]{Biswas:2012bd}%
  \BibitemOpen
  \bibfield  {author} {\bibinfo {author} {\bibfnamefont {S.}~\bibnamefont
  {Biswas}}, \bibinfo {author} {\bibfnamefont {E.}~\bibnamefont {Gabrielli}}, \
  and\ \bibinfo {author} {\bibfnamefont {B.}~\bibnamefont {Mele}},\ }\href
  {\doibase 10.1007/JHEP01(2013)088} {\bibfield  {journal} {\bibinfo  {journal}
  {J.~High Energy Phys.}\ }\textbf {\bibinfo {volume} {01}},\ \bibinfo {pages}
  {088} (\bibinfo {year} {2013})},\ \Eprint {http://arxiv.org/abs/1211.0499}
  {arXiv:1211.0499 [hep-ph]} \BibitemShut {NoStop}%
\bibitem [{\citenamefont {Agrawal}\ \emph {et~al.}(2013)\citenamefont
  {Agrawal}, \citenamefont {Mitra},\ and\ \citenamefont
  {Shivaji}}]{Agrawal:2012ga}%
  \BibitemOpen
  \bibfield  {author} {\bibinfo {author} {\bibfnamefont {P.}~\bibnamefont
  {Agrawal}}, \bibinfo {author} {\bibfnamefont {S.}~\bibnamefont {Mitra}}, \
  and\ \bibinfo {author} {\bibfnamefont {A.}~\bibnamefont {Shivaji}},\ }\href
  {\doibase 10.1007/JHEP12(2013)077} {\bibfield  {journal} {\bibinfo  {journal}
  {J.~High Energy Phys.}\ }\textbf {\bibinfo {volume} {12}},\ \bibinfo {pages}
  {077} (\bibinfo {year} {2013})},\ \Eprint {http://arxiv.org/abs/1211.4362}
  {arXiv:1211.4362 [hep-ph]} \BibitemShut {NoStop}%
\bibitem [{\citenamefont {Alwall}\ \emph {et~al.}(2014)\citenamefont {Alwall}
  \emph {et~al.}}]{Alwall:2014hca}%
  \BibitemOpen
  \bibfield  {author} {\bibinfo {author} {\bibfnamefont {J.}~\bibnamefont
  {Alwall}} \emph {et~al.},\ }\href {\doibase {10.1007/JHEP07(2014)079}}
  {\bibfield  {journal} {\bibinfo  {journal} {J.~High Energy Phys.}\ ,\
  \bibinfo {pages} {079}} (\bibinfo {year} {2014})},\ \Eprint
  {http://arxiv.org/abs/1405.0301} {arXiv:1405.0301 [hep-ph]} \BibitemShut
  {NoStop}%
\bibitem [{\citenamefont {{B\"a}hr}\ \emph {et~al.}(2008)\citenamefont
  {{B\"a}hr} \emph {et~al.}}]{Bahr:2008pv}%
  \BibitemOpen
  \bibfield  {author} {\bibinfo {author} {\bibfnamefont {M.}~\bibnamefont
  {{B\"a}hr}} \emph {et~al.},\ }\href {\doibase 10.1140/epjc/s10052-008-0798-9}
  {\bibfield  {journal} {\bibinfo  {journal} {Eur.\ Phys.\ J.\ C}\ }\textbf
  {\bibinfo {volume} {58}},\ \bibinfo {pages} {639} (\bibinfo {year} {2008})},\
  \Eprint {http://arxiv.org/abs/0803.0883} {arXiv:0803.0883 [hep-ph]}
  \BibitemShut {NoStop}%
\bibitem [{\citenamefont {Djouadi}\ \emph {et~al.}(1998)\citenamefont
  {Djouadi}, \citenamefont {Kalinowski},\ and\ \citenamefont
  {Spira}}]{Djouadi:1997yw}%
  \BibitemOpen
  \bibfield  {author} {\bibinfo {author} {\bibfnamefont {A.}~\bibnamefont
  {Djouadi}}, \bibinfo {author} {\bibfnamefont {J.}~\bibnamefont {Kalinowski}},
  \ and\ \bibinfo {author} {\bibfnamefont {M.}~\bibnamefont {Spira}},\ }\href
  {\doibase 10.1016/S0010-4655(97)00123-9} {\bibfield  {journal} {\bibinfo
  {journal} {Comput.\ Phys.\ Commun.}\ }\textbf {\bibinfo {volume} {108}},\
  \bibinfo {pages} {56} (\bibinfo {year} {1998})},\ \Eprint
  {http://arxiv.org/abs/hep-ph/9704448} {arXiv:hep-ph/9704448 [hep-ph]}
  \BibitemShut {NoStop}%
\bibitem [{\citenamefont {Actis}\ \emph {et~al.}(2009)\citenamefont {Actis},
  \citenamefont {Passarino}, \citenamefont {Sturm},\ and\ \citenamefont
  {Uccirati}}]{Actis:2008ts}%
  \BibitemOpen
  \bibfield  {author} {\bibinfo {author} {\bibfnamefont {S.}~\bibnamefont
  {Actis}}, \bibinfo {author} {\bibfnamefont {G.}~\bibnamefont {Passarino}},
  \bibinfo {author} {\bibfnamefont {C.}~\bibnamefont {Sturm}}, \ and\ \bibinfo
  {author} {\bibfnamefont {S.}~\bibnamefont {Uccirati}},\ }\href {\doibase
  10.1016/j.nuclphysb.2008.11.024} {\bibfield  {journal} {\bibinfo  {journal}
  {Nucl.\ Phys.\ B}\ }\textbf {\bibinfo {volume} {811}},\ \bibinfo {pages}
  {182} (\bibinfo {year} {2009})},\ \Eprint {http://arxiv.org/abs/0809.3667}
  {arXiv:0809.3667 [hep-ph]} \BibitemShut {NoStop}%
\bibitem [{\citenamefont {Lai}\ \emph {et~al.}(2010)\citenamefont {Lai} \emph
  {et~al.}}]{cteq10}%
  \BibitemOpen
  \bibfield  {author} {\bibinfo {author} {\bibfnamefont {H.-L.}\ \bibnamefont
  {Lai}} \emph {et~al.},\ }\href {\doibase 10.1103/PhysRevD.82.074024}
  {\bibfield  {journal} {\bibinfo  {journal} {Phys.\ Rev.\ D}\ }\textbf
  {\bibinfo {volume} {82}},\ \bibinfo {pages} {074024} (\bibinfo {year}
  {2010})},\ \Eprint {http://arxiv.org/abs/1007.2241} {arXiv:1007.2241
  [hep-ph]} \BibitemShut {NoStop}%
\bibitem [{\citenamefont {Pumplin}\ \emph {et~al.}(2002)\citenamefont {Pumplin}
  \emph {et~al.}}]{cteq6}%
  \BibitemOpen
  \bibfield  {author} {\bibinfo {author} {\bibfnamefont {J.}~\bibnamefont
  {Pumplin}} \emph {et~al.},\ }\href {\doibase 10.1088/1126-6708/2002/07/012}
  {\bibfield  {journal} {\bibinfo  {journal} {J.~High Energy Phys.}\ }\textbf
  {\bibinfo {volume} {07}},\ \bibinfo {pages} {012} (\bibinfo {year}
  {2002})}\BibitemShut {NoStop}%
\bibitem [{\citenamefont {de~Florian}\ \emph {et~al.}(2012)\citenamefont
  {de~Florian}, \citenamefont {Ferrera}, \citenamefont {Grazzini},\ and\
  \citenamefont {Tommasini}}]{hres_1}%
  \BibitemOpen
  \bibfield  {author} {\bibinfo {author} {\bibfnamefont {D.}~\bibnamefont
  {de~Florian}}, \bibinfo {author} {\bibfnamefont {G.}~\bibnamefont {Ferrera}},
  \bibinfo {author} {\bibfnamefont {M.}~\bibnamefont {Grazzini}}, \ and\
  \bibinfo {author} {\bibfnamefont {D.}~\bibnamefont {Tommasini}},\ }\href
  {\doibase 10.1007/JHEP06(2012)132} {\bibfield  {journal} {\bibinfo  {journal}
  {J.~High Energy Phys.}\ }\textbf {\bibinfo {volume} {06}},\ \bibinfo {pages}
  {132} (\bibinfo {year} {2012})},\ \Eprint {http://arxiv.org/abs/1203.6321}
  {arXiv:1203.6321 [hep-ph]} \BibitemShut {NoStop}%
\bibitem [{\citenamefont {Grazzini}\ and\ \citenamefont
  {Sargsyan}(2013)}]{hres_2}%
  \BibitemOpen
  \bibfield  {author} {\bibinfo {author} {\bibfnamefont {M.}~\bibnamefont
  {Grazzini}}\ and\ \bibinfo {author} {\bibfnamefont {H.}~\bibnamefont
  {Sargsyan}},\ }\href {\doibase 10.1007/JHEP09(2013)129} {\bibfield  {journal}
  {\bibinfo  {journal} {J.~High Energy Phys.}\ }\textbf {\bibinfo {volume}
  {09}},\ \bibinfo {pages} {129} (\bibinfo {year} {2013})},\ \Eprint
  {http://arxiv.org/abs/1306.4581} {arXiv:1306.4581 [hep-ph]} \BibitemShut
  {NoStop}%
\bibitem [{\citenamefont {Boughezal}\ \emph {et~al.}(2014)\citenamefont
  {Boughezal}, \citenamefont {Liu}, \citenamefont {Petriello}, \citenamefont
  {Tackmann},\ and\ \citenamefont {Walsh}}]{Boughezal:2013oha}%
  \BibitemOpen
  \bibfield  {author} {\bibinfo {author} {\bibfnamefont {R.}~\bibnamefont
  {Boughezal}}, \bibinfo {author} {\bibfnamefont {X.}~\bibnamefont {Liu}},
  \bibinfo {author} {\bibfnamefont {F.}~\bibnamefont {Petriello}}, \bibinfo
  {author} {\bibfnamefont {F.~J.}\ \bibnamefont {Tackmann}}, \ and\ \bibinfo
  {author} {\bibfnamefont {J.~R.}\ \bibnamefont {Walsh}},\ }\href {\doibase
  10.1103/PhysRevD.89.074044} {\bibfield  {journal} {\bibinfo  {journal}
  {Phys.\ Rev.\ D}\ }\textbf {\bibinfo {volume} {89}},\ \bibinfo {pages}
  {074044} (\bibinfo {year} {2014})},\ \Eprint {http://arxiv.org/abs/1312.4535}
  {arXiv:1312.4535 [hep-ph]} \BibitemShut {NoStop}%
\bibitem [{\citenamefont {Campbell}\ \emph {et~al.}(2006)\citenamefont
  {Campbell}, \citenamefont {Ellis},\ and\ \citenamefont
  {Zanderighi}}]{minlo_hjj}%
  \BibitemOpen
  \bibfield  {author} {\bibinfo {author} {\bibfnamefont {J.~M.}\ \bibnamefont
  {Campbell}}, \bibinfo {author} {\bibfnamefont {R.~K.}\ \bibnamefont {Ellis}},
  \ and\ \bibinfo {author} {\bibfnamefont {G.}~\bibnamefont {Zanderighi}},\
  }\href {\doibase 10.1088/1126-6708/2006/10/028} {\bibfield  {journal}
  {\bibinfo  {journal} {J.~High Energy Phys.}\ }\textbf {\bibinfo {volume}
  {10}},\ \bibinfo {pages} {028} (\bibinfo {year} {2006})},\ \Eprint
  {http://arxiv.org/abs/hep-ph/0608194} {arXiv:hep-ph/0608194 [hep-ph]}
  \BibitemShut {NoStop}%
\bibitem [{\citenamefont {Ciccolini}\ \emph
  {et~al.}(2007{\natexlab{b}})\citenamefont {Ciccolini}, \citenamefont
  {Denner},\ and\ \citenamefont {Dittmaier}}]{hawk_1}%
  \BibitemOpen
  \bibfield  {author} {\bibinfo {author} {\bibfnamefont {M.}~\bibnamefont
  {Ciccolini}}, \bibinfo {author} {\bibfnamefont {A.}~\bibnamefont {Denner}}, \
  and\ \bibinfo {author} {\bibfnamefont {S.}~\bibnamefont {Dittmaier}},\ }\href
  {\doibase 10.1103/PhysRevLett.99.161803} {\bibfield  {journal} {\bibinfo
  {journal} {Phys.\ Rev.\ Lett.}\ }\textbf {\bibinfo {volume} {99}},\ \bibinfo
  {pages} {161803} (\bibinfo {year} {2007}{\natexlab{b}})},\ \Eprint
  {http://arxiv.org/abs/0707.0381} {arXiv:0707.0381 [hep-ph]} \BibitemShut
  {NoStop}%
\bibitem [{\citenamefont {Ciccolini}\ \emph
  {et~al.}(2008{\natexlab{b}})\citenamefont {Ciccolini}, \citenamefont
  {Denner},\ and\ \citenamefont {Dittmaier}}]{hawk_2}%
  \BibitemOpen
  \bibfield  {author} {\bibinfo {author} {\bibfnamefont {M.}~\bibnamefont
  {Ciccolini}}, \bibinfo {author} {\bibfnamefont {A.}~\bibnamefont {Denner}}, \
  and\ \bibinfo {author} {\bibfnamefont {S.}~\bibnamefont {Dittmaier}},\ }\href
  {\doibase 10.1103/PhysRevD.77.013002} {\bibfield  {journal} {\bibinfo
  {journal} {Phys.\ Rev.\ D}\ }\textbf {\bibinfo {volume} {77}},\ \bibinfo
  {pages} {013002} (\bibinfo {year} {2008}{\natexlab{b}})},\ \Eprint
  {http://arxiv.org/abs/0710.4749} {arXiv:0710.4749 [hep-ph]} \BibitemShut
  {NoStop}%
\bibitem [{\citenamefont {Denner}\ \emph {et~al.}(2012)\citenamefont {Denner},
  \citenamefont {Dittmaier}, \citenamefont {Kallweit},\ and\ \citenamefont
  {Muck}}]{hawk_3}%
  \BibitemOpen
  \bibfield  {author} {\bibinfo {author} {\bibfnamefont {A.}~\bibnamefont
  {Denner}}, \bibinfo {author} {\bibfnamefont {S.}~\bibnamefont {Dittmaier}},
  \bibinfo {author} {\bibfnamefont {S.}~\bibnamefont {Kallweit}}, \ and\
  \bibinfo {author} {\bibfnamefont {A.}~\bibnamefont {Muck}},\ }\href {\doibase
  10.1007/JHEP03(2012)075} {\bibfield  {journal} {\bibinfo  {journal} {J.~High
  Energy Phys.}\ }\textbf {\bibinfo {volume} {03}},\ \bibinfo {pages} {075}
  (\bibinfo {year} {2012})},\ \Eprint {http://arxiv.org/abs/1112.5142}
  {arXiv:1112.5142 [hep-ph]} \BibitemShut {NoStop}%
\bibitem [{\citenamefont {Dawson}\ \emph {et~al.}(2004)\citenamefont {Dawson},
  \citenamefont {Jackson}, \citenamefont {Reina},\ and\ \citenamefont
  {Wackeroth}}]{Dawson:2003kb}%
  \BibitemOpen
  \bibfield  {author} {\bibinfo {author} {\bibfnamefont {S.}~\bibnamefont
  {Dawson}}, \bibinfo {author} {\bibfnamefont {C.}~\bibnamefont {Jackson}},
  \bibinfo {author} {\bibfnamefont {L.}~\bibnamefont {Reina}}, \ and\ \bibinfo
  {author} {\bibfnamefont {D.}~\bibnamefont {Wackeroth}},\ }\href {\doibase
  10.1103/PhysRevD.69.074027} {\bibfield  {journal} {\bibinfo  {journal}
  {Phys.\ Rev.\ D}\ }\textbf {\bibinfo {volume} {69}},\ \bibinfo {pages}
  {074027} (\bibinfo {year} {2004})},\ \Eprint
  {http://arxiv.org/abs/hep-ph/0311067} {arXiv:hep-ph/0311067 [hep-ph]}
  \BibitemShut {NoStop}%
\bibitem [{\citenamefont {Dittmaier}\ \emph {et~al.}(2004)\citenamefont
  {Dittmaier}, \citenamefont {Kramer},\ and\ \citenamefont
  {Spira}}]{Dittmaier:2003ej}%
  \BibitemOpen
  \bibfield  {author} {\bibinfo {author} {\bibfnamefont {S.}~\bibnamefont
  {Dittmaier}}, \bibinfo {author} {\bibfnamefont {M.}~\bibnamefont {Kramer}}, \
  and\ \bibinfo {author} {\bibfnamefont {M.}~\bibnamefont {Spira}},\ }\href
  {\doibase 10.1103/PhysRevD.70.074010} {\bibfield  {journal} {\bibinfo
  {journal} {Phys.\ Rev.\ D}\ }\textbf {\bibinfo {volume} {70}},\ \bibinfo
  {pages} {074010} (\bibinfo {year} {2004})},\ \Eprint
  {http://arxiv.org/abs/hep-ph/0309204} {arXiv:hep-ph/0309204 [hep-ph]}
  \BibitemShut {NoStop}%
\bibitem [{\citenamefont {Dawson}\ \emph {et~al.}(2006)\citenamefont {Dawson},
  \citenamefont {Jackson}, \citenamefont {Reina},\ and\ \citenamefont
  {Wackeroth}}]{Dawson:2005vi}%
  \BibitemOpen
  \bibfield  {author} {\bibinfo {author} {\bibfnamefont {S.}~\bibnamefont
  {Dawson}}, \bibinfo {author} {\bibfnamefont {C.}~\bibnamefont {Jackson}},
  \bibinfo {author} {\bibfnamefont {L.}~\bibnamefont {Reina}}, \ and\ \bibinfo
  {author} {\bibfnamefont {D.}~\bibnamefont {Wackeroth}},\ }\href {\doibase
  10.1142/S0217732306019256} {\bibfield  {journal} {\bibinfo  {journal} {Mod.\
  Phys.\ Lett.\ A}\ }\textbf {\bibinfo {volume} {21}},\ \bibinfo {pages} {89}
  (\bibinfo {year} {2006})},\ \Eprint {http://arxiv.org/abs/hep-ph/0508293}
  {arXiv:hep-ph/0508293 [hep-ph]} \BibitemShut {NoStop}%
\bibitem [{\citenamefont {Harlander}\ and\ \citenamefont
  {Kilgore}(2003)}]{Harlander:2003ai}%
  \BibitemOpen
  \bibfield  {author} {\bibinfo {author} {\bibfnamefont {R.~V.}\ \bibnamefont
  {Harlander}}\ and\ \bibinfo {author} {\bibfnamefont {W.~B.}\ \bibnamefont
  {Kilgore}},\ }\href {\doibase 10.1103/PhysRevD.68.013001} {\bibfield
  {journal} {\bibinfo  {journal} {Phys.\ Rev.\ D}\ }\textbf {\bibinfo {volume}
  {68}},\ \bibinfo {pages} {013001} (\bibinfo {year} {2003})},\ \Eprint
  {http://arxiv.org/abs/hep-ph/0304035} {arXiv:hep-ph/0304035 [hep-ph]}
  \BibitemShut {NoStop}%
\bibitem [{\citenamefont {Harlander}\ \emph {et~al.}(2011)\citenamefont
  {Harlander}, \citenamefont {{Kr\"amer}},\ and\ \citenamefont
  {Schumacher}}]{Harlander:2011aa}%
  \BibitemOpen
  \bibfield  {author} {\bibinfo {author} {\bibfnamefont {R.}~\bibnamefont
  {Harlander}}, \bibinfo {author} {\bibfnamefont {M.}~\bibnamefont
  {{Kr\"amer}}}, \ and\ \bibinfo {author} {\bibfnamefont {M.}~\bibnamefont
  {Schumacher}},\ }\href {http://cdsweb.cern.ch/record/1407669} {\bibfield
  {journal} {\bibinfo  {journal} {CERN-PH-TH/2011-134}\ } (\bibinfo {year}
  {2011})},\ \Eprint {http://arxiv.org/abs/1112.3478} {arXiv:1112.3478
  [hep-ph]} \BibitemShut {NoStop}%
\bibitem [{\citenamefont {{LHC Higgs Cross Section Working Group}}\ \emph
  {et~al.}(2012)\citenamefont {{LHC Higgs Cross Section Working Group}},
  \citenamefont {Dittmaier}, \citenamefont {Mariotti}, \citenamefont
  {Passarino},\ and\ \citenamefont
  {Tanaka~(Eds.)}}]{LHCHiggsCrossSectionWorkingGroup:2012vm}%
  \BibitemOpen
  \bibfield  {author} {\bibinfo {author} {\bibnamefont {{LHC Higgs Cross
  Section Working Group}}}, \bibinfo {author} {\bibfnamefont {S.}~\bibnamefont
  {Dittmaier}}, \bibinfo {author} {\bibfnamefont {C.}~\bibnamefont {Mariotti}},
  \bibinfo {author} {\bibfnamefont {G.}~\bibnamefont {Passarino}}, \ and\
  \bibinfo {author} {\bibfnamefont {R.}~\bibnamefont {Tanaka~(Eds.)}},\
  }\href@noop {} {\bibfield  {journal} {\bibinfo  {journal} {CERN-2012-002}\ }
  (\bibinfo {year} {CERN, Geneva, 2012})},\ \Eprint
  {http://arxiv.org/abs/1201.3084} {arXiv:1201.3084 [hep-ph]} \BibitemShut
  {NoStop}%
\bibitem [{\citenamefont {Gleisberg}\ \emph {et~al.}(2009)\citenamefont
  {Gleisberg} \emph {et~al.}}]{sherpa}%
  \BibitemOpen
  \bibfield  {author} {\bibinfo {author} {\bibfnamefont {T.}~\bibnamefont
  {Gleisberg}} \emph {et~al.},\ }\href {\doibase 10.1088/1126-6708/2009/02/007}
  {\bibfield  {journal} {\bibinfo  {journal} {J.~High Energy Phys.}\ }\textbf
  {\bibinfo {volume} {02}},\ \bibinfo {pages} {007} (\bibinfo {year} {2009})},\
  \Eprint {http://arxiv.org/abs/0811.4622} {arXiv:0811.4622 [hep-ph]}
  \BibitemShut {NoStop}%
\bibitem [{\citenamefont {Hoeche}\ \emph {et~al.}(2010)\citenamefont {Hoeche},
  \citenamefont {Schumann},\ and\ \citenamefont {Siegert}}]{Hoeche:2009xc}%
  \BibitemOpen
  \bibfield  {author} {\bibinfo {author} {\bibfnamefont {S.}~\bibnamefont
  {Hoeche}}, \bibinfo {author} {\bibfnamefont {S.}~\bibnamefont {Schumann}}, \
  and\ \bibinfo {author} {\bibfnamefont {F.}~\bibnamefont {Siegert}},\ }\href
  {\doibase 10.1103/PhysRevD.81.034026} {\bibfield  {journal} {\bibinfo
  {journal} {Phys.\ Rev.\ D}\ }\textbf {\bibinfo {volume} {81}},\ \bibinfo
  {pages} {034026} (\bibinfo {year} {2010})},\ \Eprint
  {http://arxiv.org/abs/0912.3501} {arXiv:0912.3501 [hep-ph]} \BibitemShut
  {NoStop}%
\bibitem [{\citenamefont {{ATLAS
  Collaboration}}(2010{\natexlab{a}})}]{simuAtlas}%
  \BibitemOpen
  \bibfield  {author} {\bibinfo {author} {\bibnamefont {{ATLAS
  Collaboration}}},\ }\href {\doibase 10.1140/epjc/s10052-010-1429-9}
  {\bibfield  {journal} {\bibinfo  {journal} {Eur.\ Phys.\ J.\ C}\ }\textbf
  {\bibinfo {volume} {70}},\ \bibinfo {pages} {823} (\bibinfo {year}
  {2010}{\natexlab{a}})},\ \Eprint {http://arxiv.org/abs/1005.4568}
  {arXiv:1005.4568 [physics.ins-det]} \BibitemShut {NoStop}%
\bibitem [{\citenamefont {Agostinelli}\ \emph {et~al.}(2003)\citenamefont
  {Agostinelli} \emph {et~al.}}]{geant4}%
  \BibitemOpen
  \bibfield  {author} {\bibinfo {author} {\bibfnamefont {S.}~\bibnamefont
  {Agostinelli}} \emph {et~al.} (\bibinfo {collaboration} {{\rm GEANT4}}),\
  }\href {\doibase 10.1016/S0168-9002(03)01368-8} {\bibfield  {journal}
  {\bibinfo  {journal} {Nucl.\ Instrum.\ Methods\ A}\ }\textbf {\bibinfo
  {volume} {506}},\ \bibinfo {pages} {250} (\bibinfo {year}
  {2003})}\BibitemShut {NoStop}%
\bibitem [{\citenamefont {{ATLAS
  Collaboration}}(2011{\natexlab{a}})}]{photonreco}%
  \BibitemOpen
  \bibfield  {author} {\bibinfo {author} {\bibnamefont {{ATLAS
  Collaboration}}},\ }\href {http://cdsweb.cern.ch/record/1345329} {\bibfield
  {journal} {\bibinfo  {journal} {ATL-PHYS-PUB-2011-007}\ } (\bibinfo {year}
  {2011}{\natexlab{a}})},\ \bibinfo {note}
  {\href{http://cdsweb.cern.ch/record/1345329}{http://cdsweb.cern.ch/record/1345329}}\BibitemShut
  {NoStop}%
\bibitem [{\citenamefont {Hoecker}\ \emph {et~al.}(2007)\citenamefont
  {Hoecker}, \citenamefont {Speckmayer}, \citenamefont {Stelzer}, \citenamefont
  {Therhaag}, \citenamefont {von Toerne},\ and\ \citenamefont {Voss}}]{TMVA}%
  \BibitemOpen
  \bibfield  {author} {\bibinfo {author} {\bibfnamefont {A.}~\bibnamefont
  {Hoecker}}, \bibinfo {author} {\bibfnamefont {P.}~\bibnamefont {Speckmayer}},
  \bibinfo {author} {\bibfnamefont {J.}~\bibnamefont {Stelzer}}, \bibinfo
  {author} {\bibfnamefont {J.}~\bibnamefont {Therhaag}}, \bibinfo {author}
  {\bibfnamefont {E.}~\bibnamefont {von Toerne}}, \ and\ \bibinfo {author}
  {\bibfnamefont {H.}~\bibnamefont {Voss}},\ }\href@noop {} {\bibfield
  {journal} {\bibinfo  {journal} {PoS}\ }\textbf {\bibinfo {volume} {ACAT}},\
  \bibinfo {pages} {040} (\bibinfo {year} {2007})},\ \Eprint
  {http://arxiv.org/abs/physics/0703039} {arXiv:physics/0703039} \BibitemShut
  {NoStop}%
\bibitem [{\citenamefont {{ATLAS
  Collaboration}}(2012{\natexlab{d}})}]{PhotonID}%
  \BibitemOpen
  \bibfield  {author} {\bibinfo {author} {\bibnamefont {{ATLAS
  Collaboration}}},\ }\href {http://cdsweb.cern.ch/record/1473426} {\bibfield
  {journal} {\bibinfo  {journal} {ATLAS-CONF-2012-123}\ } (\bibinfo {year}
  {2012}{\natexlab{d}})},\ \bibinfo {note}
  {\href{http://cdsweb.cern.ch/record/1473426}{http://cdsweb.cern.ch/record/1473426}}\BibitemShut
  {NoStop}%
\bibitem [{\citenamefont {Lampl}\ \emph {et~al.}(2008)\citenamefont {Lampl},
  \citenamefont {Laplace}, \citenamefont {Lelas}, \citenamefont {Loch},
  \citenamefont {Ma}, \citenamefont {Menke}, \citenamefont {Rajagopalan},
  \citenamefont {Rousseau}, \citenamefont {Snyder},\ and\ \citenamefont
  {Unal}}]{clustering}%
  \BibitemOpen
  \bibfield  {author} {\bibinfo {author} {\bibfnamefont {W.}~\bibnamefont
  {Lampl}}, \bibinfo {author} {\bibfnamefont {S.}~\bibnamefont {Laplace}},
  \bibinfo {author} {\bibfnamefont {D.}~\bibnamefont {Lelas}}, \bibinfo
  {author} {\bibfnamefont {P.}~\bibnamefont {Loch}}, \bibinfo {author}
  {\bibfnamefont {H.}~\bibnamefont {Ma}}, \bibinfo {author} {\bibfnamefont
  {S.}~\bibnamefont {Menke}}, \bibinfo {author} {\bibfnamefont
  {S.}~\bibnamefont {Rajagopalan}}, \bibinfo {author} {\bibfnamefont
  {D.}~\bibnamefont {Rousseau}}, \bibinfo {author} {\bibfnamefont
  {S.}~\bibnamefont {Snyder}}, \ and\ \bibinfo {author} {\bibfnamefont
  {G.}~\bibnamefont {Unal}},\ }\href {http://cdsweb.cern.ch/record/1099735}
  {\bibfield  {journal} {\bibinfo  {journal} {ATL-LARG-PUB-2008-002}\ }
  (\bibinfo {year} {2008})},\ \bibinfo {note}
  {\href{http://cdsweb.cern.ch/record/1099735}{http://cdsweb.cern.ch/record/1099735}}\BibitemShut
  {NoStop}%
\bibitem [{\citenamefont {Cacciari}\ \emph {et~al.}(2010)\citenamefont
  {Cacciari}, \citenamefont {Salam},\ and\ \citenamefont {Sapeta}}]{Salam}%
  \BibitemOpen
  \bibfield  {author} {\bibinfo {author} {\bibfnamefont {M.}~\bibnamefont
  {Cacciari}}, \bibinfo {author} {\bibfnamefont {G.~P.}\ \bibnamefont {Salam}},
  \ and\ \bibinfo {author} {\bibfnamefont {S.}~\bibnamefont {Sapeta}},\ }\href
  {\doibase {10.1007/JHEP04(2010)065}} {\bibfield  {journal} {\bibinfo
  {journal} {J.~High Energy Phys.}\ }\textbf {\bibinfo {volume} {{04}}},\
  \bibinfo {pages} {{065}} (\bibinfo {year} {2010})}\BibitemShut {NoStop}%
\bibitem [{\citenamefont {{ATLAS
  Collaboration}}(2011{\natexlab{b}})}]{ATLASPromptPhoton}%
  \BibitemOpen
  \bibfield  {author} {\bibinfo {author} {\bibnamefont {{ATLAS
  Collaboration}}},\ }\href {\doibase 10.1103/PhysRevD.83.052005} {\bibfield
  {journal} {\bibinfo  {journal} {Phys.\ Rev.\ D}\ }\textbf {\bibinfo {volume}
  {83}},\ \bibinfo {pages} {052005} (\bibinfo {year} {2011}{\natexlab{b}})},\
  \Eprint {http://arxiv.org/abs/1012.4389} {arXiv:1012.4389 [hep-ex]}
  \BibitemShut {NoStop}%
\bibitem [{\citenamefont {{ATLAS
  Collaboration}}(2010{\natexlab{b}})}]{atlasprimaryvertexperf}%
  \BibitemOpen
  \bibfield  {author} {\bibinfo {author} {\bibnamefont {{ATLAS
  Collaboration}}},\ }\href {http://cdsweb.cern.ch/record/1281344} {\bibfield
  {journal} {\bibinfo  {journal} {ATLAS-CONF-2010-069}\ } (\bibinfo {year}
  {2010}{\natexlab{b}})},\ \bibinfo {note}
  {\href{http://cdsweb.cern.ch/record/1281344}{http://cdsweb.cern.ch/record/1281344}}\BibitemShut
  {NoStop}%
\bibitem [{\citenamefont {{ATLAS
  Collaboration}}(2014{\natexlab{c}})}]{ElectronIDLikelihood}%
  \BibitemOpen
  \bibfield  {author} {\bibinfo {author} {\bibnamefont {{ATLAS
  Collaboration}}},\ }\href {http://cdsweb.cern.ch/record/1706245} {\bibfield
  {journal} {\bibinfo  {journal} {ATLAS-CONF-2014-032}\ } (\bibinfo {year}
  {2014}{\natexlab{c}})},\ \bibinfo {note}
  {\href{http://cdsweb.cern.ch/record/1706245}{http://cdsweb.cern.ch/record/1706245}}\BibitemShut
  {NoStop}%
\bibitem [{\citenamefont {{ATLAS
  Collaboration}}(2014{\natexlab{d}})}]{Aad:2014fxa}%
  \BibitemOpen
  \bibfield  {author} {\bibinfo {author} {\bibnamefont {{ATLAS
  Collaboration}}},\ }\href {\doibase 10.1140/epjc/s10052-014-2941-0}
  {\bibfield  {journal} {\bibinfo  {journal} {Eur.\ Phys.\ J.\ C}\ }\textbf
  {\bibinfo {volume} {74}},\ \bibinfo {pages} {2941} (\bibinfo {year}
  {2014}{\natexlab{d}})},\ \Eprint {http://arxiv.org/abs/1404.2240}
  {arXiv:1404.2240 [hep-ex]} \BibitemShut {NoStop}%
\bibitem [{\citenamefont {{ATLAS
  Collaboration}}(2014{\natexlab{e}})}]{Aad:2014rra}%
  \BibitemOpen
  \bibfield  {author} {\bibinfo {author} {\bibnamefont {{ATLAS
  Collaboration}}},\ }\href@noop {} {\bibfield  {journal} {\bibinfo  {journal}
  {CERN-PH-EP-2014-151}\ } (\bibinfo {year} {2014}{\natexlab{e}})},\ \Eprint
  {http://arxiv.org/abs/1407.3935} {arXiv:1407.3935 [hep-ex]} \BibitemShut
  {NoStop}%
\bibitem [{\citenamefont {Cacciari}\ \emph {et~al.}(2008)\citenamefont
  {Cacciari}, \citenamefont {Salam},\ and\ \citenamefont {Soyez}}]{JetAlgo}%
  \BibitemOpen
  \bibfield  {author} {\bibinfo {author} {\bibfnamefont {M.}~\bibnamefont
  {Cacciari}}, \bibinfo {author} {\bibfnamefont {G.~P.}\ \bibnamefont {Salam}},
  \ and\ \bibinfo {author} {\bibfnamefont {G.}~\bibnamefont {Soyez}},\ }\href
  {\doibase 10.1088/1126-6708/2008/04/063} {\bibfield  {journal} {\bibinfo
  {journal} {J.~High Energy Phys.}\ }\textbf {\bibinfo {volume} {04}},\
  \bibinfo {pages} {063} (\bibinfo {year} {2008})},\ \Eprint
  {http://arxiv.org/abs/0802.1189} {arXiv:0802.1189 [hep-ph]} \BibitemShut
  {NoStop}%
\bibitem [{\citenamefont {{ATLAS
  Collaboration}}(2014{\natexlab{f}})}]{jetcal_2011}%
  \BibitemOpen
  \bibfield  {author} {\bibinfo {author} {\bibnamefont {{ATLAS
  Collaboration}}},\ }\href@noop {} {\bibfield  {journal} {\bibinfo  {journal}
  {CERN-PH-EP-2013-222}\ } (\bibinfo {year} {2014}{\natexlab{f}})},\ \Eprint
  {http://arxiv.org/abs/1406.0076} {arXiv:1406.0076 [hep-ex]} \BibitemShut
  {NoStop}%
\bibitem [{\citenamefont {Cacciari}\ and\ \citenamefont
  {Salam}(2008)}]{Cacciari:2007fd}%
  \BibitemOpen
  \bibfield  {author} {\bibinfo {author} {\bibfnamefont {M.}~\bibnamefont
  {Cacciari}}\ and\ \bibinfo {author} {\bibfnamefont {G.~P.}\ \bibnamefont
  {Salam}},\ }\href {\doibase 10.1016/j.physletb.2007.09.077} {\bibfield
  {journal} {\bibinfo  {journal} {Phys.\ Lett.\ B}\ }\textbf {\bibinfo {volume}
  {659}},\ \bibinfo {pages} {119} (\bibinfo {year} {2008})},\ \Eprint
  {http://arxiv.org/abs/0707.1378} {arXiv:0707.1378 [hep-ph]} \BibitemShut
  {NoStop}%
\bibitem [{\citenamefont {{ATLAS
  Collaboration}}(2013{\natexlab{c}})}]{jetPileupCorr}%
  \BibitemOpen
  \bibfield  {author} {\bibinfo {author} {\bibnamefont {{ATLAS
  Collaboration}}},\ }\href {http://cdsweb.cern.ch/record/1570994} {\bibfield
  {journal} {\bibinfo  {journal} {ATLAS-CONF-2013-083}\ } (\bibinfo {year}
  {2013}{\natexlab{c}})},\ \bibinfo {note}
  {\href{http://cdsweb.cern.ch/record/1570994}{http://cdsweb.cern.ch/record/1570994}}\BibitemShut
  {NoStop}%
\bibitem [{\citenamefont {{ATLAS
  Collaboration}}(2011{\natexlab{c}})}]{ATLAS-CONF-2011-102}%
  \BibitemOpen
  \bibfield  {author} {\bibinfo {author} {\bibnamefont {{ATLAS
  Collaboration}}},\ }\href {http://cdsweb.cern.ch/record/1369219} {\bibfield
  {journal} {\bibinfo  {journal} {ATLAS-CONF-2011-102}\ } (\bibinfo {year}
  {2011}{\natexlab{c}})},\ \bibinfo {note}
  {\href{http://cdsweb.cern.ch/record/1369219}{http://cdsweb.cern.ch/record/1369219}}\BibitemShut
  {NoStop}%
\bibitem [{\citenamefont {{ATLAS
  Collaboration}}(2014{\natexlab{g}})}]{ATLAS-CONF-2014-046}%
  \BibitemOpen
  \bibfield  {author} {\bibinfo {author} {\bibnamefont {{ATLAS Collaboration}}}
  (\bibinfo {collaboration} {ATLAS}),\ }\href
  {http://cdsweb.cern.ch/record/1741020} {\bibfield  {journal} {\bibinfo
  {journal} {ATLAS-CONF-2014-046}\ } (\bibinfo {year} {2014}{\natexlab{g}})},\
  \bibinfo {note}
  {\href{http://cdsweb.cern.ch/record/1741020}{http://cdsweb.cern.ch/record/1741020}}\BibitemShut
  {NoStop}%
\bibitem [{\citenamefont {{ATLAS
  Collaboration}}(2013{\natexlab{d}})}]{met_perf}%
  \BibitemOpen
  \bibfield  {author} {\bibinfo {author} {\bibnamefont {{ATLAS
  Collaboration}}},\ }\href {http://cdsweb.cern.ch/record/1570993} {\bibfield
  {journal} {\bibinfo  {journal} {ATLAS-CONF-2013-082}\ } (\bibinfo {year}
  {2013}{\natexlab{d}})},\ \bibinfo {note}
  {\href{http://cdsweb.cern.ch/record/1570993}{http://cdsweb.cern.ch/record/1570993}}\BibitemShut
  {NoStop}%
\bibitem [{\citenamefont {{ATLAS
  Collaboration}}(2012{\natexlab{e}})}]{ATLAS:2012wna}%
  \BibitemOpen
  \bibfield  {author} {\bibinfo {author} {\bibnamefont {{ATLAS
  Collaboration}}},\ }\href {http://cdsweb.cern.ch/record/1463915} {\bibfield
  {journal} {\bibinfo  {journal} {ATLAS-CONF-2012-101}\ } (\bibinfo {year}
  {2012}{\natexlab{e}})},\ \bibinfo {note}
  {\href{http://cdsweb.cern.ch/record/1463915}{http://cdsweb.cern.ch/record/1463915}}\BibitemShut
  {NoStop}%
\bibitem [{\citenamefont {{ATLAS
  Collaboration}}(2014{\natexlab{h}})}]{ttHpaper}%
  \BibitemOpen
  \bibfield  {author} {\bibinfo {author} {\bibnamefont {{ATLAS
  Collaboration}}},\ }\href@noop {} {\bibfield  {journal} {\bibinfo  {journal}
  {to be submitted to Phys.\ Lett.\ B.}\ } (\bibinfo {year}
  {2014}{\natexlab{h}})}\BibitemShut {NoStop}%
\bibitem [{pTt()}]{pTtendnote}%
  \BibitemOpen
  \href@noop {} {}\bibinfo {note} {{The quantity \pTt\ is defined as
  $p_{\rm{Tt}} = |(\vec{p}_\mathrm{T}^{\gamma_1} +
  \vec{p}_\mathrm{T}^{\gamma_2}) \times \hat{\bf t}|$, where $\hat{\bf t} = {
  {(\vec{p}_\mathrm{T}^{\gamma_1}} -{\vec{p}_\mathrm{T}^{\gamma_2}})}/{
  |{\vec{p}_\mathrm{T}^{\gamma_1}} - {\vec{p}_\mathrm{T}^{\gamma_2}}|}$ denotes
  the thrust axis in the transverse plane, and
  ${\vec{p}_\mathrm{T}^{\gamma_1}}$, ${\vec{p}_\mathrm{T}^{\gamma_2}}$ are the
  transverse momenta of two photons $\gamma_1$ and $\gamma_2$.}}\BibitemShut
  {Stop}%
\bibitem [{\citenamefont {{Rainwater}}\ \emph {et~al.}(1996)\citenamefont
  {{Rainwater}}, \citenamefont {{Szalapski}},\ and\ \citenamefont
  {{Zeppenfeld}}}]{1996PhRvD..54.6680R}%
  \BibitemOpen
  \bibfield  {author} {\bibinfo {author} {\bibfnamefont {D.}~\bibnamefont
  {{Rainwater}}}, \bibinfo {author} {\bibfnamefont {R.}~\bibnamefont
  {{Szalapski}}}, \ and\ \bibinfo {author} {\bibfnamefont {D.}~\bibnamefont
  {{Zeppenfeld}}},\ }\href {\doibase 10.1103/PhysRevD.54.6680} {\bibfield
  {journal} {\bibinfo  {journal} {Phys.\ Rev.\ D}\ }\textbf {\bibinfo {volume}
  {54}},\ \bibinfo {pages} {6680} (\bibinfo {year} {1996})},\ \Eprint
  {http://arxiv.org/abs/hep-ph/9605444} {arXiv:hep-ph/9605444} \BibitemShut
  {NoStop}%
\bibitem [{\citenamefont {Oreglia}(1980)}]{crystalball}%
  \BibitemOpen
  \bibfield  {author} {\bibinfo {author} {\bibfnamefont {M.}~\bibnamefont
  {Oreglia}},\ }\href@noop {} {\bibfield  {journal} {\bibinfo  {journal}
  {SLAC-R-0236}\ ,\ \bibinfo {pages} {Appendix D}} (\bibinfo {year}
  {1980})}\BibitemShut {NoStop}%
\bibitem [{\citenamefont {{ATLAS
  Collaboration}}(2012{\natexlab{f}})}]{Aad:2011mh}%
  \BibitemOpen
  \bibfield  {author} {\bibinfo {author} {\bibnamefont {{ATLAS
  Collaboration}}},\ }\href {\doibase 10.1103/PhysRevD.85.012003} {\bibfield
  {journal} {\bibinfo  {journal} {Phys.\ Rev.\ D}\ }\textbf {\bibinfo {volume}
  {85}},\ \bibinfo {pages} {012003} (\bibinfo {year} {2012}{\natexlab{f}})},\
  \Eprint {http://arxiv.org/abs/1107.0581} {arXiv:1107.0581 [hep-ex]}
  \BibitemShut {NoStop}%
\bibitem [{\citenamefont {{ATLAS
  Collaboration}}(2013{\natexlab{e}})}]{diphoton2011}%
  \BibitemOpen
  \bibfield  {author} {\bibinfo {author} {\bibnamefont {{ATLAS
  Collaboration}}},\ }\href {\doibase 10.1007/JHEP01(2013)086} {\bibfield
  {journal} {\bibinfo  {journal} {J.~High Energy Phys.}\ }\textbf {\bibinfo
  {volume} {01}},\ \bibinfo {pages} {086} (\bibinfo {year}
  {2013}{\natexlab{e}})},\ \Eprint {http://arxiv.org/abs/1211.1913}
  {arXiv:1211.1913 [hep-ex]} \BibitemShut {NoStop}%
\bibitem [{\citenamefont {{Cowan}}\ \emph {et~al.}(2011)\citenamefont
  {{Cowan}}, \citenamefont {{Cranmer}}, \citenamefont {{Gross}},\ and\
  \citenamefont {{Vitells}}}]{stat}%
  \BibitemOpen
  \bibfield  {author} {\bibinfo {author} {\bibfnamefont {G.}~\bibnamefont
  {{Cowan}}}, \bibinfo {author} {\bibfnamefont {K.}~\bibnamefont {{Cranmer}}},
  \bibinfo {author} {\bibfnamefont {E.}~\bibnamefont {{Gross}}}, \ and\
  \bibinfo {author} {\bibfnamefont {O.}~\bibnamefont {{Vitells}}},\ }\href
  {\doibase 10.1140/epjc/s10052-011-1554-0} {\bibfield  {journal} {\bibinfo
  {journal} {European Physical Journal C}\ }\textbf {\bibinfo {volume} {71}},\
  \bibinfo {pages} {1554} (\bibinfo {year} {2011})},\ \Eprint
  {http://arxiv.org/abs/1007.1727} {arXiv:1007.1727 [physics.data-an]}
  \BibitemShut {NoStop}%
\bibitem [{\citenamefont {Bernstein}(1912)}]{Bernstein}%
  \BibitemOpen
  \bibfield  {author} {\bibinfo {author} {\bibfnamefont {S.~N.}\ \bibnamefont
  {Bernstein}},\ }\href@noop {} {\bibfield  {journal} {\bibinfo  {journal}
  {Comm. Soc. Math. Kharkov}\ }\textbf {\bibinfo {volume} {13}},\ \bibinfo
  {pages} {1} (\bibinfo {year} {1912})}\BibitemShut {NoStop}%
\bibitem [{\citenamefont {{ATLAS Collaboration}}(2013{\natexlab{f}})}]{lumi}%
  \BibitemOpen
  \bibfield  {author} {\bibinfo {author} {\bibnamefont {{ATLAS
  Collaboration}}},\ }\href {\doibase 10.1140/epjc/s10052-013-2518-3}
  {\bibfield  {journal} {\bibinfo  {journal} {Eur.\ Phys.\ J.\ C}\ }\textbf
  {\bibinfo {volume} {73}},\ \bibinfo {pages} {2518} (\bibinfo {year}
  {2013}{\natexlab{f}})},\ \Eprint {http://arxiv.org/abs/1302.4393}
  {arXiv:1302.4393 [hep-ex]} \BibitemShut {NoStop}%
\bibitem [{\citenamefont {{ATLAS
  Collaboration}}(2012{\natexlab{g}})}]{Aad:2012xs}%
  \BibitemOpen
  \bibfield  {author} {\bibinfo {author} {\bibnamefont {{ATLAS
  Collaboration}}},\ }\href {\doibase 10.1140/epjc/s10052-011-1849-1}
  {\bibfield  {journal} {\bibinfo  {journal} {Eur.\ Phys.\ J.\ C}\ }\textbf
  {\bibinfo {volume} {72}},\ \bibinfo {pages} {1849} (\bibinfo {year}
  {2012}{\natexlab{g}})},\ \Eprint {http://arxiv.org/abs/1110.1530}
  {arXiv:1110.1530 [hep-ex]} \BibitemShut {NoStop}%
\bibitem [{\citenamefont {Stewart}\ and\ \citenamefont
  {Tackmann}(2012)}]{PhysRevD.85.034011}%
  \BibitemOpen
  \bibfield  {author} {\bibinfo {author} {\bibfnamefont {I.~W.}\ \bibnamefont
  {Stewart}}\ and\ \bibinfo {author} {\bibfnamefont {F.~J.}\ \bibnamefont
  {Tackmann}},\ }\href {\doibase 10.1103/PhysRevD.85.034011} {\bibfield
  {journal} {\bibinfo  {journal} {Phys.\ Rev.\ D}\ }\textbf {\bibinfo {volume}
  {85}},\ \bibinfo {pages} {034011} (\bibinfo {year} {2012})},\ \Eprint
  {http://arxiv.org/abs/1107.2117} {arXiv:1107.2117 [hep-ph]} \BibitemShut
  {NoStop}%
\bibitem [{\citenamefont {Gangal}\ and\ \citenamefont
  {Tackmann}(2013)}]{STmethod}%
  \BibitemOpen
  \bibfield  {author} {\bibinfo {author} {\bibfnamefont {S.}~\bibnamefont
  {Gangal}}\ and\ \bibinfo {author} {\bibfnamefont {F.~J.}\ \bibnamefont
  {Tackmann}},\ }\href {\doibase 10.1103/PhysRevD.87.093008} {\bibfield
  {journal} {\bibinfo  {journal} {Phys.\ Rev.\ D}\ }\textbf {\bibinfo {volume}
  {87}},\ \bibinfo {pages} {093008} (\bibinfo {year} {2013})},\ \Eprint
  {http://arxiv.org/abs/1302.5437} {arXiv:1302.5437 [hep-ph]} \BibitemShut
  {NoStop}%
\bibitem [{\citenamefont {Campbell}\ \emph {et~al.}(2010)\citenamefont
  {Campbell}, \citenamefont {Ellis},\ and\ \citenamefont {Williams}}]{MCFM}%
  \BibitemOpen
  \bibfield  {author} {\bibinfo {author} {\bibfnamefont {J.~M.}\ \bibnamefont
  {Campbell}}, \bibinfo {author} {\bibfnamefont {R.~K.}\ \bibnamefont {Ellis}},
  \ and\ \bibinfo {author} {\bibfnamefont {C.}~\bibnamefont {Williams}},\
  }\href {\doibase 10.1103/PhysRevD.81.074023} {\bibfield  {journal} {\bibinfo
  {journal} {Phys.\ Rev.\ D}\ }\textbf {\bibinfo {volume} {81}},\ \bibinfo
  {pages} {074023} (\bibinfo {year} {2010})},\ \Eprint
  {http://arxiv.org/abs/1001.4495} {arXiv:1001.4495 [hep-ph]} \BibitemShut
  {NoStop}%
\bibitem [{\citenamefont {{ATLAS
  Collaboration}}(2010{\natexlab{c}})}]{jet_reco}%
  \BibitemOpen
  \bibfield  {author} {\bibinfo {author} {\bibnamefont {{ATLAS
  Collaboration}}},\ }\href {http://cdsweb.cern.ch/record/1281310} {\bibfield
  {journal} {\bibinfo  {journal} {ATLAS-CONF-2010-053}\ } (\bibinfo {year}
  {2010}{\natexlab{c}})},\ \bibinfo {note}
  {\href{http://cdsweb.cern.ch/record/1281310}{http://cdsweb.cern.ch/record/1281310}}\BibitemShut
  {NoStop}%
\bibitem [{\citenamefont {{ATLAS
  Collaboration}}(2012{\natexlab{h}})}]{Aad:2012an}%
  \BibitemOpen
  \bibfield  {author} {\bibinfo {author} {\bibnamefont {{ATLAS
  Collaboration}}},\ }\href {\doibase 10.1103/PhysRevD.86.032003} {\bibfield
  {journal} {\bibinfo  {journal} {Phys.\ Rev.\ D}\ }\textbf {\bibinfo {volume}
  {86}},\ \bibinfo {pages} {032003} (\bibinfo {year} {2012}{\natexlab{h}})},\
  \Eprint {http://arxiv.org/abs/1207.0319} {arXiv:1207.0319 [hep-ex]}
  \BibitemShut {NoStop}%
\bibitem [{\citenamefont {{M.\ H.\ Quenouille}}(1949)}]{jk1}%
  \BibitemOpen
  \bibfield  {author} {\bibinfo {author} {\bibnamefont {{M.\ H.\
  Quenouille}}},\ }\href@noop {} {\bibfield  {journal} {\bibinfo  {journal}
  {{Journal of the Royal Statistical Society. Series B (Methodological)}}\
  }\textbf {\bibinfo {volume} {{11}}},\ \bibinfo {pages} {68} (\bibinfo {year}
  {{1949}})}\BibitemShut {NoStop}%
\bibitem [{\citenamefont {{J. W. Tukey}}()}]{jk2}%
  \BibitemOpen
  \bibfield  {author} {\bibinfo {author} {\bibnamefont {{J. W. Tukey}}},\
  }\href@noop {} {\bibinfo  {journal} {{Annals of Math. Statist. 29 (1958)
  614}}\ }\BibitemShut {NoStop}%
\bibitem [{\citenamefont {{ATLAS
  Collaboration}}(2014{\natexlab{i}})}]{ggfiducial_xs}%
  \BibitemOpen
\bibfield  {journal} {  }\bibfield  {author} {\bibinfo {author} {\bibnamefont
  {{ATLAS Collaboration}}},\ }\href@noop {} {\bibfield  {journal} {\bibinfo
  {journal} {CERN-PH-EP-2014-148}\ } (\bibinfo {year} {2014}{\natexlab{i}})},\
  \bibinfo {note} {submitted to J.~High Energy Phys.},\ \Eprint
  {http://arxiv.org/abs/1407.4222} {arXiv:1407.4222 [hep-ex]} \BibitemShut
  {NoStop}%
\end{thebibliography}

\clearpage

\begin{widetext}
% ATLAS Collaboration author list
% Data extracted on 27-Aug-2014 for paper reference HIGG-2013-08
%\documentclass[11pt]{article}
%\usepackage{a4wide}\begin{document}
\begin{flushleft}
{\Large The ATLAS Collaboration}

\bigskip

G.~Aad$^{\rm 85}$,
B.~Abbott$^{\rm 113}$,
J.~Abdallah$^{\rm 153}$,
S.~Abdel~Khalek$^{\rm 117}$,
O.~Abdinov$^{\rm 11}$,
R.~Aben$^{\rm 107}$,
B.~Abi$^{\rm 114}$,
M.~Abolins$^{\rm 90}$,
O.S.~AbouZeid$^{\rm 160}$,
H.~Abramowicz$^{\rm 155}$
% ...
\bigskip
\\
$^{1}$ Department of Physics, University of Adelaide, Adelaide, Australia\\
% ...
$^{11}$ Institute of Physics, Azerbaijan Academy of Sciences, Baku, Azerbaijan\\
% ...
$^{85}$ CPPM, Aix-Marseille Universit{\'e} and CNRS/IN2P3, Marseille, France\\
% ...
$^{90}$ Department of Physics and Astronomy, Michigan State University, East Lansing MI, United States of America\\
% ...
$^{107}$ Nikhef National Institute for Subatomic Physics and University of Amsterdam, Amsterdam, Netherlands\\
% ...
$^{113}$ Homer L. Dodge Department of Physics and Astronomy, University of Oklahoma, Norman OK, United States of America\\
$^{114}$ Department of Physics, Oklahoma State University, Stillwater OK, United States of America\\
% ...
$^{117}$ LAL, Universit{\'e} Paris-Sud and CNRS/IN2P3, Orsay, France\\
% ...
$^{153}$ Institute of Physics, Academia Sinica, Taipei, Taiwan\\
% ...
$^{155}$ Raymond and Beverly Sackler School of Physics and Astronomy, Tel Aviv University, Tel Aviv, Israel\\
% ...
$^{160}$ Department of Physics, University of Toronto, Toronto ON, Canada\\
% ...
\end{flushleft}

%\end{document}
% Created with xml2latex.py

\end{widetext}

\end{document}